\documentclass{article}
\usepackage{amssymb}
\usepackage{amsmath}
\usepackage{amsthm}
\usepackage{amsfonts}
\usepackage{bm}
\usepackage{natbib}
\usepackage{dsfont}
\usepackage{url}
\usepackage{mathrsfs}
\usepackage{array}
\newcolumntype{P}[1]{>{\centering\arraybackslash}p{#1}}
\usepackage[margin=2.5cm]{geometry}
\usepackage{setspace}
\usepackage{comment}
\usepackage{graphicx}
\usepackage{tikz}
\usetikzlibrary{matrix,arrows}
\usepackage{xcolor}
\usepackage{hyperref}
\usepackage[colorinlistoftodos,prependcaption,textsize=tiny]{todonotes}
\usepackage{listings}
\definecolor{links}{RGB}{0,0,128}

\hypersetup{
    unicode=true,          %
    pdftitle={Smooth bootstrapping of copula functionals},    %
    pdfauthor={Maximilian Coblenz, Oliver Grothe, Klaus Herrmannm, Marius Hofert},     %
    colorlinks=true,       %
    linkcolor=links,          %
    citecolor=links,        %
    filecolor=links,      %
    urlcolor=links           %
}

\def\P{\mathbb{P}}

\def\bU{\bm{U}}
\def\bX{\bm{X}}
\def\bY{\bm{Y}}
\def\bZ{\bm{Z}}

\def\bB{\bm{B}}
\def\ba{\bm{a}}
\def\bb{\bm{b}}
\def\bd{\bm{d}}
\def\bz{\bm{z}}
\def\by{\bm{y}}

\def\E{\mathbb{E}}

\def\bu{\bm{u}}
\def\bx{\bm{x}}
\def\bZero{\bm{0}}
\DeclareMathOperator{\diag}{diag}
\def\bMu{\bm{\mu}}
\def\corrMat{\bm{R}}
\def\dispMat{\bm{\Sigma}}

\def\covMat{\bm{S}}
\def\empcovMat{\widehat{\covMat}}

\def\bt{\bm{t}}
\def\Spherical{S}
\def\Ellip{E}
\def\charGen{\psi}
\def\dGen{g}
\def\dim{d}
\def\Bessel{K}
\def\Copula{\bm{C}}

\def\CopulaCondKDE{\widehat{\Copula}^{|\sample}_{\norig}}

\def\kernel{\mathbf{k}}
\def\Kernel{\mathbf{K}}

\def\BWMatrix{H}

\def\BW{h}

\def\R{\mathbb{R}}
\def\C{\mathbb{C}}
\def\binfty{\bm{\infty}}

\def\JointDist{\bm{F}}
\def\JointDens{\bm{f}}

\def\kde{\widehat{\JointDens}}
\def\KDE{\widehat{\JointDist}}
\def\KDECop{\widehat{\Copula}}

\def\KDET{\widetilde{\JointDist}}

\def\Normal{\mathcal{N}}
\def\IdMat{\bm{I}}

\def\sample{\mathbb{X}}
\def\samplebs{\sample^*}
\def\samplesbs{\mathbb{S}^*}

\newcommand{\Eval}[1]{\mathbb{E}\left[ #1 \right]}
\newcommand{\Prob}[1]{\mathbb{P}\left[ #1 \right]}
\DeclareMathOperator{\covv}{cov}
\DeclareMathOperator{\corrr}{corr}
\newcommand{\cov}[1]{\covv\left[ #1 \right]}
\newcommand{\corr}[1]{\corrr\left[ #1 \right]}

\DeclareMathOperator{\varr}{var}
\newcommand{\var}[1]{\varr\left[ #1 \right]}

\newcommand{\ind}[1]{\mathds{1}_{#1}}
\newcommand{\xNorm}[2]{\left\Vert #1 \right\Vert_{#2}}

\DeclareMathOperator{\tr}{\top}
\newtheorem{Proposition}{Proposition}[section]
\newtheorem{Theorem}{Theorem}[section]
\newtheorem{Definition}{Definition}[section]

\newtheorem{Corollary}{Corollary}[section]

\newtheorem{Algorithm}{Algorithm}[section]
\newtheorem{Remark}{Remark}[section]
\newtheorem{Example}{Example}[section]

\def\inprob{\stackrel{\P}{\rightarrow}}
\def\equalindist{\stackrel{d}{=}}
\DeclareMathOperator{\as}{a.s.}
\def\almostsurly{\stackrel{\as}{\rightarrow}}

\def\conv{*}

\def\nbs{B}
\def\norig{n}
\def\naugm{m}
\def\dirac{\delta}

\def\mdensI{f}
\def\mdistI{F}
\def\mdensII{g}
\def\mdistII{G}
\def\densI{\bm{f}}
\def\distI{\bm{F}}
\def\densII{\bm{g}}
\def\distII{\bm{G}}

\def\kdeMargDist{\widehat{F}}

\def\func{T}
\def\funcbs{\func^{*}}

\def\SigAlg{\mathscr{F}}
\DeclareMathOperator{\trace}{tr}

\def\expkde{\kde_{\norig}^{\E}}
\def\Expkde{\KDE_{\norig}^{\E}}

\def\condkde{\kde_{\norig}^{|\sample}}
\def\condKDE{\KDE_{\norig}^{|\sample}}

\def\ExpCop{\widehat{\Copula}^{\E}_{\norig}}

\def\Hesse{\mathcal{H}}

\def\Landau{O}%
\def\landau{o}

\definecolor{theory}{RGB}{150, 150, 150}
\definecolor{observed}{RGB}{0, 0, 0}

\newcommand{\abs}[1]{\left|#1\right|}

\DeclareMathOperator{\range}{range}
\DeclareMathOperator{\rank}{rank}
\def\CF{\phi}

\def\radialPart{R}
\def\sphericalPart{\bm{S}}
\DeclareMathOperator{\sd}{sd}

\def\ginv{-1}

\DeclareMathOperator{\metricd}{d}
\newcommand{\metric}[2]{\metricd\left( #1, #2\right)}
\DeclareMathOperator{\KS}{KS}
\def\dKS{\metricd_{\KS}}

\DeclareMathOperator{\Hausdorff}{H}
\DeclareMathOperator{\TV}{TV}
\newcommand{\dH}[2]{\metricd_{\Hausdorff}\left( #1, #2\right)}
\newcommand{\dTV}[2]{\metricd_{\TV}\left( #1, #2\right)}

\DeclareMathOperator{\Spearman}{S}
\def\SpearmansRho{\rho_{\Spearman}}

\def\imu{\mathrm{i}}
\def\d{\,\mathrm{d}}

\def\df{\nu}

\def\abbrData{\bX}
\def\abbrKernel{\bY}
\def\abbrSmooth{\bZ}

\def\Error{E}

\def\LeaveOneOut{\KDE_{-i}}
\DeclareMathOperator{\CV}{CV}
\DeclareMathOperator{\MISE}{WISE}

\def\({\left(}
\def\){\right)}

\def\weightFunction{w}
\def\intConst{D}

\def\LevelSet{\mathbb{L}}

\newcommand{\footremember}[2]{%
    \footnote{#2}
    \newcounter{#1}
    \setcounter{#1}{\value{footnote}}%
}

\begin{document}

\title{Smooth bootstrapping of copula functionals}
\author{Maximilian Coblenz\footremember{LW}{Department of Services and Consulting, Ludwigshafen University of Business and Society, Ludwigshafen, Germany}, Oliver Grothe\footremember{KIT}{Institute of Operations Research, Karlsruhe Institute of Technology, Karlsruhe, Germany}, Klaus Herrmann\footremember{UdeS}{D\'{e}partement de math\'{e}matiques, Universit\'{e} de Sherbrooke, Sherbrooke, Canada}, Marius Hofert\footremember{UW}{Department of Statistics and Actuarial Science, University of Waterloo, Waterloo, Canada}}
\maketitle

\abstract{
  The smooth bootstrap for estimating copula functionals in small samples is
  investigated. It can be used both to gauge the distribution of the estimator
  in question and to augment the data. Issues arising from kernel density and
  distribution estimation in the copula domain are addressed, such as how to
  avoid the bounded domain, which bandwidth matrix to choose, and how the
  smoothing can be carried out. Furthermore, we investigate how the smooth
  bootstrap impacts the underlying dependence structure or the functionals in
  question and under which conditions it does not.  We provide specific examples
  and simulations that highlight advantages and caveats of the approach.
}


\section{Introduction and notation}
Contrary to resampling from the observed values, resampling in the smooth bootstrap introduced in \cite{Efron1982} is done from a smoothed version of the empirical distribution function.
We consider the multivariate smooth bootstrap for
functionals $\func$ defined on the set (or possibly only a subset) of copulas which represent the possible dependence structures at hand.
Examples for such functionals include measures of association such as Kendall's
tau or Spearman's rho, the upper and lower tail dependence coefficients, or
level sets that are used to quantify the risk inherent in joint events; see, for
example, \cite{SalvadoriDuranteDeMicheleBernardiPetrella2016}, and \cite{CoblenzDyckerhoffGrothe2017} where in the latter the smooth bootstrap has recently been used to estimate level sets of copulas.

Our investigation is motivated by the question of how much the smoothing aspect of
smooth bootstrap influences the underlying dependence structure in a
multivariate framework. The whole procedure can also be seen as a form of data augmentation -- a topic which recently has drawn considerable attention in the machine learning community, see, e.g., \cite{Shorten2019}, \cite{Wong2016}, and \cite{Taylor2018}.
In general the extent of this \emph{dependence distortion} introduced by the smooth bootstrap may
depend on the functional, the smoothing kernel or the sample size.

The smooth bootstrap used in this paper is different from jittering as, e.g., investigated in \cite{Nagler2018}.
Usually, jittering is used for making discrete variables continuous and is not a resampling scheme in the classical sense.
In addition to that, jittering is carried out in the data space, whereas we employ the smooth bootstrap in a transformed data space and focus specifically on the dependence structure, cf. Figure~\ref{fig_schematicRelationships}.

Furthermore, we want to point out that there are links to other empirical versions of the copula such as the empirical beta copula, see \cite{SegersSibuyaTsukahara2017}, and the checkerboard copula.
In contrast to the empirical copula, a sample from the latter two is not a simple resampling of the initial observations, but produces previously unseen observations.
However, the sampling schemes are different from the resampling scheme employed in this paper. While the empirical beta copula smoothes based on the distribution of rank orders, the smoothing considered here relies on adding a smoothing kernel from a given family.
Further, the 'smoothing' implied by drawing from the checkerboard copula is related to latin hypercube sampling, see \cite{PackhamSchmidt2010}, and thus also different in basic principle from the kernel based smoothing analyzed here.

We contribute to the existing literature in the following ways.
Firstly, we provide theoretical details on the smooth bootstrap for copula functionals.
We focus on elliptical distributions at the population level where the bandwidth matrices are obtained by the commonly used sphering approach.
In this framework we show that the dependence distortion of the underlying elliptical copula is solely due to a distortion of the associated characteristic generator, whereas the associated parameter matrix remains unaffected.
Our investigation reveals surprising cases where kernel smoothing has no impact on the underlying dependence structure and we identify and characterize the responsible mechanism in detail.
We also discuss that this fact holds true for certain functionals of copulas which remain unaffected by kernel smoothing.
Concerning the rate of convergence in our framework, we connect the absolute value difference between the characteristic functions with the regular variation property of the characteristic generator of the smoothing kernel and we discuss related examples and practical implications. While most motivating
examples are bivariate in nature, our discussion is valid in arbitrary
dimensions $\dim$ making the results viable for high-dimensional settings and
data science applications in general.

Secondly, when considering a fixed sample we provide practical details concerning the evaluation of the copula of the smoothed distribution estimator, where our approach is not limited to product kernels when elliptical smoothing kernels are used.
Furthermore, in Algorithm~\ref{algo_smoothed_copula_bootstrap} we discuss a sampling strategy for constructing smooth bootstrap samples. The resulting smooth bootstrap sample can then be used to gauge the variability of a functional $\func_{\norig}(\bx_1,\ldots,\bx_{\norig})$ based on a sample of size $\norig$.
Furthermore, as in \cite{CoblenzDyckerhoffGrothe2017}, the suggested algorithm can be used for data augmentation, for example, to obtain smoothed results in order to facilitate the numerical computation of $\func$ or to circumvent problems with ties in the original or an alternative (ordinary) bootstrap sample.
Data augmentation in this sense is particularly valuable for small sample sizes.

Thirdly, we generalise the bandwidth selection procedure of \cite{BowmanHallPrvan1998} for kernel distribution estimation to the multivariate case.
While bandwidth selection is a crucial part of kernel distribution estimation it is generally discussed for diagonal bandwidth matrices, see, for example, \cite{LiuYang2008}.
We introduce a cross-validation-based bandwidth matrix selection procedure that is not limited to diagonal bandwidth matrices or product kernels to overcome the aforementioned limitations and can select non-diagonal bandwidth matrices in a data driven and optimal way.

Lastly we illustrate the smooth bootstrap for copula functionals with a number of theoretical examples and simulation studies.

The remainder of this paper is structured as follows: In the next section we introduce our notation and provide necessary preliminaries on copulas and multivariate kernel smoothing. Additionally, we give an overview of the smooth bootstrap setting and the estimators involved. Section~\ref{section_kernel_smoothing} comprises our theoretical considerations. We start in  \ref{section_population_version} by reviewing basic facts about the connection of multivariate kernel smoothing to convolution and thus sums of independent random vectors.
Section~\ref{section_dependence_distortion_elliptical} introduces dependence distortion at the population level for elliptical distributions and smoothing kernels.
Aside from deriving conditions under which the dependence distortion can be quantified precisely, we also discuss examples of functionals that do not exhibit dependence distortion.
In Section~\ref{section_convergence_rates} we link our results to regular variation properties of the characteristic generator associated with the elliptical kernel distribution.
We discuss convergence rates of the distorted characteristic function towards the undistorted characteristic function and hence dependence structure in terms of regular variation of the radial distribution associated with the smoothing kernel.
Section~\ref{section_exceptions} identifies special classes of distributions for which the dependence structure is unaffected by smoothing at the population level.
To complete the picture we also present examples for which the dependence distortion effects are present and can be worked out in detail.
In Section~\ref{section_bootstrap} we review the multivariate smooth bootstrap for a given sample. We introduce an algorithm to obtain a smooth bootstrap sample from the copula.
Furthermore, we not only introduce and discuss a method to select appropriate non-diagonal bandwidth matrices via cross-validation but also cover some asymptotic considerations of our approach. A simulation study to show the impact of dependence distortion for popular functionals is conducted in Section~\ref{section_simulation_studies}. Final conclusions are given in Section~\ref{section_conclusion}. Appendix~\ref{section_properties_of_CF} and Appendix~\ref{section_elliptical_random_vectors} provide details about characteristic functions and elliptical random vectors, respectively.

\section{Notation, preliminaries, and setting}\label{section_prelim}

In this section we provide necessary preliminaries and introduce our setting. Also, we introduce the notation as needed. Much of the attention of this article is devoted to the impact of kernel smoothing and bootstrapping on the dependence structure between random variables which we represent by copulas. Copulas are one way to model the dependence between components of random vectors.
While we only review the notions relevant for our exposition, textbook introductions to dependence modeling via copulas can be found, e.g., in \cite{Nelsen2006,MaiScherer2012,Joe2015,DuranteSempi2016,HofertKojadinovicMachlerYan2018}.

A function $\Copula \colon [0,1]^{\dim} \to [0,1]$ is called a $\dim$-\emph{copula} (or \emph{copula}), if $\Copula$ is the distribution function of a $\dim$-dimensional random vector $\bU = (U_1,\ldots,U_d)$ with standard uniform univariate marginals, that is, $\Prob{U_k \leq u_k} = u_k$ for all $k \in \{1,\ldots,\dim\}$ and $u_k \in [0,1]$.
The importance of copulas stems from Sklar's Theorem, see \cite{Sklar1959}, which states that any $\dim$-dimensional distribution function $\JointDist$ with margins $\mdistI_1,\ldots,\mdistI_{\dim}$ can be decomposed as (where $\overline{\R} = \R \cup \{-\infty,\infty\}$)
\begin{equation}\label{eq:JointDistributionCopula}
	\JointDist(x_1,\ldots,x_{\dim}) = \Copula(\mdistI_1(x_1),\ldots,\mdistI_{\dim}(x_{\dim})),\quad \bx\in\overline{\R}^{\dim},
\end{equation}
where $\Copula$ is a copula.
If $\mdistI_1,\ldots,\mdistI_{\dim}$ are all continuous, then $\Copula$ is unique.
Otherwise, $\Copula$ is uniquely determined on $\times_{j=1}^{\dim} \range \mdistI_j$, where $\range \mdistI_j = \mdistI_j(\overline{\R})$ for $j=1,\ldots,\dim$.
Conversely, if $\Copula$ is a $\dim$-copula and $\mdistI_1,\ldots,\mdistI_{\dim}$ are distribution functions, then the function $\JointDist$ defined by \eqref{eq:JointDistributionCopula} is a $\dim$-dimensional distribution function with margins $\mdistI_1,\ldots,\mdistI_{\dim}$.

\begin{figure}
	\centering

	\begin{tikzpicture}[thick,scale=1.25, every node/.style={scale=1.25}]
		\node (X) {\color{theory}{$\bX$}};
		\node (G8) [above of=X] {};

		\node (S) [above of=G8] {\color{theory}{$(\bX_1,\ldots,\bX_{\norig})$}};
		\node (G3) [above of=S] {};

		\node (s) [above of=G3] {\color{observed}{$(\bx_1,\ldots,\bx_{\norig})$}};
		\node (G2) [right of=s] {};
		\node (G6) [right of=G2] {};

		\node (G5) [right of=G6] {};

		\node (CEn) [right of=G5] {\color{observed}{$\CopulaCondKDE$}};
		\node (G) [right of=X] {};
		\node (G7) [right of=G] {};

		\node (G4) [right of=G7] {};

		\node (K) [right of=G4] {\color{observed}{$\Kernel_{\BWMatrix_{\norig}}$}};
		\node (CE) [below of=G7] {\color{theory}{$\ExpCop$}};
		\node (CX) [below of=X,left of=X] {\color{theory}{$\Copula_{\bX}$}};
		\node (SBS) [above of=CEn] {\color{observed}{$(\bu_1^*,\ldots,\bu^*_{\naugm})$}};
		\node (TCX) [below of=CX] {\color{theory}{$\func\(\Copula_{\bX}\)$}};
		\node (TCE) [below of=CE] {\color{theory}{$\func\(\ExpCop\)$}};
		\node (G9) [right of=CEn] {};
		\node (G10) [right of=G9] {};
		\node (TCEn) [right of=G10] {\color{observed}{$\func\(\CopulaCondKDE\)$}};
		\node (Tz) [above of=TCEn] {\color{observed}{$\func_{\naugm}\(\bu_1^*,\ldots,\bu^*_{\naugm}\)$}};
		\node (Tx) [above of=s] {\color{observed}{$\func_{\norig}\(\bx_1,\ldots,\bx_{\norig}\)$}};

		\draw[->] (X) to node[left] {} (CX);
		\draw[->] (X) to node[left] {} (CE);
		\draw[->] (K) to node[right] {} (CE);
		\draw[->] (X) to node[left] {{\tiny{iid}}} (S);
		\draw[->] (S) to node[left] {$\bx_i = \bX_i(\omega)$} (s);
		\draw[->] (s) to node{} (CEn);
		\draw[->] (K) to node[right] {} (CEn);
		\draw[->] (CEn) to node[right] {\tiny{Algorithm~\ref{algo_smoothed_copula_bootstrap}}} (SBS);
		\draw[->] (CX) to node {} (TCX);
		\draw[->] (CE) to node {} (TCE);
		\draw[->] (CEn) to node {} (TCEn);
		\draw[->] (SBS) to node {} (Tz);
		\draw[->] (s) to node {} (Tx);

	\end{tikzpicture}
	\caption{Schematic relationships between the objects under consideration. Objects in gray are at the (inaccessible) population level, while objects in black are tangible to the statistician. $\Copula_{\bX}$ is the theoretical copula, $\ExpCop$ the theoretical copula constructed by convolving the underlying joint distribution $\JointDist_{\bX}$ with the kernel $\Kernel_{\BWMatrix_{\norig}}$, and  $\CopulaCondKDE$ the smoothed copula. $\func$ and $\func_{\norig}$ denote the population version and estimated version of the considered functional, respectively.}
	\label{fig_schematicRelationships}
\end{figure}
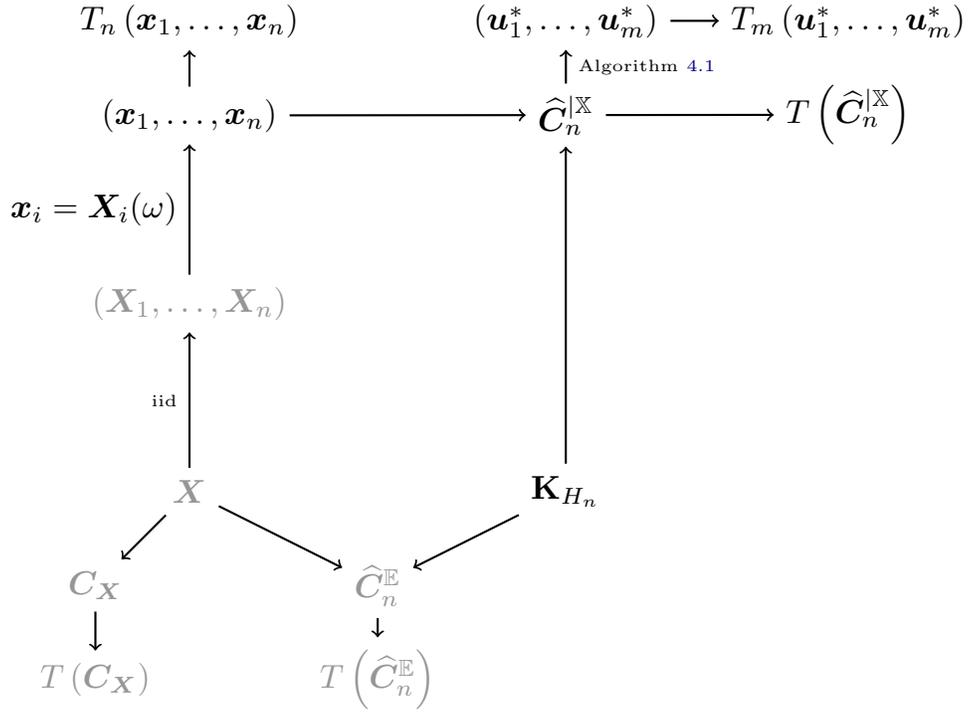

Throughout we generally consider a $\dim$-dimensional random vector $\bX$ defined on a probability space $(\Omega,\SigAlg,\P)$.
In our presentation we draw clear distinctions between (i) the (theoretical) underlying copula $\Copula_{\bX}$ related to the joint distribution $\JointDist_{\bX}$ of $\bX$ via Sklar's theorem, (ii) the (theoretical) copula $\ExpCop$ constructed by convolving the underlying joint distribution $\JointDist_{\bX}$ with the kernel $\Kernel$ with bandwidth matrix $\BWMatrix_{\norig}$, and (iii) the smoothed copula $\CopulaCondKDE$ based on a sample $\{\bx_1,\ldots,\bx_{\norig}\}$ of $\bX$ that is used to generate a smooth bootstrap sample $\{\bu_1^*,\ldots,\bu_{\naugm}^*\}$ of size $\naugm$ based on Algorithm~\ref{algo_smoothed_copula_bootstrap}. Figure~\ref{fig_schematicRelationships} shows the schematic relationship between the considered objects; gray entries represent the inaccessible objects at the population level, black entries are observable or under the direct control of the statistician.

To understand the ideas behind the smooth bootstrap later on, we
need to briefly review multivariate kernel density estimation.
We follow the notation of \cite{WandJones1995}; other references are
\cite{DevroyeGyorfi1985,Silverman1986,HardleMullerSperlichWerwatz2012} and \cite{Scott2015}.
We denote the $\dim$-dimensional identity matrix by $\IdMat_{\dim}$.
Vectors are generally understood as column vectors, and $^{\tr}$ is used to denote the transpose when required.

The key idea in kernel density estimation is to smooth out the empirical mass function
by means of a kernel which is defined as follows.
\begin{Definition}[Kernel]
	A function $\kernel$ is called a \emph{$\dim$-dimensional kernel}, if
	\begin{enumerate}
		\item $\kernel$ is the density function of an absolutely continuous
		$\dim$-dimensional random vector $\bY$, i.e., $\kernel(\bx) \geq 0$ for all
		$\bx \in \R^{\dim}$, $\kernel$ is integrable and $\int_{\R^{\dim}} \kernel(\bx) \d\bx = 1$;
		\item $\kernel$ has zero mean, i.e., $\Eval{\bY} = \int_{\R^{\dim}} \bx \kernel(\bx) \d\bx = \bZero$; and
		\item $\kernel$ has uncorrelated components with finite and identical marginal second moments, i.e.,
		$\cov{\bY} = \int_{\R^{\dim}} \bx \bx^{\tr} \kernel(\bx) \d\bx =
		\mu_{2,\kernel}\IdMat_{\dim}$, where the marginal second moment
		$\mu_{2,\kernel} = \int_{\R^{\dim}} x_j^2 \kernel(\bx) \d\bx$ is equal for each
		$j$.
	\end{enumerate}
\end{Definition}
More general
definitions of kernels (such as higher order kernels) are available in the
literature, but are not considered here.

For a given kernel $\kernel$ we define a rescaled version via
\begin{align}\label{eq_rescaled_kernel}
	\kernel_{\BWMatrix}(\bx) = \frac{1}{\sqrt{\det(\BWMatrix)}}\kernel\left(\BWMatrix^{-\frac{1}{2}}\bx\right),\quad \bx\in\R^{\dim},
\end{align}
where $\BWMatrix$ is a symmetric positive definite (spd) matrix called the \emph{bandwidth matrix}.
Accordingly, $\Kernel$ and $\Kernel_{\BWMatrix}$ denote the corresponding distribution functions of the kernel and rescaled kernel.
Note that if $\bY \sim \Kernel$, then the density of $\bY_{\BWMatrix} = \BWMatrix^{1/2} \bY$ is $\kernel_{\BWMatrix}$.
Based on the properties of $\kernel$ it is then straightforward to see that $\kernel_{\BWMatrix}$ has zero mean, i.e., $\Eval{\bY_{\BWMatrix}} = \bZero$, and covariance matrix $\cov{\bY_{\BWMatrix}} = \mu_{2,\kernel} \BWMatrix$.

For a random sample $\sample = (\bX_i)_{i=1}^{\norig}$ from a $\dim$-dimensional random vector $\bX$, the kernel density estimate of the underlying $\dim$-dimensional density $\JointDens_{\bX}$ is defined by
\begin{align}\label{eq_kernel_density}
	\kde_{\norig}(\bx) = \kde_{\norig}(\bx;\sample) = \frac{1}{\norig} \sum^{\norig}_{i=1} \kernel_{\BWMatrix_{\norig}}(\bx-\bX_i),
\end{align}
where the bandwidth matrix $\BWMatrix_{\norig}$ typically only depends on the sample size $\norig$.
However, it is possible to consider a local bandwidth matrix $\BWMatrix_{\norig}(\bx)$ that (possibly) changes with the evaluation point $\bx \in \R^{\dim}$.
In a practical setting, $\BWMatrix_{\norig}$ can also depend on $\sample$, for example, when considering a scaled version of the empirical covariance matrix; we discuss the selection of $\BWMatrix_{\norig}$ for our problem setting in Section~\ref{section_bandwidth_selection}.
The estimate $\KDE_{\norig}$ of the corresponding joint distribution function $\JointDist_{\bX}$ is thus given by
\begin{align*}
	\KDE_{\norig}(\bx) = \KDE_{\norig}(\bx;\sample) = \int_{(-\binfty,\bx]}\kde_{\norig}(\by)\d\by = \frac{1}{\norig} \sum^{\norig}_{i=1}  \int_{(-\binfty,\bx]} \kernel_{\BWMatrix_{\norig}}(\by-\bX_i) \d\by = \frac{1}{\norig} \sum^{\norig}_{i=1} \Kernel_{\BWMatrix_{\norig}}(\bx-\bX_i),
\end{align*}
where $(-\binfty,\bx] = \times_{j=1}^{\dim} (-\infty,x_j]$. Note that by Sklar's Theorem, see \cite{Sklar1959}, there is a unique (random) copula associated to $\KDE_{\norig}$ which is given by
\begin{align}\label{eq_random_copula}
	\KDECop_{\norig}(\bu) = \KDE_{\norig}\left(\kdeMargDist_{\norig 1}^{-1}(u_1),\ldots,\kdeMargDist_{\norig \dim}^{-1}(u_{\dim})\right),
\end{align}
where $\kdeMargDist_{\norig 1}^{-1},\ldots,\kdeMargDist_{\norig \dim}^{-1}$ are the marginal quantile functions associated to $\KDE_{\norig}$.

While \eqref{eq_rescaled_kernel} provides a direct relationship to evaluate $\kernel_{\BWMatrix}$ in terms of $\kernel$, there is in general no such relationship between $\Kernel_{\BWMatrix}$ and $\Kernel$.
For certain kernel distributions $\Kernel$, such as the multivariate normal distribution or, more generally, elliptical distributions, the corresponding rescaled distribution function $\Kernel_{\BWMatrix}$ can be identified and is (at least numerically) accessible.
Special cases, such as diagonal bandwidth matrices and product kernels, allow in general for a direct evaluation of $\Kernel_{\BWMatrix}$ in terms of $\Kernel$.
In the case of product kernels and a strictly stationary and geometrically strongly mixing data generating process, \cite{LiuYang2008} derive pointwise first and second order asymptotics of $\KDE_{\norig}$.

The kernel density estimate in \eqref{eq_kernel_density} can also be understood from the viewpoint of convolutions.
The \emph{convolution} $\densI \conv \densII$, see \citealt[p.~170]{Rudin1991}, of two $\dim$-dimensional functions $\densI, \densII \in L^1(\R^{\dim})$ at $\bx \in \R^{\dim}$ is defined by
\begin{align}\label{eq_density_convolution}
	(\densI \conv \densII)(\bx) = \int_{\R^{\dim}} \densI(\bx-\by)\densII(\by) \d\by.
\end{align}

If $\densI$ and $\densII$ are density functions, the convolution also has a probabilistic interpretation.
When two independent random vectors $\bX$ and $\bY$ have densities $\densI$ and $\densII$, the density $\JointDens_{\bX+\bY}$ of $\bX+\bY$ is given by $\JointDens_{\bX+\bY} = \densI\conv\densII$.
To also cover probability mass functions in $\densI \conv \densII$ one can extend the convolution formula in \eqref{eq_density_convolution} in terms of a Lebesgue-Stieltjes integral to
$(\densI \conv \densII)(\bx) = \int_{\R^{\dim}} \densI(\bx-\by) \d\distII(\by)$ when $\densII$ is a probability mass function.
The kernel density estimator in \eqref{eq_kernel_density} can then be seen as the convolution of the (rescaled) kernel density $\kernel_{\BWMatrix_{\norig}}$ with the point masses $\sum^{\norig}_{i=1} \frac{1}{\norig}\dirac_{\bX_i}$ belonging to the empirical distribution function; here $\dirac_{\ba}$ denotes a point mass at $\ba \in \R^{\dim}$, i.e., $\dirac_{\ba}(\bx)=1$ if $\bx=\ba$ and zero otherwise.
For specific observations $\{\bx_i\}_{i=1}^{\norig}$, so realizations $\bx_i = \bX_i(\omega)$ for some $\omega \in \Omega$ (or, equivalently, given $\sample$), the kernel density estimate is a mixture density where the $i$th mixing density has mean $\bx_i$ and is given by $\kernel_{\BWMatrix_{\norig}}(\bx - \bx_i)$ while the mixing weights are $\frac{1}{\norig}$.

The bias and variance of $\kde_{\norig}$ can be derived under additional
assumptions on $\kernel$, $\BWMatrix_{\norig}$ and $\JointDens_{\bX}$; see
\citealt[Chapter~4.3]{WandJones1995}. To state the results, we denote by
$\trace(\BWMatrix) = \sum^{\dim}_{j=1} \BWMatrix_{jj}$ the trace of a
$\dim\times\dim$ matrix $\BWMatrix$.  Furthermore, for a twice continuously
differentiable function $f \colon \R^{\dim} \to \R$, the Hessian matrix of second
order partial derivatives at $\bx \in \R^{\dim}$ is denoted by
$\Hesse_{\bx}(f) = \left(\frac{\partial^2 f}{\partial x_i \partial
	x_j}(\bx)\right)_{i,j=1,\ldots,\dim}$.  To simplify limits we utilize the Landau
symbols $\landau$ and $\Landau$.
Specifically, we assume
the following conditions to hold:
\begin{enumerate}
	\item $\xNorm{\kernel}{2}^2 = \int_{\R^{\dim}} \kernel^2(\bx) \d\bx < \infty$;
	\item each entry of the Hesse matrix $\Hesse_{\bx}(\JointDens_{\bX})$ is piecewise continuous and square integrable;
	\item $\left(\BWMatrix_{\norig}\right)_{\norig \geq 1}$ is a sequence of bandwidth matrices such that $1/(\norig \sqrt{\det(\BWMatrix_{\norig})})$ and all entries of $\BWMatrix_{\norig}$ approach zero as $\norig \to \infty$;
	\item the ratio of the largest and smallest eigenvalue of $\BWMatrix_{\norig}$ is bounded for all $\norig$; and
	\item $\kernel$ is a bounded and compactly supported $\dim$-dimensional kernel.
\end{enumerate}
Under these conditions the bias and variance of the kernel density estimate $\kde_{\norig}$ can be computed as a function of the sample size $\norig$ and the function $\JointDens_{\bX}$ via
\begin{align*}
	\Eval{\kde_{\norig}(\bx)} - \JointDens_{\bX}(\bx) = \frac{1}{2}\mu_{2,\kernel}\trace\left(\BWMatrix_{\norig} \Hesse_{\bx}(\JointDens_{\bX})\right) + \landau(\trace\left(\BWMatrix_{\norig})\right)\approx \frac{1}{2}\mu_{2,\kernel}\trace\left(\BWMatrix_{\norig} \Hesse_{\bx}(\JointDens_{\bX}) \right),\\
	\var{\kde_{\norig}(\bx)} = \frac{1}{\norig \sqrt{\det(\BWMatrix_{\norig})}} \xNorm{\kernel}{2}^2 \JointDens_{\bX}(\bx) + \landau\left(\frac{1}{n \sqrt{\det(\BWMatrix_{\norig})}}\right)\approx \frac{1}{\norig \sqrt{\det(\BWMatrix_{\norig})}} \xNorm{\kernel}{2}^2 \JointDens_{\bX}(\bx).
\end{align*}
Furthermore, under the above conditions the estimate is consistent at any fixed point $\bx$, that is, $\kde_{\norig}(\bx) \inprob \JointDens_{\bX}(\bx)$ for $\norig \to \infty$.

In the following section we discuss kernel smoothing and the dependence distortion it (possibly) introduces. This will be important for the subsequent analysis of the smooth bootstrap.
\section{Kernel smoothing and dependence distortion}\label{section_kernel_smoothing}

This section investigates the effects of kernel smoothing on the dependence structure on the population level. We start by considering a fixed sample size.

\subsection{Population version of kernel smoothing for a fixed sample size}\label{section_population_version}

To prepare the presentation of the smooth bootstrap later on we now discuss the dependence distortion introduced by kernel density estimation for a fixed sample size $\norig$.
This links to the bootstrap in that the number of sample points $\norig$ is fixed, but the number of (bootstrap) samples $\nbs$ is effectively unlimited.
If we \emph{had} access to $\nbs$ independent samples $(\sample_b)_{b=1}^{\nbs}$, where $\sample_b = \left\{ \bX_{b1},\ldots,\bX_{b\norig} \right\}$ is a collection of iid random vectors with common density $\JointDens_{\bX}$, we could indeed average over our samples and obtain from the strong law of large numbers that
\begin{align*}
\frac{1}{\nbs} \sum^{\nbs}_{b=1} \kde_{\norig}(\bx;\sample_b) \almostsurly \Eval{\kde_{\norig}(\bx)},\quad (\nbs \to \infty).
\end{align*}
When considering the smooth bootstrap in Section~\ref{section_bootstrap}, bootstrap samples $\samplebs_b$ will replace the unavailable samples $\sample_b$.
When comparing (as a function of $\bx$) the expected density and distribution function estimators
\begin{align}\label{eq_expected_kernel_density_estimate}
\expkde(\bx) &=
\Eval{\kde_{\norig}(\bx)} = \int_{\R^{\dim}} \kernel_{\BWMatrix_{\norig}}(\bx - \by) \JointDens_{\bX}(\by) \d\by,\\
\Expkde(\bx) &=
\Eval{\KDE_{\norig}(\bx)} = \int_{\R^{\dim}} \Kernel_{\BWMatrix_{\norig}}(\bx - \by) \JointDens_{\bX}(\by) \d\by
\end{align}
with the convolution formula \eqref{eq_density_convolution}, we see that $\expkde$ coincides with the density of $\bZ = \bX + \bY_{\BWMatrix_{\norig}}$, where $\bX$ is distributed with density $\JointDens_{\bX}$ and $\bY_{\BWMatrix_{\norig}}$ with density $\kernel_{\BWMatrix_{\norig}}$, and $\bX$ and $\bY_{\BWMatrix_{\norig}}$ are independent.
The mean of $\bZ$ is therefore $\Eval{\bX}$ and hence undistorted compared to $\JointDens_{\bX}$.
However, the covariance matrix is $\cov{\bZ} = \cov{\bX} + \cov{\bY_{\BWMatrix_{\norig}}} = \cov{\bX} + \mu_{2,\kernel}\BWMatrix_{\norig}$ by independence of $\bX$ and $\bY_{\BWMatrix_{\norig}}$ and the properties of $\kernel_{\BWMatrix_{\norig}}$, and thus in general not equal to $\cov{\bX}$.
The independence between $\bX$ and $\bY_{\BWMatrix_{\norig}}$ also allows us to compute the characteristic function of $\expkde$ as $\CF_{\expkde}(\bt) = \CF_{\bX}(\bt)\CF_{\kernel}\left(\BWMatrix_{\norig}^{1/2}\bt\right)$, see Appendix~\ref{section_properties_of_CF} where we provide further details on properties of characteristic functions.

According to Sklar's Theorem there is a unique copula associated to $\Expkde$ which is given by
\begin{align*}
\ExpCop(\bu) = \Expkde\left(\kdeMargDist_{\norig 1}^{-1}(u_1),\ldots,\kdeMargDist_{\norig \dim}^{-1}(u_{\dim})\right),
\end{align*}
where $\kdeMargDist_{\norig 1}^{-1},\ldots,\kdeMargDist_{\norig \dim}^{-1}$ are the marginal quantile functions associated to $\Expkde$.
It is important to notice here that $\ExpCop$, just like $\expkde$ and $\Expkde$, does not depend on the sample or the sample size $\norig$ directly.
However, the bandwidth matrix $\BWMatrix_{\norig}$ typically depends on the sample size (and possibly the sample) which establishes an indirect connection and justifies the subscript $\norig$ in the notation.
Further properties of $\ExpCop$ and the differences between $\ExpCop$ and $\Copula_{\bX}$ in the case of elliptical distributions are the subject of the following section.

\subsection{Population version of dependence distortion for elliptical random vectors and elliptical smoothing kernels}\label{section_dependence_distortion_elliptical}
Combining the observations from Section~\ref{section_population_version} with the
properties of elliptical random vectors gives a first idea on how
smoothing distorts the underlying dependence structure for the statistically
important class of elliptical distributions.
An elliptical random vector can be described by three components: a location vector $\bMu$, a dispersion matrix $\dispMat$ and a characteristic generator $\charGen$, which is a real-valued function such that $\CF(\bt) = \exp\left(\imu\bt^{\tr}\bMu\right)\charGen\left(\bt^{\tr}\dispMat \bt\right)$ where $\CF$ denotes the multivariate characteristic function. Appendix~\ref{section_elliptical_random_vectors} provides further details on elliptical random vectors.
In the notation of \citealt[Chapter~6]{McNeilFreyEmbrechts2015}, consider $\bX \sim \Ellip_{\dim}(\bMu,\dispMat,\charGen_{\abbrData})$ and $\bY \sim \Ellip_{\dim}(\bZero,\IdMat_{\dim},\charGen_{\abbrKernel})$ (related to the kernel density $\kernel$) to be independent and
elliptically distributed, where we directly have $\mu_{2,\kernel} = -2 \charGen_{\abbrKernel}'(0)$.
In general, see Theorem~\ref{theorem_elliptical_moments}, we have for an elliptical random vector $\bZ \sim \Ellip_{\dim}(\bMu,\dispMat,\charGen_{\bZ})$ with an associated radial part $\radialPart\geq 0$ with finite second moment that $\E[\bZ]=\bMu$ and $\cov{\bZ} = \frac{\E[\radialPart^2]}{\rank(\dispMat)}\dispMat = -2 \charGen_{\bZ}'(0)\dispMat$.
However, the associated correlation matrix is independent of $\charGen_{\bZ}$ and give as $\corr{\bZ} = \diag\left(1/\sqrt{\dispMat_{11}},\ldots,1/\sqrt{\dispMat_{\dim\dim}}\right) \dispMat \diag\left(1/\sqrt{\dispMat_{11}},\ldots,1/\sqrt{\dispMat_{\dim\dim}}\right)$ so that $\corr{Z_i,Z_j} = \dispMat_{ij} / \sqrt{\dispMat_{ii}\dispMat_{jj}}$.

Furthermore, if $\bX$ and $\bY$ are independent, the distribution of $\bZ= \bX + \bY_{\BWMatrix_{\norig}}$ with density $\expkde$ is elliptical if $\BWMatrix_{\norig} = c_{\norig} \dispMat$ for some $c_{\norig} > 0$; see \citealt[Remark~6.26]{McNeilFreyEmbrechts2015}.
In this case $\bY_{\BWMatrix_{\norig}} \sim \Ellip_{\dim}(\bZero,c_{\norig}\dispMat,\charGen_{\abbrKernel})$.
In Corollary~\ref{corollary_summation_elliptical_random_vectors} we derive general conditions that imply $\bZ \sim \Ellip_{\dim}(\bMu,\dispMat,\charGen_{\abbrSmooth})$ in the given context, where $\charGen_{\abbrSmooth}(u) = \charGen_{\abbrData}(u)\charGen_{\abbrKernel}(c_{\norig} u)$.
This shows that if $\BWMatrix_{\norig} = c_{\norig} \dispMat$, then, on average (i.e., when considering the expected density estimate $\expkde$), the distortion introduced by kernel smoothing affects the rescaling of the characteristic generator of $\bX$ by a factor of $\charGen_{\abbrKernel}(c_{\norig} u)$ for every $u \geq 0$.

While the specific choice $\BWMatrix_{\norig} = c_{\norig} \dispMat$ seems limiting at first, it corresponds to the population version of the \emph{sphering} approach commonly used in multivariate kernel density estimation; see \citealt[Chapter~4.6]{WandJones1995} and references therein.
In this case the bandwidth matrix is determined by a one-dimensional parameter $\BW$ by setting $\BWMatrix = \BW \empcovMat_{\norig}$.
This links to our theoretical discussion since $\empcovMat \inprob \cov{\bX}$ and hence $\BWMatrix \approx -2 \BW \charGen'(0) \dispMat$ for elliptical random vectors.
As we will see, in the context of elliptical random vectors, sphering will allow us to derive theoretical results without being limited to product kernels or diagonal bandwidth matrices.
We return to a general discussion of bandwidth matrices in Section~\ref{section_bandwidth_selection}.

Furthermore, since $\bZ \sim \Ellip_{\dim}(\bMu,\dispMat,\charGen_{\abbrSmooth})$,
the covariance matrix of $\bZ$ is given by
\begin{align}\label{eq_var_covar_Z}
\cov{\bZ} = -2\charGen_{\abbrSmooth}'(0)\dispMat = \left( 1 + c_{\norig} \frac{\charGen_{\abbrKernel}'(0)}{\charGen_{\abbrData}'(0)}\right) \cov{\bX},
\end{align}
where $\cov{\bX} = -2\charGen_{\abbrData}'(0)\dispMat$.
As a consequence of $\cov{\bZ}$ being a re-scaling of $\cov{\bX}$, the correlation matrix of $\bZ$
is thus
\begin{align*}
  \corr{\bZ} = \corr{\bX}
\end{align*}
and hence, although the covariance matrix is distorted in general, the correlation matrix is not.
Note that the $j$th margin of $\bX$ is
$X_j \sim \Ellip_1(\bMu_j,\dispMat_{jj},\charGen_{\abbrData})$,
see \eqref{eq_univariate_margins_elliptical}, so the margins of $\bZ$ are
$Z_j \sim \Ellip_1(\bMu_j,\dispMat_{jj},\charGen_{\abbrSmooth})$,
i.e., their characteristic generators have also been altered.
While we still have that $\Eval{Z_j} = \bMu_j$, the variance changes to $\var{Z_j} = -2\charGen_{\abbrSmooth}'(0)\dispMat_{jj} = \left( 1 + c_{\norig} \frac{\charGen_{\abbrKernel}'(0)}{\charGen_{\abbrData}'(0)}\right) \var{X_j}$ and so do the marginal quantile functions when going from $\bX$ to $\bZ$.

In summary, the correlation structure remains unchanged when going from $\bX$ to
$\bZ$, but the altered characteristic generator affects the marginal distributions.
Differences in the resulting elliptical copulas, see Definition~\ref{definition_elliptical_copula}, are hence not due to the respective correlation matrices (which enter the copulas as parameters), but are solely due to differences of the characteristic generators, i.e., the distributions of the underlying radial parts.

Contrasting the copula of $\bZ$ given by
$\ExpCop = \Copula_{\corr{\bZ},\charGen_{\abbrSmooth}}$ with the original
copula of interest $\Copula_{\bX} = \Copula_{\corr{\bX},\charGen_{\abbrData}}$,
$\corr{\bZ} = \corr{\bX}$ implies that the only difference between $\ExpCop$ associated to the average kernel density estimate based on samples of size $\norig$ and the copula $\Copula_{\bX}$ underlying the
data generating process is due to
the difference in the characteristic generators $\charGen_{\abbrSmooth}$ versus
$\charGen_{\abbrData}$.  At this point it is worth noticing that
$\lim_{c_{\norig} \to 0} \charGen_{\abbrKernel}(c_{\norig} u) = 1$ for all
$u \geq 0$ due to uniform continuity of characteristic functions, see
\citealt[Theorem~1.1.2]{Sasvari2013}, and thus for $c_{\norig} \to 0$ we have that
$\charGen_{\abbrSmooth}(u) \to \charGen_{\abbrData}(u)$ for all points
$u \geq 0$.  Generally this implies, even in the absence of estimation of
$\charGen_{\abbrData}$ and $\corr{\bX}$, that the average estimated density $\expkde$ has
a different copula $\ExpCop$ than the original sample $\Copula_{\bX}$.  Interestingly this is not always the case and we discuss
conditions for such exceptions and examples in
Section~\ref{section_exceptions}. Furthermore, when $\ExpCop$ is used to
estimate functionals of $\Copula_{\bX}$, the distortion
introduced may or may not affect the result depending on the functional under
consideration.
We illustrate this point by considering a number of popular
copula functionals and their properties in the case of elliptical distributions in the following paragraphs.

In the case of absolutely continuous meta-elliptical distributions, Kendall's tau does not depend on the characteristic generator, but solely on the entries of the dispersion matrix $\dispMat$; see \cite{LindskogMcNeilSchmock2003}.
As shown in \citealt[Proposition~8]{SchmidSchmidt2007}, the same is true for Blomqvist's beta which equals Kendall's tau for elliptical models.
On the population level in the elliptical class these measures of association hence coincide surprisingly for the underlying random vector $\bX$ and the smooth version $\bZ$ as long as $\BWMatrix_{\norig} = c_{\norig} \dispMat$.

However, Spearman's rho may depend on the density generator of the elliptical density; see \cite{AbdousGenestRemillard2005} and \cite{HultLindskog2002} for an example showing that Spearman's rho is not invariant among (meta) elliptical models with a common correlation structure.
For absolutely continuous multivariate elliptical random vectors the corresponding density generator provides a simple description of the multivariate density in terms of a univariate function, see Theorem~\ref{theorem_density_generator} for the details.
The density generator in turn depends on the associated characteristic generator $\charGen$ as the characteristic function fully describes the joint distribution.
The same is true for the tail dependence coefficient of elliptical distributions for which
\cite{Schmidt2002} shows that it depends on the regular variation property of the density generator.
As expected, smoothing thus leads to different values of these functionals for the underlying random vector $\bX$ and the smooth version $\bZ$ even if $\BWMatrix_{\norig} = c_{\norig} \dispMat$.

For spherical distributions it is possible to give a precise condition when a test statistic will be invariant under changes of the underlying spherical distribution.
This complements the earlier discussion on the invariance of certain functionals of elliptical copulas and it potentially opens an alternative route of investigation.
\begin{Theorem}[{\citealt[Theorem~2.22]{FangKotzNg1990}}]
The distribution of a statistic $T(\bX)$ remains unchanged as long as $\bX \sim \Ellip_{\dim}(\bZero,\IdMat_{\dim},\charGen_{\abbrData})$ with $\Prob{\bX = \bZero} = 0$, provided that
\begin{align*}
T(\alpha \bX) \equalindist T(\bX)
\end{align*}
for each $\alpha > 0$.
In this case $T(\bX) \equalindist T(\bY)$ where $\bY \sim \Normal(\bZero,\IdMat_{\dim})$.
\end{Theorem}

Next, we investigate the convergence of the characteristic generator and the characteristic function.
\subsection{Convergence rates of \texorpdfstring{$\charGen_{\abbrSmooth}(u) \to \charGen_{\abbrData}(u)$}{} and \texorpdfstring{$\CF_{\abbrSmooth}(\bt) \to \CF_{\abbrData}(\bt)$}{}}
\label{section_convergence_rates}
In this section, we investigate the difference between the characteristic generator $\charGen_{\abbrSmooth}(u) = \charGen_{\abbrData}(u) \charGen_{\abbrKernel}(c_{\norig} u)$ and the original $\charGen_{\abbrData}$ and the impact on the difference between the respective characteristic functions.
The difference between $\charGen_{\abbrSmooth}$ and $\charGen_{\abbrData}$ only depends on the convergence rate $\charGen_{\abbrKernel}(c_{\norig} u) \to 1$ as $c_{\norig} \to 0$, since
\begin{align*}
\abs{\charGen_{\abbrData}(u) - \charGen_{\abbrSmooth}(u)} = \abs{\charGen_{\abbrData}(u)}(1-\charGen_{\abbrKernel}(c_{\norig} u)) \leq 1-\charGen_{\abbrKernel}(c_{\norig} u).
\end{align*}
The equality as well as the inequality hold since $\abs{ \charGen(z) } \leq 1$ for all $z \in [0,\infty)$ and all characteristic generators $\charGen$.
Also the relative error hence conveniently takes the form
\begin{align}\label{eq_relative_error}
\abs{
\frac{\charGen_{\abbrData}(u) - \charGen_{\abbrSmooth}(u)}{\charGen_{\abbrData}(u)}
}
 = \abs{ 1 - \charGen_{\abbrKernel}(c_{\norig} u) } = 1 - \charGen_{\abbrKernel}(c_{\norig} u).
\end{align}
For certain generators $\charGen_{\abbrKernel}$ this rate may be slower than for others.
For example, in case of a Gaussian kernel with $\charGen_{\abbrKernel}(u) = \exp(-u/2)$ or a Laplace kernel with $\charGen_{\abbrKernel}(u) = (1+u/2)^{-1}$ we have in both cases
\begin{align*}
1 - \charGen_{\abbrKernel}(c_{\norig}u) = c_{\norig} \frac{u}{2} + \Landau\(c_{\norig}^2\), \quad c_{\norig} \to 0.
\end{align*}
For the Cauchy distribution with  $\charGen_{\abbrKernel}(u) = \exp\(-\sqrt{u}\)$ we have
\begin{align*}
1 - \charGen_{\abbrKernel}(c_{\norig}u) = \sqrt{c_{\norig}} \sqrt{u} + \Landau\(c_{\norig}\), \quad c_{\norig} \to 0.
\end{align*}
The convergence rate of the relative error in \eqref{eq_relative_error}, that simultaneously acts as an upper bound for the absolute error, is hence slower in the latter case.
For a given characteristic generator $\charGen_{\abbrKernel}$ the behaviour of $1-\charGen_{\abbrKernel}(y)$ for $y \to 0$ is analyzed in \cite{Bingham1972}. It is connected to the behavior of the corresponding radial distribution as follows.
Denote by $\mdistI_{\radialPart}$ the distribution function of the radial part corresponding to the spherical distribution $\Ellip_{\dim}(\bZero,\IdMat_{\dim},\charGen_{\abbrKernel})$.
For $0 < \alpha < 2$ we then have that
\begin{align}\label{equation_Bingham_1}
1-\charGen_{\abbrKernel}(y) \sim y^{\alpha/2}L\left(1/\sqrt{y}\right), \quad\quad\quad (y \to 0),
\end{align}
for a function $L$ varying slowly at infinity,
if and only if
\begin{align}\label{equation_Bingham_2}
1-\mdistI_{\radialPart}(y) \sim \frac{L(y)}{y^{\alpha}} \frac{2^{\alpha}\Gamma\left((\dim+\alpha)/2\right)}{\Gamma\left(\dim/2\right)\Gamma\left(1-\alpha/2\right)}, \quad\quad\quad (y \to \infty);
\end{align}
see \cite{Bingham1972}, where one can also find a discussion of the cases $\alpha=0$ and $\alpha\geq 2$.
It is important to note that we adapted the result in \cite{Bingham1972} to our convention concerning the characteristic generators, whereas \cite{Bingham1972} works with comparable functions $\widetilde{\charGen}_{\abbrKernel}$ such that $\widetilde{\charGen}_{\abbrKernel}\left(\xNorm{\bt}{2}\right) = \E\left[\exp\left(\imu \bt^{\tr}\abbrKernel\right)\right] = \CF_{\abbrKernel}(\bt)$.
However, neither convention impacts the role of the radial distribution $\mdistI_{\radialPart}$.
The convergence rate of the relative error in \eqref{eq_relative_error} representing the discrepancy between the dependence structures is therefore directly linked to the asymptotic behavior of the survival function $1-\mdistI_{\radialPart}$ of the radial distribution of $\bY$ representing the kernel.

We can connect the previous discussion to the absolute difference $\abs{\CF_{\bX}\left(\bt \right) - \CF_{\bZ}\left(\bt \right)}$, $\bt\in\R^{\dim}$, between the characteristic functions of the initial random vector $\bX$ and the smoothed version $\bZ$ as follows.
With $\abs{e^{i\bMu^{\tr}\bt}}=1$ we have
\begin{align*}
  \abs{\CF_{\bX}(\bt)-\CF_{\bZ}(\bt)} &= \abs{e^{i\bMu^{\tr}\bt}\charGen_{\bX}\left(\bt^{\tr}\dispMat\bt\right) - e^{i\bMu^{\tr}\bt}\charGen_{\bZ}\left(\bt^{\tr}\dispMat\bt\right)} = \abs{\charGen_{\bX}\left(\bt^{\tr}\dispMat\bt\right) - \charGen_{\bZ}\left(\bt^{\tr}\dispMat\bt\right)}\\
                                      &\leq 1 - \charGen_{\bY}\left(c_{\norig}\bt^{\tr}\dispMat\bt\right),
\end{align*}
implying pointwise convergence for $c_{\norig}\to 0$.
We also have a bound on the relative error given by
\begin{align*}
\abs{\frac{\CF_{\bX}(\bt)-\CF_{\bZ}(\bt)}{\CF_{\bX}(\bt)}} \leq 1 - \charGen_{\bY}\left(c_{\norig}\bt^{\tr}\dispMat\bt\right).
\end{align*}
The absolute difference $\abs{\CF_{\bX}\left(\bt \right) - \CF_{\bZ}\left(\bt \right)}$ can be used to construct an upper bound on the uniform distance $\xNorm{\JointDist_{\bX}-\JointDist_{\bZ}}{\infty}$ via the smoothing inequality and its higher dimensional analogues, see Theorem~\ref{theorem_Berry_Essen_inequalty_2D} in Appendix~\ref{section_properties_of_CF}.
For ease of presentation we proceed with a univariate example.
\begin{Example}
For simplicity, we discuss this approach in the one-dimensional case where a Laplace kernel with $\charGen_Y(u) = (1+u/2)^{-1}$ is used to smooth a one dimensional elliptical random variable $X\sim\Ellip_1(\mu,\sigma^2,\charGen_X)$.
Combining the fact that
\begin{align*}
\abs{\CF_X(t)-\CF_Z(t)} \leq 1 - \charGen_Y\left(c_n\sigma^2t^2\right) = \frac{c_n\sigma^2t^2/2}{c_n\sigma^2t^2/2+1}
\end{align*}
with the smoothing inequality we have for $T>0$ that
\begin{align}\label{eq_uniform_bound_1D}
\xNorm{\mdistI_X-\mdistI_Z}{\infty}
&\leq \frac{1}{\pi} \int_{-T}^T \abs{\frac{\CF_X(t)-\CF_Z(t)}{t}} \d t + \frac{24}{\pi T} \sup_{x\in\R}\abs{\mdensI_Z(x)}\nonumber\\
&\leq \frac{2}{\pi} \int_{0}^T \frac{1-\charGen_Y(c_n\sigma^2t^2)}{t} \d t + \frac{24}{\pi T} \sup_{x\in\R}\abs{\mdensI_Z(x)} = \frac{1}{\pi} \int_{0}^T \frac{c_n\sigma^2t}{c_n\sigma^2t^2/2+1} \d t + \frac{24}{\pi T} \sup_{x\in\R}\abs{\mdensI_Z(x)}\nonumber\\
&= \frac{\log(c_n\sigma^2T^2/2+1)}{\pi} + \frac{24}{\pi T} \sup_{x\in\R}\abs{\mdensI_Z(x)}\nonumber\\
&\leq \frac{\log(c_n\sigma^2T^2/2+1)}{\pi} + \frac{M}{\pi T},
\end{align}
where $M$ is such that $24 \sup_{x\in\R}\abs{\mdensI_Z(x)}\leq M$.
Given that $\mdensI_Z$ is the convolution of the kernel density $\mdensI_Y$ with $\mdensI_X$, a simple upper bound independent of $c_n$ is given by $M = 24 \sup_{x\in\R}\abs{\mdensI_X(x)}$.

Optimizing the bound in \eqref{eq_uniform_bound_1D} with respect to $T$ we obtain the first order condition $c_n\sigma^2T^3 - M c_n\sigma^2T^2/2-M = 0$ with the unique (real) solution
\begin{align*}
  T^* = \frac{1}{6} \left(\frac{ c_n\sigma^2 M^2}{a^{1/3}} + \frac{a^{1/3}}{c_n\sigma^2} + M\right),
\end{align*}
where
\begin{align*}
  a = c_n^3 \sigma^6 M^3 + 108 c_n^2\sigma^4 M + 6 \sqrt{6} \sqrt{c_n^5\sigma^{10} M^4 + 54 c_n^4\sigma^8 M^2}.
\end{align*}
Given that $T^*$ is the unique stationary point and $\lim_{T\to 0}\frac{\log(c_n\sigma^2T^2/2+1)}{\pi} + \frac{M}{\pi T} = \lim_{T\to\infty}\frac{\log(c_n\sigma^2T^2/2+1)}{\pi} + \frac{M}{\pi T} = \infty$ it is clear that $T^*$ leads to the minimal upper bound.
Considering $T_n = T^*(c_n)$ as a function of $c_n$ we have $T_n = \Landau\left(c_n^{-1/3}\right)$ for $c_n\to 0$ and hence $c_n T_n^2 = \Landau\left(c_n^{1/3}\right)$ which leads to
\begin{align*}
\xNorm{\mdistI_X-\mdistI_Z}{\infty} = \xNorm{\mdistI_X-\mdistI_{X+Y_{c_n\sigma^2}}}{\infty} = \Landau\left(c_n^{1/3}\right).
\end{align*}
It is important to note that the rate $\Landau\left(c_n^{1/3}\right)$ is universal in the sense that it only depends on the kernel distribution $Y$ and holds simultaneously for all absolutely continuous univariate elliptical random variables $X\sim\Ellip_1(\mu,\sigma^2,\charGen_X)$ with a bounded density.
The dependence on $X$ is in fact only visible in the chosen bound $M$.
\end{Example}
In the next section, we provide cases for which the dependence structure remains unaffected by smoothing at the population level.

\subsection{Exceptions to dependence distortions at the population level}\label{section_exceptions}

Section~\ref{section_dependence_distortion_elliptical} has established how smoothing impacts the dependence structure for elliptical models and kernels at the population level.
In this section we give conditions under which these effects surprisingly do not impact the dependence structure.
As can be expected, these examples are rather artificial but serve the purpose of establishing a comprehensive view of dependence distortion in kernel density estimation.
To complete the picture we also discuss in Section~\ref{section_dependence_distortion_student_t} and \ref{section_dependence_distortion_Laplace} examples for which the effects of dependence distortion are present and can be worked out in detail.

\subsubsection{Multivariate normal distribution}
We start by considering the Gaussian random vector $\bX$, assume that the smoothing random vector $\bY$ is also Gaussian and that the bandwidth matrix $\BWMatrix_{\norig} = c_{\norig} \dispMat$ is a rescaled version of the dispersion matrix $\dispMat$ of $\bX$.
In this case, see \citealt[Example~2.3, page 28]{FangKotzNg1990}, the characteristic generators of $\bX$ and $\bY$ are given by $\charGen_{\abbrData}(u) = \charGen_{\abbrKernel}(u) = \exp(-u/2)$, and hence we can compute the characteristic generator of the expected density $\expkde$ of the smoothed random vector $\bZ = \bX + \bY$ since
\begin{align*}
\charGen_{\abbrSmooth}(u) = \charGen_{\abbrData}(u)\charGen_{\abbrKernel}(c_{\norig}u) = e^{-u/2}e^{-c_{\norig} u/2} = e^{-(1+c_{\norig})u/2} = \charGen_{\abbrData}((1+c_{\norig})u).
\end{align*}
This allows us to represent $\bZ$ in two different ways, namely as $\bZ \sim \Ellip_{\dim}(\bMu,\dispMat,\charGen_{\abbrSmooth})$ as in the previous section, and as $\bZ \sim \Ellip_{\dim}(\bMu,(1+c_{\norig})\dispMat,\charGen_{\abbrData})$ by invoking a re-parameterization; see Remark~\ref{remark_reparameterization}.
Recalling from \eqref{eq_var_covar_Z} that $\corr{\bZ} = \corr{\bX}$, the second parameterization in fact yields that
\begin{align*}
\ExpCop = \Copula_{\corr{\bZ},\charGen_{\abbrSmooth}} = \Copula_{\corr{\bX},\charGen_{\bX}} = \Copula_{\bX}.
\end{align*}
This is due to the specific properties of the characteristic generator that allows us to shift the effects of multiplying the two characteristic generators $\charGen_{\abbrData}$ and $\charGen_{\abbrKernel}$ into a rescaling of the dispersion matrix.
This rescaling in turn gets lost when considering the associated correlation matrix.
The effect of reverting $\charGen_{\abbrSmooth}$ back to the standard form $\charGen_{\abbrData}$, however, is persisting.

The key property we have used to derive this result is that for all $u\geq 0$ and $\beta \geq 0$ we have
\begin{align}\label{eq_charGen_key_property}
\charGen_{\abbrData}(u)\charGen_{\abbrKernel}(\beta u) = \charGen_{\abbrData}(\gamma u)
\end{align}
for some $\gamma > 0$.
The functional equation in \eqref{eq_charGen_key_property} is reminiscent of a characterization of the exponential function stating that, for non-zero continuous functions, the property
\begin{align}\label{eq_expo_key_property}
  \phi(x)\phi(y) = \phi(x+y)
\end{align}
uniquely characterizes the exponential function; see, for example, \citealt[Exercise~6 of Chapter~8]{Rudin1976}.

\subsubsection{Multivariate elliptical stable distributions}\label{section_multivariate_elliptical_stable_distributions}
Since the requirement in \eqref{eq_charGen_key_property} is less strict than \eqref{eq_expo_key_property}, it is possible to find solutions other than the normal distribution discussed above.
One class of distributions that allows one to shift multiplicative scalars as in \eqref{eq_charGen_key_property} is the class of multivariate elliptical stable distributions; see \cite{Nolan2013} for an overview (note that this reference calls this class of distributions multivariate elliptically contoured stable distributions).
Multivariate elliptical stable distributions are at the intersection of stable and elliptical distributions.
\begin{Definition}[Multivariate elliptical stable distributions]
For $0 < \alpha \leq 2$ a $\dim$-dimensional random vector $\bX$ has a \emph{multivariate elliptical stable distribution} if its characteristic function takes the form
\begin{align*}
\CF_{\bX}(\bt) = \exp\left(\imu \bMu^{\tr} \bt - \left( \bt^{\tr} \dispMat \bt \right)^{\alpha/2} \right),
\end{align*}
where $\dispMat \in \R^{\dim\times\dim}$ is symmetric positive definite and $\bMu \in \R^{\dim}$.
\end{Definition}

The characteristic generator of a multivariate elliptical stable distribution is $\charGen(u) = \exp(-u^{\alpha/2})$. For $\beta \geq 0$, we thus have that
\begin{align*}
\charGen(u) \charGen(\beta u) = \exp(- (1 + \beta^{\alpha/2}) u^{\alpha/2}) =  \charGen(\gamma u)
\end{align*}
with $\gamma = (1 + \beta^{\alpha/2})^{2/\alpha}$.
If the original random vector and the smoothing kernel both belong to the class of multivariate elliptical stable distributions with the same shape parameter $\alpha$ and dispersion matrices $\dispMat$ and $c_{\norig} \dispMat$, it is thus possible to shift the scaling factor $c_{\norig}$ into the dispersion matrix analogously to the case of the multivariate normal distribution.
This in turn leaves the underlying dependence structure undistorted, i.e.,  $\ExpCop = \Copula_{\bX}$.
As a consequence, functionals that only depend on the copula are  unaffected when smoothing.

One application of this result are the values of the upper and lower tail dependence coefficients which are derived in \cite{Schmidt2002} for elliptical random vectors.
For multivariate elliptical stable distributions we have for $0< \alpha \leq 2$ that
\begin{align*}
	1-\charGen(u) = \sum^{\infty}_{k=1} (-1)^{k+1} \frac{u^{k \alpha/2}}{k!} \sim u^{\alpha/2}, \quad\quad\quad (u \to 0),
\end{align*}
and thus by the results of \cite{Bingham1972} in \eqref{equation_Bingham_1} and \eqref{equation_Bingham_2} that the distribution function of the associated radial part is regularly varying with index $-\alpha$ for $0<\alpha<2$.
Following Theorem~5.2 and (5.2) of \cite{Schmidt2002}, this implies a non-zero upper and lower tail dependence coefficient for multivariate elliptical stable distributions, except for the Gaussian case ($\alpha=2$) where there is no tail dependence.
However, since the tail dependence coefficients are functionals of the copula, our calculations show that the tail dependence will not be distorted if the kernel and bandwidth matrix are chosen appropriately.
\subsubsection{Multivariate stable distributions}
To complement the discussion in Section~\ref{section_multivariate_elliptical_stable_distributions} we briefly touch upon the general case of multivariate stable distributions.
We follow \citealt[Chapter~2]{SamorodnitskyTaqqu1994}, with the following definition.
\begin{Definition}[Multivariate stable distribution]\label{definition_multivariate_stable}
A random vector $\bX$ is said to be \emph{stable} in $\R^{\dim}$ if for any positive numbers $b_1>0$ and $b_2>0$ there exists a vector $\bd \in \R^{\dim}$ such that
\begin{align}\label{eq_multivariate_stable}
b_1 \bX_1 + b_2 \bX_2 \equalindist (b_1^{\alpha} + b_2^{\alpha})^{1/\alpha} \bX + \bd,
\end{align}
where $\bX_1$ and $\bX_2$ are independent copies of $\bX$, and $0 < \alpha \leq 2$ does not depend on $b_1$ and $b_2$.
A stable random vector is called \emph{strictly stable}, if \eqref{eq_multivariate_stable} holds with $\bd = \bZero$ for any $b_1>0$ and $b_2>0$.
\end{Definition}
It is straightforward to link Definition~\ref{definition_multivariate_stable} to the framework of multivariate kernel smoothing by choosing a smoothing kernel $\Kernel = \JointDist_{\bX}$.
For $\bY \sim \Kernel$ we then have for $\BW > 0$ that
\begin{align*}
\bZ = \bX + \BW \bY \sim \JointDist_{\bX}\( (1 + h^{\alpha})^{-1/\alpha} (\bx - \bd) \).
\end{align*}
While it is straightforward to show that we have under the current assumptions $\ExpCop = \Copula_{\bX}$, the result is not of much practical use.
On the one hand, if the underlying distribution function $\JointDist_{\bX}$ is known and hence can be used as the kernel distribution no estimation is necessary.
On the other hand, if a smoothing kernel in the class of multivariate stable distributions is chosen, the chances that the underlying data generating process follows the same distribution are (without additional knowledge) slim.

From a theoretical point of view it is, however, noteworthy that multivariate stable distributions are presumably the largest class of distributions for which no dependence distortion occurs if the smoothing kernel is chosen appropriately.

\subsubsection{Multivariate Student $t$ distribution}\label{section_dependence_distortion_student_t}
For the Student $t$ family with $\df>0$ degrees of freedom, the characteristic generator is derived in \cite{Sutradhar1986}.
We utilize the form derived in \cite{JoarderAli1996} and \cite{SongParkKim2014}, given by
\begin{align*}
  \charGen_{\df}(x) = \frac{\Bessel_{\df/2}\left(\sqrt{\df x}\right)\left(\sqrt{\df x}\right)^{\df/2}}{\Gamma\left(\df/2\right)2^{\df/2-1}},
\end{align*}
where $\Bessel_{\alpha}$ denotes the modified Bessel function of the second kind (\cite{JoarderAli1996} refer to it as the Mcdonald function, \cite{SongParkKim2014} and \cite{KotzKozubowskiPodgorski2001} refer to it as the modified Bessel function of the third kind) which can be represented as
\begin{align}\label{eq_definition_modified_bessel_function_2nd_kind}
\Bessel_{\alpha}(t) = \left(\frac{2}{t}\right)^{\alpha} \frac{\Gamma\left(\alpha+1/2\right)}{\sqrt{\pi}} \int_0^{\infty} \left(1+u^2\right)^{-\left(\alpha+\frac{1}{2}\right)} \cos(tu)\d u
\end{align}
for $t > 0$ and $\alpha > -1/2$; see, for example, \citealt[Equation~10.32.11]{NIST:DLMF}.
For parameters of the form $\alpha = r + 0.5$ with $r \in \{0,1,2,\ldots\}$ we have the explicit formula
\begin{align}\label{eq_Bessel_integerPlusHalf}
\Bessel_{\alpha}(t) = \sqrt{\frac{\pi}{2t}}e^{-t} \sum^{r}_{k=0} \frac{(r+k)!}{(r-k)! k!} (2t)^{-k},
\end{align}
see \citealt[Equation~(A.0.10)]{KotzKozubowskiPodgorski2001}.
For the special case $\alpha=1/2$ we consequently have
\begin{align*}
\Bessel_{\frac{1}{2}}(t) = \sqrt{\frac{\pi}{2t}}e^{-t},
\end{align*}
see also \citealt[Equation~10.39.2]{NIST:DLMF}.
For the special case of the multivariate Cauchy distribution we have $\df = 1$ and therefore obtain the characteristic generator as
\begin{align*}
\charGen_1(u) = \frac{\Bessel_{\frac{1}{2}}\left(\sqrt{u}\right)\left(\sqrt{u}\right)^{1/2}}{\Gamma\left(\frac{1}{2}\right)2^{-\frac{1}{2}}}
= \sqrt{\frac{2}{\pi}} u^{\frac{1}{4}}\sqrt{\frac{\pi}{2\sqrt{u}}}e^{-\sqrt{u}}
= e^{-\sqrt{u}}.
\end{align*}
If the smoothing kernel is also a multivariate Cauchy distribution we consequently have for $\beta \geq 0$ that,
\begin{align*}
  \charGen_1(u)\charGen_1(\beta u) = e^{-\sqrt{u}} e^{-\sqrt{\beta u}}
= e^{-\sqrt{u\left(1+\sqrt{\beta}\right)^2}} = \charGen_1\left(u\left(1+\sqrt{\beta}\right)^2\right).
\end{align*}
This is as expected since the Cauchy distribution is an elliptical stable distribution with index $\alpha = 1$.

For general $\df \neq 1$ the modified Bessel function of the second kind does not reduce to the exponential function.
A similar rescaling is therefore in general not possible for the Student $t$ distribution.
This implies that for $\df \neq 1$ the dependence structure will be distorted even when the smoothing kernel is chosen to match the distribution of the original data.
For a concrete example we set $\df = 3$ and obtain via \eqref{eq_Bessel_integerPlusHalf} that
\begin{align*}
\charGen_3(u) = e^{-\sqrt{3u}}\(1 + \sqrt{3u}\).
\end{align*}
Consequently we have
\begin{align*}
\charGen_3(u)\charGen_3(\beta u) = e^{-\sqrt{3u\(1+\sqrt{\beta}\)^2}}\(1 + \sqrt{3u\(1+\sqrt{\beta}\)^2} + 3u\sqrt{\beta}\) = \charGen_3(\gamma u) e^{-\sqrt{3 \gamma u}}3u\sqrt{\beta},
\end{align*}
where $\gamma = \(1+\sqrt{\beta}\)^2$.
Hence we can identify the multiplicative term $e^{-\sqrt{3 \gamma u}}3u\sqrt{\beta}$ on the right as being responsible for the dependence distortion.

\subsubsection{Elliptical distributions not elliptical stable}\label{section_dependence_distortion_Laplace}
As a last example, we consider an elliptical distribution which is not
elliptical stable. We focus on the multivariate Laplace distribution, in which
case \eqref{eq_charGen_key_property} will not hold.
For this model the effect of the
dependence distortion can be worked out explicitly.
A general introduction to the multivariate Laplace distribution can be found in \cite{KotzKozubowskiPodgorski2001}.

Denote by $\bX \sim \Ellip_{\dim}(\bMu,\dispMat,\charGen_L)$ a $\dim$-dimensional random vector with characteristic generator
\begin{align*}
\charGen_L(u) = \frac{1}{1+u/2}.
\end{align*}
For such a random vector $\bX$ the density generator is given by
\begin{align*}
\mdensII_L(t) = \frac{2}{(2\pi)^{\dim/2}}\left(\frac{t}{2}\right)^{(2-\dim)/4}\Bessel_{(2-\dim)/2}\left(\sqrt{2t}\right),
\end{align*}
see \citealt[Equation~(5.2.2)]{KotzKozubowskiPodgorski2001}, and $\bX$ follows a multivariate Laplace distribution; see \citealt[Equations~(5.2.1) and (5.2.2)]{KotzKozubowskiPodgorski2001}. The corresponding radial distribution $\radialPart_L$ of $\bX$ has the density
\begin{align*}
  \mdensI_{\radialPart_L}(x) = \frac{2x^{\dim/2}\Bessel_{\dim/2-1}\left(x\sqrt{2}\right)}{\left(\sqrt{2}\right)^{\dim/2-1}\Gamma(\dim/2)},\quad x > 0;
\end{align*}
see \citealt[Proposition~6.3.1]{KotzKozubowskiPodgorski2001}. In this case, the
product of the characteristic generators related to kernel smoothing for
$\beta \geq 0$ is given by
\begin{align*}
\charGen_{\abbrSmooth}(u) =
\charGen_L(u)\charGen_L(\beta u)
= \frac{1}{1+u/2} \frac{1}{1+\beta u/2}
= \frac{1}{1+u(1+\beta)/2 + \beta u^2/4},
\end{align*}
where the term $\beta u^2/4$ in the denominator prevents a simplification as in the Gauss and Cauchy (and general multivariate elliptical stable) cases discussed before.
In this example it is therefore clearly not possible to convert the effects of dependence distortion into a rescaling of the dispersion matrix.
When trying to identify the radial distribution connected to the generator $\charGen_{\abbrSmooth}$ resulting from smoothing with a Laplace kernel (with an appropriate bandwidth matrix) we can use partial fraction decomposition for $\beta\neq 1$ to get
\begin{align*}
\charGen_{\abbrSmooth}(u) &= \frac{2}{1 - \beta}\left(\frac{1}{u+2} - \frac{\beta}{\beta u + 2}\right)
= \frac{1}{1 - \beta}\charGen_L(u) - \frac{\beta}{1 - \beta}\charGen_L(\beta u).
\end{align*}
To identify the radial distribution connected to $\charGen_{\abbrSmooth}$ we can draw on the connection between characteristic generators and radial distributions; see \citealt[Section~2.1]{FangKotzNg1990}.
Specifically, for any spherical distribution $\Ellip_{\dim}(\bZero,\IdMat_{\dim},\charGen)$ there exists a distribution function $F_R$ of an a.s. positive random variable $R$ (the radial part) such that
\begin{align*}
\charGen(u) = \int_0^{\infty} \Omega_{\dim}\left(ut^2\right)\d F_R(t),
\end{align*}
where $\Omega_{\dim}$ is the characteristic generator of a random vector $\sphericalPart$ uniformly distributed on the unit sphere $\{\bx \in \R^{\dim} : \xNorm{\bx}{2} = 1\}$, i.e., $\CF_{\sphericalPart}(\bt) = \E\left[e^{\imu \bt^{\tr}\sphericalPart}\right] = \Omega_{\dim}\left(\xNorm{\bt}{2}^2\right)$, see \citealt[Theorem~2.2, page~29]{FangKotzNg1990}.
Concerning $\charGen_{\abbrSmooth}$, we consequently have that
\begin{align*}
\charGen_L(u) &= \int_0^{\infty} \Omega_{\dim}\left(ut^2\right) \mdensI_{\radialPart_L}(t) \d t,\\
\charGen_L(\beta u) &= \int_0^{\infty} \Omega_{\dim}\left(\beta ut^2\right)\mdensI_{\radialPart_L}(t) \d t = \int_0^{\infty} \Omega_{\dim}\left(ut^2\right)\mdensI_{\radialPart_L}\left(t / \sqrt{\beta}\right) / \sqrt{\beta} \d t,
\end{align*}
where the last equality is obtained from a substitution.
Using linearity and the integral representation of $\charGen_L$, we now have that
\begin{align*}
\charGen_{\abbrSmooth}(u)
&= \frac{1}{1 - \beta}\charGen_L(u) - \frac{\beta}{1 - \beta}\charGen_L(\beta u)\\
&= \frac{1}{1 - \beta} \int_0^{\infty} \Omega_{\dim}\left(ut^2\right) \mdensI_{\radialPart_L}(t) \d t - \frac{\beta}{1 - \beta}\int_0^{\infty} \Omega_{\dim}\left(ut^2\right)\mdensI_{\radialPart_L}\left(t / \sqrt{\beta}\right) / \sqrt{\beta} \d t\\
&= \int_0^{\infty} \Omega_{\dim}\left(ut^2\right) \mdensI_{\radialPart_{\abbrSmooth}}(t) \d t,
\end{align*}
where the radial density $\mdensI_{\radialPart_{\abbrSmooth}}$ connected to $\charGen_{\abbrSmooth}$ can be identified as
\begin{align}\label{eq_radial_distortion_Laplace}
\mdensI_{\radialPart_{\abbrSmooth}}(x) &= \frac{1}{1-\beta}\left(  \mdensI_{\radialPart_L}(x) - \sqrt{\beta} \mdensI_{\radialPart_L}\left(x / \sqrt{\beta} \right) \right),\quad x > 0.
\end{align}
When smoothing the considered multivariate Laplace distribution with a matching multivariate Laplace kernel and an appropriate bandwidth matrix $\BWMatrix = \beta \dispMat$, the smoothing thus affects the radial distribution as shown in \eqref{eq_radial_distortion_Laplace}.
The difference between $\mdensI_{\radialPart_{\abbrSmooth}}$ and the original radial density $\mdensI_{\radialPart_L}$ distorts the joint distribution and hence the implied copula.

This concludes our theoretical discussion and investigation of dependence structure distortion at the population level. In the next section, we link this to the smooth bootstrap.

\section{Smooth bootstrap}\label{section_bootstrap}

Section~\ref{section_kernel_smoothing} focuses on dependence distortion at the population level for kernel based estimators.
In this section, we connect our previous results to bootstrapping.
While we discuss asymptotic theory where appropriate, we mainly focus on the smooth bootstrap as a computational tool.
A general introduction to the theory of the non-parametric bootstrap can be found in \citealt[Chapter~1]{Hall1992}, and \cite{ShaoTu1995}.

The smooth bootstrap is motivated by, and closely related to, kernel density estimation as discussed in Section~\ref{section_prelim}.
To introduce the smooth bootstrap we denote by $\norig$ the original sample size, while $\nbs$ denotes the overall number of smooth bootstrap replications.
Each bootstrap sample is of size $\naugm$.
While $\naugm = \norig$ is a typical choice, it is possible to use $\naugm \gg \norig$ in a data augmentation situation.
Furthermore, $\tau_b$, $b\in\{1,\ldots,\nbs\}$, denotes a random vector uniformly distributed on $\times_{i=1}^{\naugm} \{1,\ldots,\norig\}$ with components $\tau_{bi}$ for $i \in \{1,\ldots,\naugm\}$ independent of the sample $\sample = \{\bX_i\}_{i=1}^{\norig}$.
In this case the individual components of $\tau_b$ are clearly independent and uniformly distributed on $\{1,\ldots,\norig\}$.
A bootstrap sample $\samplebs_b = (\bX^*_{bi})_{i=1}^{\naugm}$, $b \in \{1,\ldots,\nbs\}$, is now generated via $\bX^*_{bi} = \bX_{\tau_{bi}}$.
The draws $\tau_1,\ldots,\tau_{\nbs}$ are assumed to be independent which thus carries over to the (non-parametric) bootstrap samples $\{\samplebs_b\}_{b=1}^{\nbs}$.

The smooth bootstrap sample $\samplesbs_b = \{\bZ^*_{bi}\}_{i=1}^{\naugm}$ is obtained by setting $\bZ^*_{bi} = \bX^*_{bi} + \bY_{bi}$, where $\bY_{bi}$ is distributed according to a smoothing kernel $\Kernel_{\BWMatrix_{\norig}}$ independently of $\{\bX_i\}_{i=1}^{\norig}$ and $\{\tau_b\}_{b=1}^{\nbs}$.
This again leads to independence among all the components of $\{\samplesbs_b\}_{b=1}^{\nbs}$.
The smooth bootstrap sampling scheme can also be interpreted from the kernel smoothing perspective discussed in Sections~\ref{section_prelim} and \ref{section_kernel_smoothing}.
Considering the unconditional distribution of $\bZ^*_{bi}$ it can readily be seen that $\bZ^*_{bi} \equalindist \bX + \bY$, where $\bX$ and $\bY$ are independent random vectors with densities $\JointDens_{\bX}$ and $\kernel_{\BWMatrix_{\norig}}$.
The (unconditional) density of $\bZ^*_{bi}$ is therefore given by $\expkde$ defined in \eqref{eq_expected_kernel_density_estimate}, the expected kernel density estimate, and the observations in Sections~\ref{section_population_version}--\ref{section_exceptions} apply accordingly.

Conditionally on $\sample$, the smooth bootstrap is equivalent to sampling from the mixture density obtained from the kernel density estimation for a given dataset.
We denote this mixture density by
\begin{align}\label{eq_cond_kernel_density}
  \condkde(\bx) = \kde_{\norig}(\bx\,|\,\sample) = \frac{1}{\norig} \sum^{\norig}_{i=1} \kernel_{\BWMatrix_{\norig}}(\bx-\bx_i),
\end{align}
where $\bx_i = \bX_i(\omega)$, $i\in\{1,\dots,\norig\}$, for some fixed $\omega \in \Omega$.
Along the same lines we denote the joint distribution function conditional on $\sample$ by
\begin{align}\label{eq_cond_kernel_distribution}
  \condKDE(\bx) = \KDE_{\norig}(\bx\,|\,\sample) = \frac{1}{\norig} \sum^{\norig}_{i=1} \Kernel_{\BWMatrix_{\norig}}(\bx-\bx_i).
\end{align}
In general, sampling from a mixture density proceeds in two steps.
First, one of $\norig$ possible groups is selected with equal probability $n^{-1}$.
The second step consist of drawing a random vector from the corresponding mixing density, which in our case is represented by a random vector that follows $\Kernel_{\BWMatrix_{\norig}}$ centered at the randomly selected $\bx_i$ representing the $i$th group.
Given $\sample$, we thus equivalently have that $\bZ^*_{bi} = \bx^*_{bi} + \bY_{bi}$ in the smooth bootstrap, where $\bx^*_{bi}$ is selected uniformly from $\{\bx_1,\ldots,\bx_{\norig}\}$ according to $\tau_b$.

In the following section, we apply the smooth bootstrap to copula functionals and provide an algorithm that draws a smoothed sample from given observations.
\subsection{The smooth bootstrap for copula functionals}\label{section_smooth_bootstrap_copula_functionals}

In Section~\ref{section_dependence_distortion_elliptical} we have discussed how smoothing impacts the joint distribution and dependence structure of the expected kernel density estimate $\expkde$ with a special focus on elliptical distributions and kernels.
In this section we discuss properties of $\condkde$ with a focus on the implied dependence structure.

If the marginal quantile functions $\kdeMargDist_{\norig 1}^{-1},\ldots,\kdeMargDist_{\norig \dim}^{-1}$ associated to the joint distribution function $\condKDE$ implied by the density $\condkde$ are known, Sklar's Theorem can be used to extract the corresponding copula (conditional on the data $\sample$) via
\begin{align}\label{eq_cond_copula}
\CopulaCondKDE(\bu) = \condKDE\left(\kdeMargDist_{\norig 1}^{-1}(u_1),\ldots,\kdeMargDist_{\norig \dim}^{-1}(u_{\dim})\right).
\end{align}
However, recovering the marginal distribution and quantile functions from $\condkde$ defined in \eqref{eq_cond_kernel_density} is not an easy task in general.
In the following discussion we will thus limit ourselves to kernels that are elliptical. Aside from elliptical kernels, it is straightforward to extract the marginal distributions in the case of product kernels. There, the marginal distributions are given by standard univariate kernel distribution estimates where the respective bandwidths are selected individually by virtue of a diagonal bandwidth matrix $\BWMatrix_{\norig} = \diag(\BW_{\norig 1},\ldots,\BW_{\norig \dim})$.

If the kernel $\kernel$ is the density of an elliptical random vector $\bY \sim \Ellip_{\dim}(\bZero,\IdMat_{\dim},\charGen_{\abbrKernel})$, the rescaled and shifted kernel $\kernel_{\BWMatrix_{\norig}}(\bx - \bx_i)$, $\bx \in \R^{\dim} $, can be identified as the density of $\bZ_i = \bx_i + \BWMatrix_{\norig}^{1/2} \bY \sim \Ellip_{\dim}(\bx_i,\BWMatrix_{\norig},\charGen_{\abbrKernel})$; see Theorem~\ref{theorem_elliptical_linear_combinations}.
Considering \eqref{eq_univariate_margins_elliptical}, the $j$th marginal distribution of $\bZ_i$ is hence given by $Z_{ij} \sim \Ellip_1(\bx_{ij},\BWMatrix_{\norig jj},\charGen_{\abbrKernel})$.
Denoting by $\mdistI_{\charGen_{\abbrKernel}}$ the common univariate marginal distribution function of $\bY$, the distribution function of $Z_{ij}$, $j \in \{1,\ldots,\dim\}$, is then given in terms of a location-scale model taking the form
\begin{align*}
  \mdistI_{Z_{ij}}(z) =
  \Prob{ Z_{ij} \leq z } = \mdistI_{\charGen_{\abbrKernel}}\left(\frac{z - \bx_{ij}}{\sqrt{\BWMatrix_{\norig jj}}}\right), \quad z \in \R.
\end{align*}
In this specific setup, the marginal distribution functions of $\condKDE$ are given by mixture distribution functions of location-scale models of $\mdistI_{\charGen_{\abbrKernel}}$.
For the $j$th margin we have that
\begin{align}\label{eq_univariate_margins_condKDE}
  \kdeMargDist_{\norig j}(x) = \frac{1}{\norig}\sum^{\norig}_{i=1} \mdistI_{\charGen_{\abbrKernel}}\left(\frac{x - \bx_{ij}}{\sqrt{\BWMatrix_{\norig jj}}}\right), \quad x \in \R.
\end{align}
The marginal distributions of $\condKDE$ are thus given by univariate kernel distribution estimates with bandwidth $\BW_{\norig j} = \sqrt{\BWMatrix_{\norig jj}}$ and kernel density $\mdensI_{\charGen_{\abbrKernel}} = \mdistI_{\charGen_{\abbrKernel}}'$.

To make use of Sklar's Theorem, the corresponding quantile functions $\kdeMargDist_{\norig j}^{-1}$ need to be available.
For $p \in (0,1)$ the quantile function $\kdeMargDist_{\norig j}^{-1}(p)$ is (here) defined via the inverse relationship
\begin{align}\label{eq_univariate_quantile_inversion}
\kdeMargDist_{nj}(x) = p.
\end{align}
The probabilistic behaviour of univariate quantile functions for $\norig \to \infty$ in this case is studied in \cite{Nadaraya1964, Nadaraya1964RU} and \cite{Azzalini1981}.
Since in our case the marginal distributions are given in closed form in \eqref{eq_univariate_margins_condKDE}, the corresponding quantile functions $\kdeMargDist_{\norig j}^{-1}$ can be computed for a fixed $\norig$ via numerical inversion.
This leads to a tractable numerical evaluation of $\CopulaCondKDE$ for an argument $\bu$ in the case of elliptical kernels.

Concerning the numerical inversion, the value of a sample quantile can serve as a starting point for numerical algorithms as suggested in \cite{Azzalini1981}.
When a starting interval instead of a starting point is required for the numerical search, the value of $\kdeMargDist_{nj}$ at the first and last order statistic (in component $j$) can be used to obtain a first estimate of the relevant search region.

In \cite{Azzalini1981}, an optimal bandwidth for deriving a quantile via the implicit definition in \eqref{eq_univariate_quantile_inversion} is given.
In this case the asymptotic mean square optimal bandwidth is proportional to $\norig^{-\frac{1}{3}}$.
However, the optimal bandwidth for smoothing the joint density (or distribution) will in general depend on the dimension $\dim$.
This makes it necessary to compromise either on the marginal or joint distributional level when selecting the bandwidth.
When different bandwidths are chosen for the marginal smoothing and quantile computation via \eqref{eq_univariate_quantile_inversion}, and for the smoothing of the joint distribution via \eqref{eq_cond_kernel_distribution}, the resulting combination in
\eqref{eq_cond_copula} is not a proper copula since the resulting margins are not adapted to the joint distribution.
It is important to point out that in an asymptotic $\norig \to \infty$ consideration different bandwidth choices for the margins and joint distribution might not pose any problems as long as the usual conditions are obeyed.
In a setting with a fixed $\norig$, it is, however, not possible to mix different bandwidths and obtain a proper copula, even if they might be optimal when considered individually.

If an elliptical kernel (or any other multivariate kernel with accessible marginal distributions) is used to construct the kernel density estimate, it is possible to simulate from the implied conditional copula in \eqref{eq_cond_copula}.
This is done by combining the simulation of mixture distributions discussed at the end of the previous section with the marginal distributions given in \eqref{eq_univariate_margins_condKDE}.
This leads to the following algorithm to draw a random sample from $\CopulaCondKDE$.
\begin{Algorithm}[Smooth bootstrap sample from $\CopulaCondKDE$]\label{algo_smoothed_copula_bootstrap}
Denote by $\{\bx_i\}_{i=1}^{\norig}$ a given set of observations and assume that
a sensible bandwidth matrix $\BWMatrix_{\norig}$ has been determined (see, e.g., the discussion in Section~\ref{section_bandwidth_selection}).
To draw a pseudo-random sample $(\bu^*_{\ell})_{\ell=1}^\naugm$ from $\CopulaCondKDE$ of size $\naugm$ repeat the following steps for $\ell\in\{1,\ldots,\naugm\}$:
\begin{enumerate}
\item Draw a pseudo-random $\by_{\ell}$ distributed according to the kernel density $\kernel$.
\item Draw an index $i$ uniformly from $\{1,\ldots,\norig\}$ and set $\bz^*_{\ell} = \bx_i + \BWMatrix_{\norig}^{1/2} \by_{\ell}$.
\item Return $\bu^*_{\ell} = \left(\kdeMargDist_{\norig 1}(z^*_{\ell 1}),\ldots,\kdeMargDist_{\norig \dim}(z^*_{\ell \dim})\right)$, where $\kdeMargDist_{\norig j}$ is the $j$th marginal distribution of $\condKDE$.
\end{enumerate}
\end{Algorithm}
\begin{Remark}
  Algorithm~\ref{algo_smoothed_copula_bootstrap} is based on a direct
  application of the kernel density estimate.  However, in certain situations a
  straightforward application of the kernel density estimation might not be
  possible.  The most common situation of the latter type is when the multivariate
  joint distribution is only supported on a compact set, leading to a boundary
  bias in the estimation.  Specifically relevant to our investigation are the
  cases when the original sample $\sample$ consists (i) of observations of a
  copula $\Copula$, or (ii) of copula pseudo-observations obtained by applying
  the marginal empirical distribution functions to the component samples.  To avoid boundary
  issues in such cases it is possible to transform the observations from
  $[0,1]^{\dim}$ to $\R^{\dim}$ by an appropriate (bijective) marginal
  transformation.  Although any continuous marginal distribution function can be
  used for this transformation, it is most common to use the standard normal
  distribution function, see, for example, \cite[Section~5.10.3]{Joe2015}.
Transformation re-transformation
  approaches in the context of non-parametric copula estimation have also been
  studied in \cite{OmelkaGijbelsVeraverbeke2009} and
  \cite{GeenensCharpentierPaindaveine2018}.  In the new domain the smoothing can
  then be carried out according to
  Algorithm~\ref{algo_smoothed_copula_bootstrap}.  Since the copula is invariant
  under strictly increasing marginal transformations, the resulting sample is still representative
  of the underlying copula.
\end{Remark}
The smooth bootstrap outlined in Algorithm~\ref{algo_smoothed_copula_bootstrap} can now be used in two situations when dealing with a copula functional $\func$ and its empirical version $\func_{\norig}$ defined for samples $\sample$ (of arbitrary size $\norig$). First, the smooth bootstrap can be used to gauge the distribution (and other characteristics) of $\func_{\norig}$ for a fixed sample size $\norig$.
While the original sample $\sample$ only allows for one realization of $\func_{\norig}$, one can use the smooth bootstrap to gauge for example the distribution of $\func_{\norig}$.
Via Algorithm~\ref{algo_smoothed_copula_bootstrap} one can draw $\nbs$ smooth bootstrap samples $(\samplesbs_b)_{b=1}^{\nbs}$ of size $\norig$.
This leads to $\nbs$ smooth bootstrap observations $(\func_{\norig}^b)_{b=1}^{\nbs}$ of $\func_{\norig}$, which are based on $\condKDE$ with underlying copula $\CopulaCondKDE$.
While the number of bootstrap samples $\nbs$ is under the control of the statistician, it is crucial to verify whether the resulting bootstrap distribution is (asymptotically for $\norig\to\infty$) representative for the distribution of $\func_{\norig}$ at the population level.
A general discussion of this issue for the smooth bootstrap can be found in \citealt[Chapter~3.5]{ShaoTu1995}.

Second, the bootstrap can also be used as a method of data augmentation.
This method can come into play when an approximation of $\func(\Copula_{\bX})$ is constructed by replacing the unknown copula $\Copula_{\bX}$ with either the empirical copula or the smooth version $\CopulaCondKDE$.
This can either be necessary when the computations based on the empirical copula, i.e., the original sample $\{\bx_1,\ldots,\bx_{\norig}\}$, are too coarse to be useful,
 or to facilitate the (numerical) approximation of $\func\left(\CopulaCondKDE\right)$.
For certain functionals $\func$, such as level sets, or Kendall's tau and Spearman's rho which are given as multivariate integrals, the computation of $\func\left(\CopulaCondKDE\right)$ might pose (numerical) challenges even if $\CopulaCondKDE$ is in principle known and can be evaluated via \eqref{eq_cond_copula}.
From a practical point of view it can then be easier to use an approximation
\begin{align*}
\lim_{\naugm \to \infty} \func_{\naugm}\(\bu^*_1,\ldots,\bu^*_{\naugm}\) = \func\left(\CopulaCondKDE\right),
\end{align*}
if a suitable sample version $\func_{\naugm}$ is available.
To make sense of the limit we need to formally define a distance between $\func_{\naugm}\(\bu^*_1,\ldots,\bu^*_{\naugm}\)$ and $\func\left(\CopulaCondKDE\right)$ and we circle back to this issue in Section~\ref{section_bs_consistency}.

In both cases, instead of creating $\nbs$ samples of size $\norig$ to assess the distribution of $\func_{\norig}$, only one smooth bootstrap sample $\samplesbs_{\naugm} = (\bu^*_{i})_{i=1}^{\naugm}$ of size $\naugm \gg \norig$ is created.
Contrary to the approximation based on the original sample $\func_{\norig}\(\bx_1,\ldots,\bx_{\norig}\)$, Algorithm~\ref{algo_smoothed_copula_bootstrap} allows us to sample an arbitrary number $\naugm$ of pseudo-observations.
If the functional is well behaved, the resulting $\funcbs_{\naugm} = \func\left(\samplesbs_{\naugm}\right)$ is then a close approximation to $\func\left(\CopulaCondKDE\right)$.

An open question is which bandwidth matrix to use in Algorithm~\ref{algo_smoothed_copula_bootstrap}. This is addressed in the subsequent section.
\subsection{Cross-validation bandwidth selection}\label{section_bandwidth_selection}

A crucial part of the suggested procedure is the selection of the bandwidth matrix $\BWMatrix$.
In the univariate case the asymptotic mean integrated squared error (AMISE) optimal bandwidth is of order $\Landau \( \norig^{-1/5} \)$ in the case of density estimation.
However, as shown in \cite{Azzalini1981} the AMISE optimal bandwidth for distribution estimation is of order $\Landau \( \norig^{-1/3} \)$.
In the multivariate case we can thus not expect that bandwidth selection techniques designed for density estimation will work well when estimating distribution functions.

While a variety of bandwidth selection methods are available in the case of multivariate kernel density estimation, this is not the case when estimating distribution functions.
When restricting oneself to product kernels, and hence diagonal bandwidth matrices in our setting, a plug-in estimator can be found in \cite{LiuYang2008}.
Given that product kernels are too restrictive in our setup we instead turn to cross-validation for bandwidth selection.

In the univariate case \cite{BowmanHallPrvan1998} adapt cross-validation for kernel distribution estimators by introducing the objective function
\begin{align*}
\CV^1_{\norig}(\BW) = \frac{1}{\norig} \sum^{\norig}_{i=1} \int_{\R} \left( \ind{(-\infty, x]}(X_i) - \widehat{\mdistI}_{-i}(x) \right)^2 \d x = \frac{1}{\norig} \sum^{\norig}_{i=1} \int_{\R} \left( \ind{[X_i,\infty)}(x) - \widehat{\mdistI}_{-i}(x) \right)^2 \d x,
\end{align*}
where $\widehat{\mdistI}_{-i}$ is the (in this case one-dimensional) leave-one-out kernel distribution estimator.
Minimization of $\CV^1_{\norig}(\BW)$ with respect to $\BW$ then leads to a sensible optimal bandwidth as argued in \cite{BowmanHallPrvan1998}.
While the integral is finite for compactly supported kernel functions, kernels with support on $\R$ can be used if they decay fast enough which can be seen when decomposing the integral as
\begin{align*}
\int_{\R} \left( \ind{[X_i,\infty)}(x) - \widehat{\mdistI}_{-i}(x) \right)^2 \d x = \int_{-\infty}^{X_i} \left(\widehat{\mdistI}_{-i}(x) \right)^2 \d x + \int_{X_i}^{\infty} \left( 1 - \widehat{\mdistI}_{-i}(x) \right)^2 \d x.
\end{align*}

However, in the multivariate case a direct generalization of $\CV^1_{\norig}$ is only valid for compactly supported kernels.
When supported over $\R^{\dim}$, the respective integrals will generally not converge.
To solve this issue we introduce a weight function $\weightFunction \colon \R^{\dim} \to [0,\infty)$ and define a weighted multivariate version of $\CV^1_{\norig}$ as
\begin{align}\label{eq_CV}
\CV^{\dim}_{\norig}(\BWMatrix;\weightFunction) = \frac{1}{\norig} \sum^{\norig}_{i=1} \int_{\R^{\dim}} \left( \ind{(-\binfty,\bx]}(\bX_i) - \LeaveOneOut(\bx) \right)^2 \weightFunction(\bx) \d \bx,
\end{align}
where $(-\binfty,\bx] = \times_{i=1}^{\dim} (-\infty,x_i]$ for $\bx = (x_1,\ldots,x_{\dim}) \in \R^{\dim}$.
Here $\LeaveOneOut$ denotes the leave-one-out kernel distribution estimate when disregarding the $i$th observation, i.e.,
\begin{align*}
\LeaveOneOut(\bx;\sample) = \LeaveOneOut(\bx) = \frac{1}{\norig-1} \sum^{\norig}_{\substack{j=1 \\ j \neq i}} \Kernel_{\BWMatrix}(\bx-\bX_j).
\end{align*}
As a measure of performance we consider the weighted mean integrated squared error ($\MISE$)
\begin{align}\label{eq_MISE}
\MISE_{\norig}^{\dim}(\BWMatrix;\weightFunction)= \Eval{\int_{\R^{\dim}} \left( \KDE_{\norig}(\bx) - \JointDist_{\bX}(\bx) \right)^2 \weightFunction(\bx) \d \bx}
\end{align}
as a function of the bandwidth matrix $\BWMatrix$.
A $\MISE$ optimal bandwidth matrix is any matrix that minimizes \eqref{eq_MISE}.

In the univariate case with $\weightFunction \equiv 1$, \cite{BowmanHallPrvan1998} show that $\Eval{\CV^1_{\norig}(\BW)} = \MISE^1_{\norig-1}(\BW)$ up to a constant shift term that is independent of $\BW$ which justifies minimizing $\CV^1_{\norig}(\BW)$ to find a sensible bandwidth.
In the multivariate case we derive the following generalization concerning the objective function in \eqref{eq_CV}.
\begin{Theorem}\label{theorem_cross_validation}
If $\int_{\R^{\dim}} \weightFunction(\bx) \d \bx < \infty$ then
\begin{align}\label{eq_expectation_cross_valiation}
\Eval{ \CV^{\dim}_{\norig}(\BWMatrix;\weightFunction) } = \MISE_{\norig-1}^{\dim}(\BWMatrix;\weightFunction) + \intConst_{\bX}(\weightFunction),
\end{align}
where $\intConst_{\bX}(\weightFunction)$ is independent of $\BWMatrix$ and $\norig$ and given by
$
\intConst_{\bX}(\weightFunction) = \Eval{\int_{\R^{\dim}} \left( \ind{(-\binfty,\bx]}(\bX) - \JointDist_{\bX}(\bx) \right)^2 \weightFunction(\bx) \d \bx }.
$
\begin{proof}
We first note that the existence of the involved integrals is guaranteed by integrability of the weight function.
Due to the iid setting we have that
$\Eval{ \ind{(-\binfty,\bx]}(\bX_i) \LeaveOneOut(\bx) } = \JointDist_{\bX}(\bx)\Eval{ \LeaveOneOut(\bx) }$
and therefore
\begin{align*}
&\phantom{{}={}}\Eval{ \CV^{\dim}_{\norig}(\BWMatrix) } - \intConst_{\bX}(\weightFunction) \\
&= \Eval{ \frac{1}{\norig} \sum^{\norig}_{i=1} \int_{\R^{\dim}} \left( \ind{(-\binfty,\bx]}(\bX_i) - \LeaveOneOut(\bx) \right)^2 \weightFunction(\bx) \d \bx } - \Eval{ \int_{\R^{\dim}} \left( \ind{(-\binfty,\bx]}(\bX) - \JointDist_{\bX}(\bx) \right)^2 \weightFunction(\bx) \d \bx }\\
&= \Eval{ \frac{1}{\norig} \sum^{\norig}_{i=1} \int_{\R^{\dim}} \left( \ind{(-\binfty,\bx]}(\bX_i) - \LeaveOneOut(\bx) \right)^2 \weightFunction(\bx) \d \bx - \frac{1}{\norig} \sum^{\norig}_{i=1} \int_{\R^{\dim}} \left( \ind{(-\binfty,\bx]}(\bX_i) - \JointDist_{\bX}(\bx) \right)^2 \weightFunction(\bx) \d \bx } \\
&= \E\Big[ \frac{1}{\norig} \sum^{\norig}_{i=1} \int_{\R^{\dim}} \Big( \ind{(-\binfty,\bx]}(\bX_i)^2 + \LeaveOneOut(\bx)^2 - 2\ind{(-\binfty,\bx]}(\bX_i) \LeaveOneOut(\bx) \\
&\phantom{= \E\Big[ \frac{1}{\norig} \sum^{\norig}_{i=1} \int_{\R^{\dim}} \Big(}-
\ind{(-\binfty,\bx]}(\bX_i)^2 - \JointDist_{\bX}(\bx)^2 + 2\ind{(-\binfty,\bx]}(\bX_i)\JointDist_{\bX}(\bx) \Big) \weightFunction(\bx) \d \bx \Big] \\
&= \frac{1}{\norig} \sum^{\norig}_{i=1} \int_{\R^{\dim}} \left( \Eval{ \LeaveOneOut(\bx)^2 } - 2 \Eval{ \ind{(-\binfty,\bx]}(\bX_i) \LeaveOneOut(\bx) } - \JointDist_{\bX}(\bx)^2 + 2 \Eval{ \ind{(-\binfty,\bx]}(\bX_i) } \JointDist_{\bX}(\bx) \right) \weightFunction(\bx) \d \bx \\
&= \frac{1}{\norig} \sum^{\norig}_{i=1} \int_{\R^{\dim}} \left( \Eval{ \LeaveOneOut(\bx)^2 } - 2 \JointDist_{\bX}(\bx)\Eval{ \LeaveOneOut(\bx) } - \JointDist_{\bX}(\bx)^2 + 2 \JointDist_{\bX}(\bx)^2 \right) \weightFunction(\bx) \d \bx\\
&= \frac{1}{\norig} \sum^{\norig}_{i=1} \Eval{ \int_{\R^{\dim}} \left( \LeaveOneOut(\bx) - \JointDist_{\bX}(\bx) \right)^2 \weightFunction(\bx) \d \bx }\\
&= \Eval{ \int_{\R^{\dim}} \left( \KDE_{\norig-1}(\bx) - \JointDist_{\bX}(\bx) \right)^2 \weightFunction(\bx) \d \bx }
= \MISE_{\norig-1}^{\dim}(\BWMatrix;\weightFunction)
\end{align*}
from which the claim follows by rearranging terms.
The step from the third to last equality to the last line is justified by the identical distribution of the $\bX_i$, leading to $\norig$ times the same expectation.
\end{proof}
\end{Theorem}
Clearly, if the kernel is compactly supported we can set $\weightFunction \equiv 1$ and obtain a direct generalization of the result in \cite{BowmanHallPrvan1998} along the same lines.
Although the constant $\intConst_{\bX}$ is typically unknown in a realistic setting, $\intConst_{\bX}$ does not depend on $\BWMatrix$ and is hence irrelevant for the minimization.
This justifies minimizing $\CV^{\dim}_{\norig}(\BWMatrix;\weightFunction)$ with respect to $\BWMatrix$ to obtain an approximation to the $\MISE$ optimal bandwidth.

In the absence of specific preferences it seems natural that the weight function $\weightFunction$ decays evenly in all directions from a central point.
While any measure of centrality can in principle be used as a central point, we will be using the (sample) mean.
A possible example of an appropriately shifted weight function is hence given by $\weightFunction(\bx) = \exp\( -\xNorm{\bx - \Eval{\bX} }{2}^2 \)$.

In terms of selecting the bandwidth matrix $\BWMatrix$ we fall back to the sphering approach introduced in Section~\ref{section_dependence_distortion_elliptical}, see also \citealt[Chapter~4.6]{WandJones1995} and references therein.
Instead of optimizing over all $\dim(\dim+1)/2$ entries in $\BWMatrix$ we instead compute the empirical covariance matrix $\empcovMat_{\norig}$ and introduce a one-dimensional optimization parameter $\BW$ by setting $\BWMatrix = \BW \empcovMat_{\norig}$.
Not only does this approach avoid the otherwise high dimensional optimization, but it also fits to the theoretical discussion in Section~\ref{section_dependence_distortion_elliptical} since $\empcovMat \inprob \cov{\bX}$ and hence $\BWMatrix \approx -2 \BW \charGen'(0) \dispMat$ for elliptical random vectors.

Finally, depending on the kernel $\KDE_{-i}$, the evaluation of the integrals in \eqref{eq_CV} is not possible in closed form.
In this case multivariate numerical integration can be used to compute the integrals.
The previously introduced weight function $\weightFunction(\bx) = \exp\( -\xNorm{\bx - \Eval{\bX} }{2}^2 \)$ fits especially well with Gauss-Hermite quadrature and is utilized in our numerical examples where $\dim=2$.
Multivariate Gauss-Hermite quadrature that is compatible with the chosen weight function can be accomplished via a tensor grid, or, for higher dimensions, by the more efficient sparse grid integration.
Sparse grid integration introduced by \cite{Smolyak1963} efficiently combines univariate quadrature rules into multivariate ones; see \cite{GerstnerGriebel1998} for an overview.

\begin{figure}
\centering
\includegraphics[width=\textwidth]{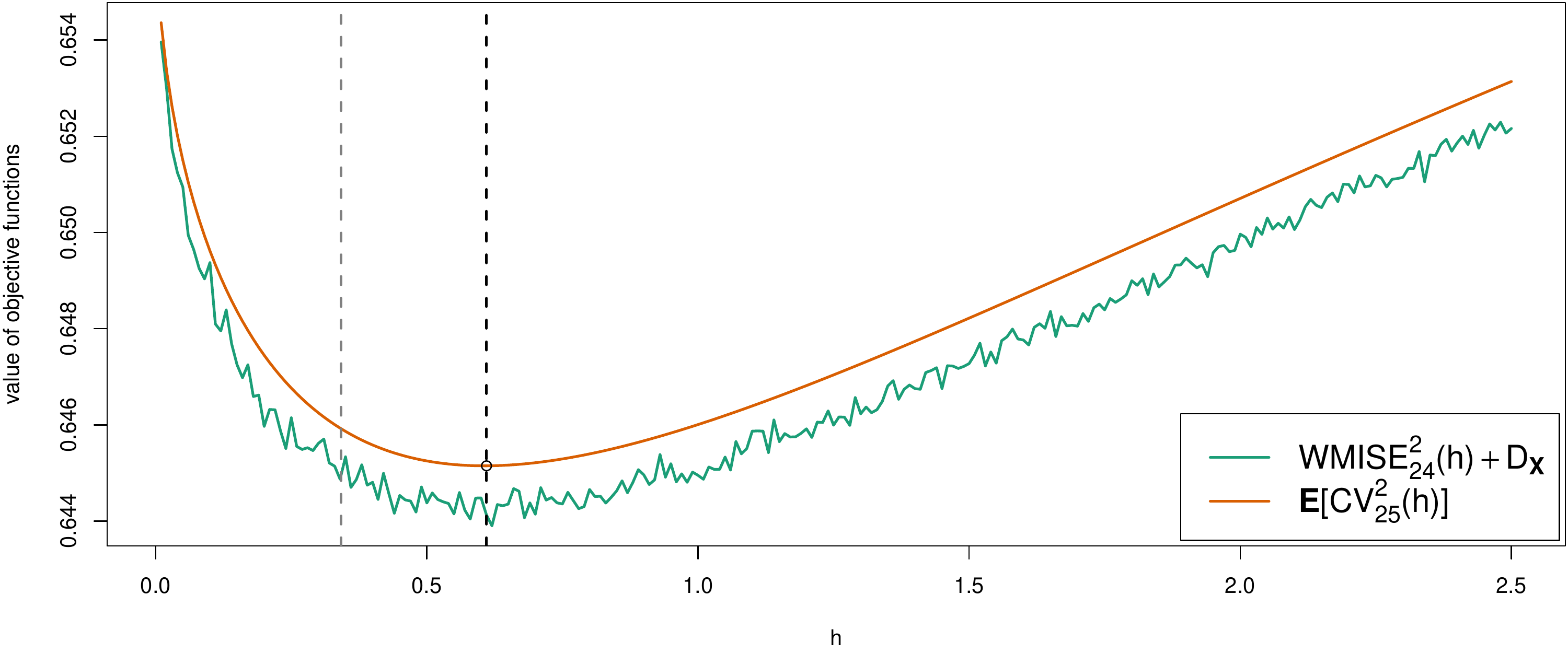}
\includegraphics[width=\textwidth]{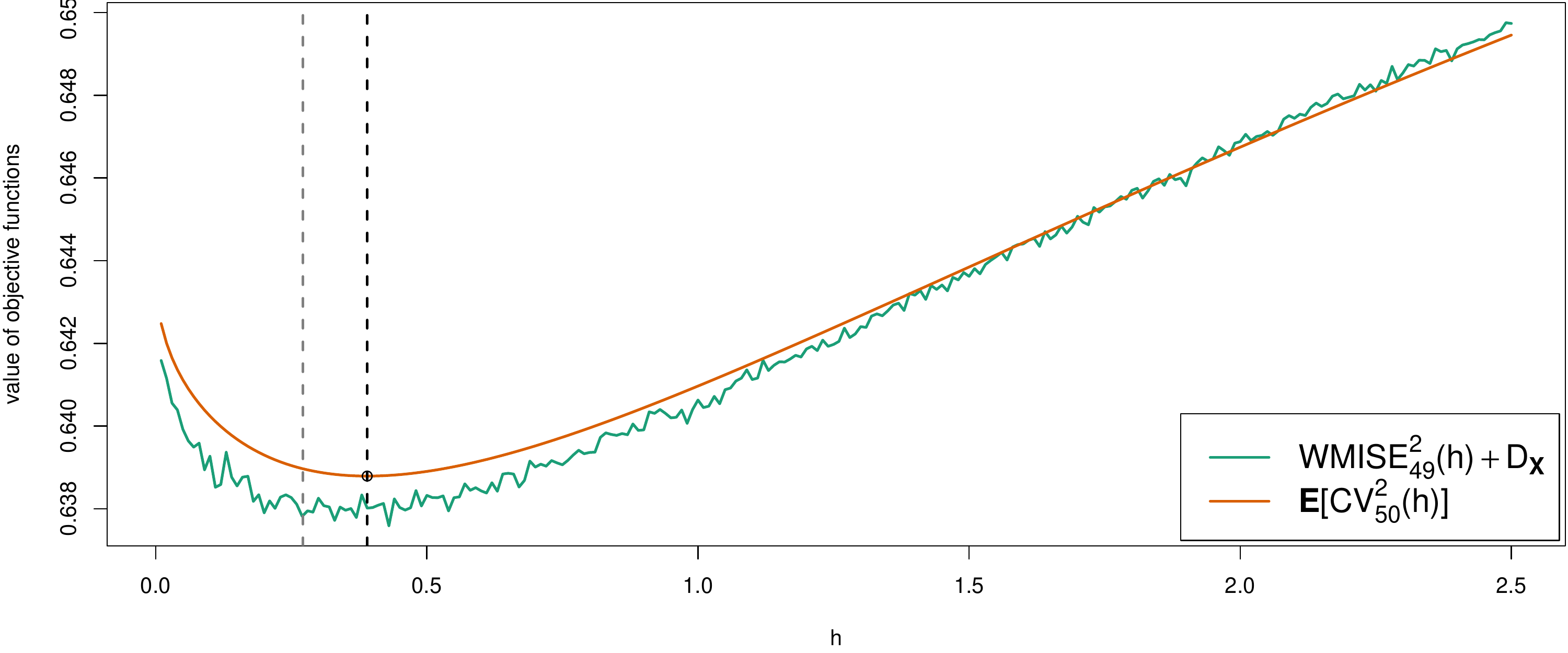}
\includegraphics[width=\textwidth]{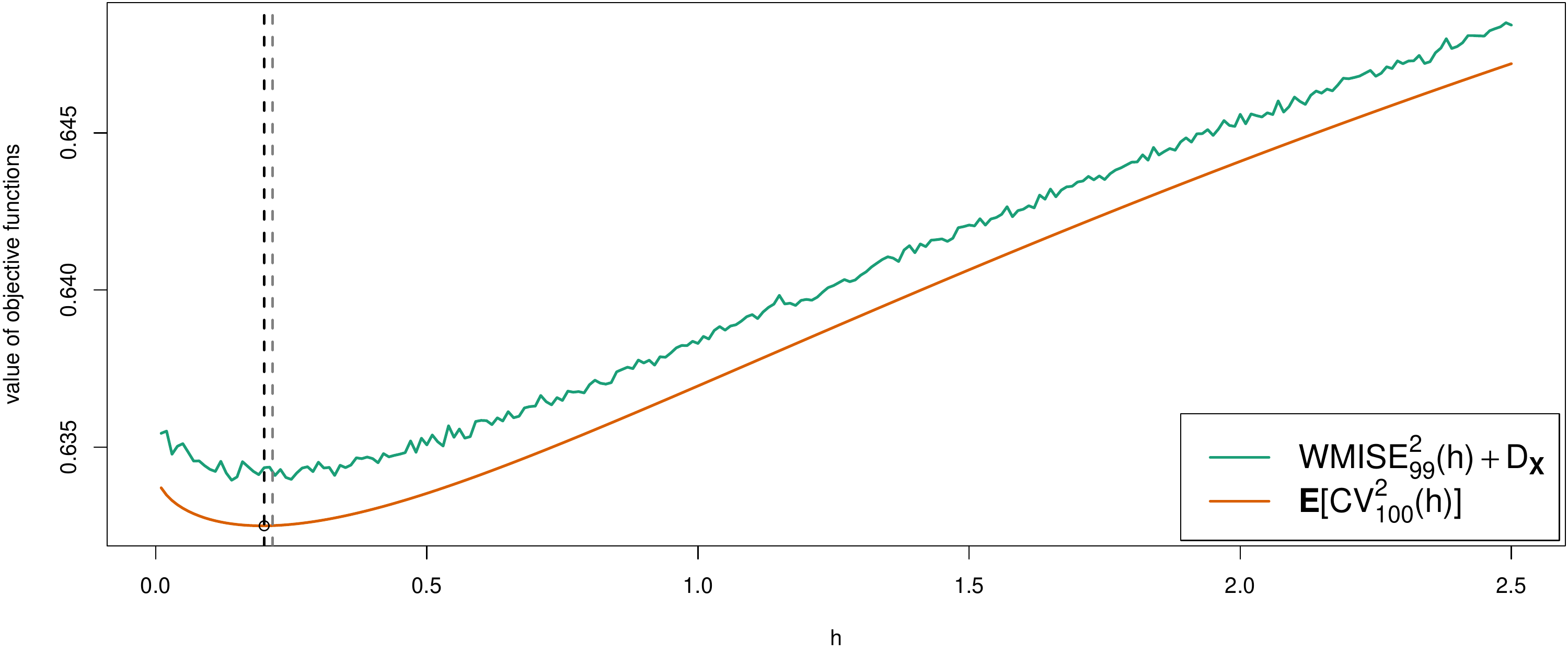}
\caption{Numerical approximation of $\Eval{\CV_{\norig}^2(\BWMatrix;\weightFunction)}$ (orange) and $\MISE_{\norig-1}^2(\BWMatrix;\weightFunction) + D_{\bX}$ (green) for $\norig=25$ (top), $\norig=50$ (middle) and $\norig=100$ (bottom) with $\BWMatrix = \BW \dispMat$ for $\BW \in \{0.01,0.02,\ldots,2.5\}$. The minimum of $\Eval{\CV_{\norig}^2(\BWMatrix;\weightFunction)}$ is indicated by a circle and the dashed vertical black line in all cases. The dashed vertical gray line indicates the choice of $\BW$ following Silverman's rule of thumb.}
\label{fig_crossValidation}
\end{figure}

Figure~\ref{fig_crossValidation} visualizes \eqref{eq_expectation_cross_valiation} by showing the approximation of $\MISE_{\norig}^{2}(\BWMatrix;\weightFunction)$ by the expectation of $\CV^{2}_{\norig}(\BWMatrix;$ $\weightFunction)$ in the case of a bivariate normal $\bX \sim \Normal(\bMu,\dispMat)$ with parameters
\begin{align*}
\bMu = \begin{pmatrix}
  -1.0 \\
  1.0
 \end{pmatrix}
\quad \mbox{and} \quad
\dispMat = \begin{pmatrix}
  1.0 & 1.05 \\
  1.05 & 1.96
 \end{pmatrix},
\end{align*}
yielding a correlation coefficient of $0.75$ between $X_1$ and $X_2$.
In our example we also use a standard bivariate normal distribution $\Normal(\bZero,\IdMat_2)$ for the kernel $\Kernel$.
Due to the lack of closed form solutions we approximate $\MISE_{\norig}^{2}(\BWMatrix;\weightFunction)$, $\intConst_{\bX}(\weightFunction)$, as well as $\Eval{\CV^{2}_{\norig}(\BWMatrix;\weightFunction)}$ numerically.
All necessary integrals inside the respective expectations are computed via a bivariate Gauss-Hermite tensor product rule with $25$ points in each dimension, totalling to $625$ evaluation points.
The weight function is always centered at $\bMu$, i.e., $\weightFunction(\bx) = \exp\( -\xNorm{\bx - \bMu }{2}^2 \)$.
Given that $\intConst_{\bX}(\weightFunction)$ is independent of $\BWMatrix$ and $\norig$ we compute it only once based on $10\,000$ independent samples from $\bX$ and find $\intConst_{\bX}(\weightFunction) =0.629732$.
For a given bandwidth matrix $\BWMatrix$ we compute $\MISE_{\norig}^{2}(\BWMatrix;\weightFunction)$ for three sample sizes $\norig \in \{24,49,99\}$.
For $\norig=24$ we approximate the outer expectation by the mean over $5\,000$ independent samples of $\bX$, for $\norig=49$ we use $2\,500$ independent samples and for $\norig=99$ the approximation is based on $1\,000$ independent samples.
The approximation of $\Eval{\CV^{2}_{\norig}(\BWMatrix;\weightFunction)}$ for the same bandwidth matrix $\BWMatrix$ is thus based on samples of size $\norig \in \{25,50,100\}$.
For $\norig=25$ the value of $\Eval{\CV^{2}_{\norig}(\BWMatrix;\weightFunction)}$ is approximated by $300$ independent samples of size $25$.
In the case of $\norig=50$ we use $60$ independent samples and for $\norig=100$ we use $20$ independent samples from $\bX$.
For $\MISE_{\norig}^{2}(\BWMatrix;\weightFunction)$ and $\Eval{\CV^{2}_{\norig}(\BWMatrix;\weightFunction)}$, the bandwidth matrices are chosen as $\BWMatrix = \BW \dispMat$ with $\BW \in \{0.01,0.02,\ldots,2.5\}$.  The resulting approximation for $\Eval{\CV^{2}_{\norig}(\BWMatrix;\weightFunction)}$ as a function of $\BW$ is shown in orange for $\norig=25$ (top), $\norig=50$ (middle) and $\norig=100$ (bottom).
The green line shows the approximation to $\MISE_{24}^2(\BWMatrix;\weightFunction) + \intConst_{\bX}(\weightFunction)$ (top), $\MISE_{49}^2(\BWMatrix;\weightFunction) + \intConst_{\bX}(\weightFunction)$ (middle) and $\MISE_{99}^2(\BWMatrix;\weightFunction) + \intConst_{\bX}(\weightFunction)$ (bottom).
The minimum of $\Eval{\CV_{\norig}^{2}(\BWMatrix;\weightFunction)}$ is indicated by a circle and the dashed vertical black line in all cases.
The settings are summarized in Table~\ref{table_example_settings}.
\begin{table}
  \centering
  \begin{tabular}{l P{3.5cm}P{3.5cm}P{3cm}}
    \hline\hline
    $\norig$ & Sample size used for $\Eval{ \CV^{2}_{\norig}(\BWMatrix;\weightFunction) }$ & Sample size used for $\MISE_{\norig-1}^{2}(\BWMatrix;\weightFunction)$ & Gauss-Hermite nodes used\\
    \hline
    $25$ & $300$ & $5\,000$ & $25^2$\\
    $50$ & $60$ & $2\,500$ & $25^2$\\
    $100$ & $20$ & $1\,000$ & $25^2$\\
    \hline\hline
  \end{tabular}
  \caption{Simulation settings for the example in Section~\ref{section_bandwidth_selection}.}
  \label{table_example_settings}
\end{table}

\begin{figure}
\centering
\includegraphics[width=\textwidth]{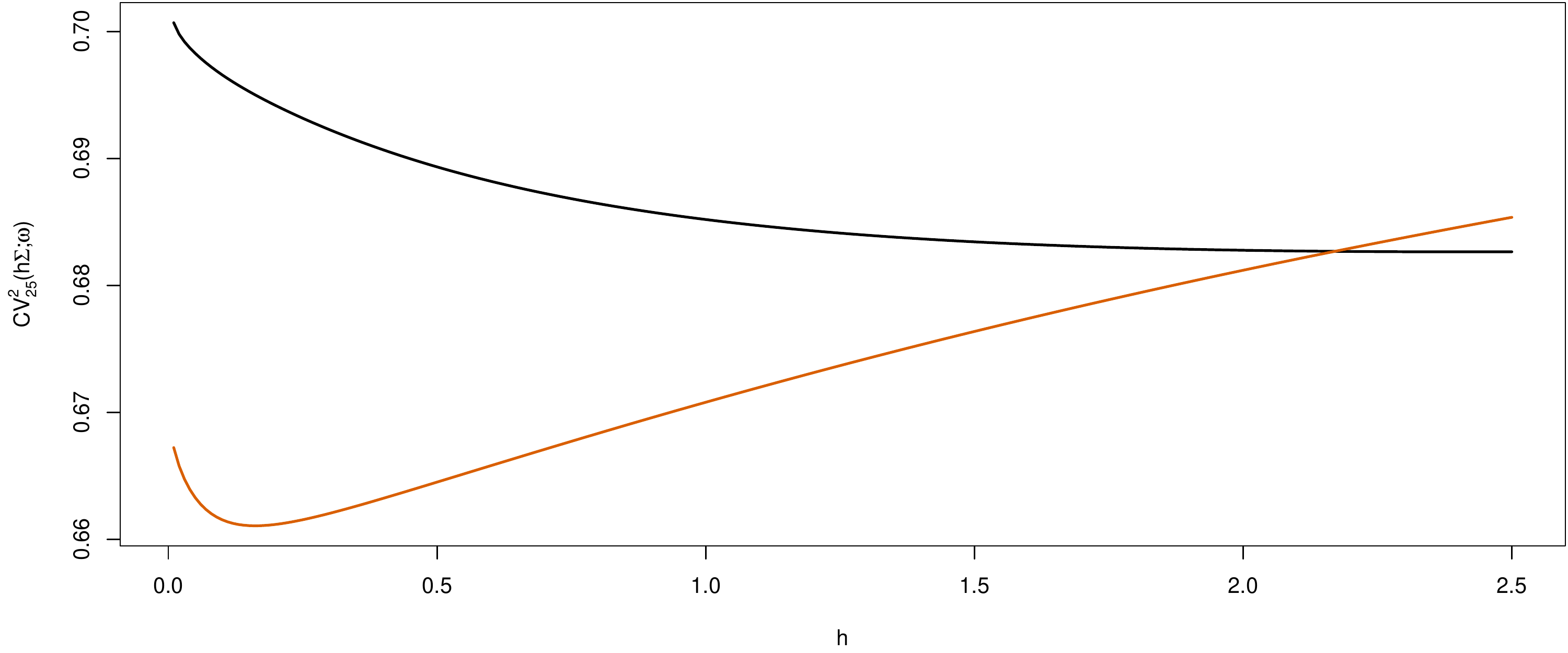}
\caption{$\CV_{25}^2(\BW \dispMat;\weightFunction)$ for two independent samples $\sample_1 = \{\bX_i\}_{i=1}^{25}$ (black) and $\sample_2 = \{\bX_i\}_{i=1}^{25}$ (orange) of $\bX$ over a grid $\BW \in \{0.01,0.02,\ldots,2.5\}$.}
\label{fig_problemCrossValidation}
\end{figure}

From Figure~\ref{fig_crossValidation} we see that the approximation to $\MISE_{\norig}^{2}(\BWMatrix;\weightFunction)$ seems to be more erratic than for $\Eval{\CV_{\norig}^{2}(\BWMatrix;\weightFunction)}$.
While the functions should perfectly match according to Theorem~\ref{theorem_cross_validation}, the remaining differences can be attributed to the limited sample sizes.
In line with intuition, the bandwidth parameter $\BW$ is decreasing with sample size.
We also see that the population version $\Eval{\CV_{\norig}^{2}(\BWMatrix;\weightFunction)}$, i.e., the average over the sample versions $\CV_{\norig}^{2}(\BWMatrix;\weightFunction)$, is a smooth function with a unique minimum.
However, as known from cross-validation in other contexts, this is not necessarily the case for a given sample $\sample = \{\bX_i\}_{i=1}^{\norig}$.
In Figure~\ref{fig_problemCrossValidation} we show two out of the $300$ curves of $\CV_{25}^{2}(\BWMatrix;\weightFunction)$ that are used in the computation of $\Eval{\CV_{25}^{2}(\BWMatrix;\weightFunction)}$.
The two underlying samples are denoted by $\sample_1 = \{\bX_i\}_{i=1}^{25}$ and $\sample_2 = \{\bX_i\}_{i=1}^{25}$.
While the orange curve generated from $\sample_2$ has a shape that is conducive to optimization, the black curve generated from $\sample_1$ is monotonically decreasing over the considered range.
When considering $\Eval{\CV_{25}^{2}(\BWMatrix;\weightFunction)}$ all $300$ curves get averaged which finally yields a reasonable target for minimization, but the individual curves might not be good optimization targets.

\begin{figure}
\centering
\includegraphics[width=\textwidth]{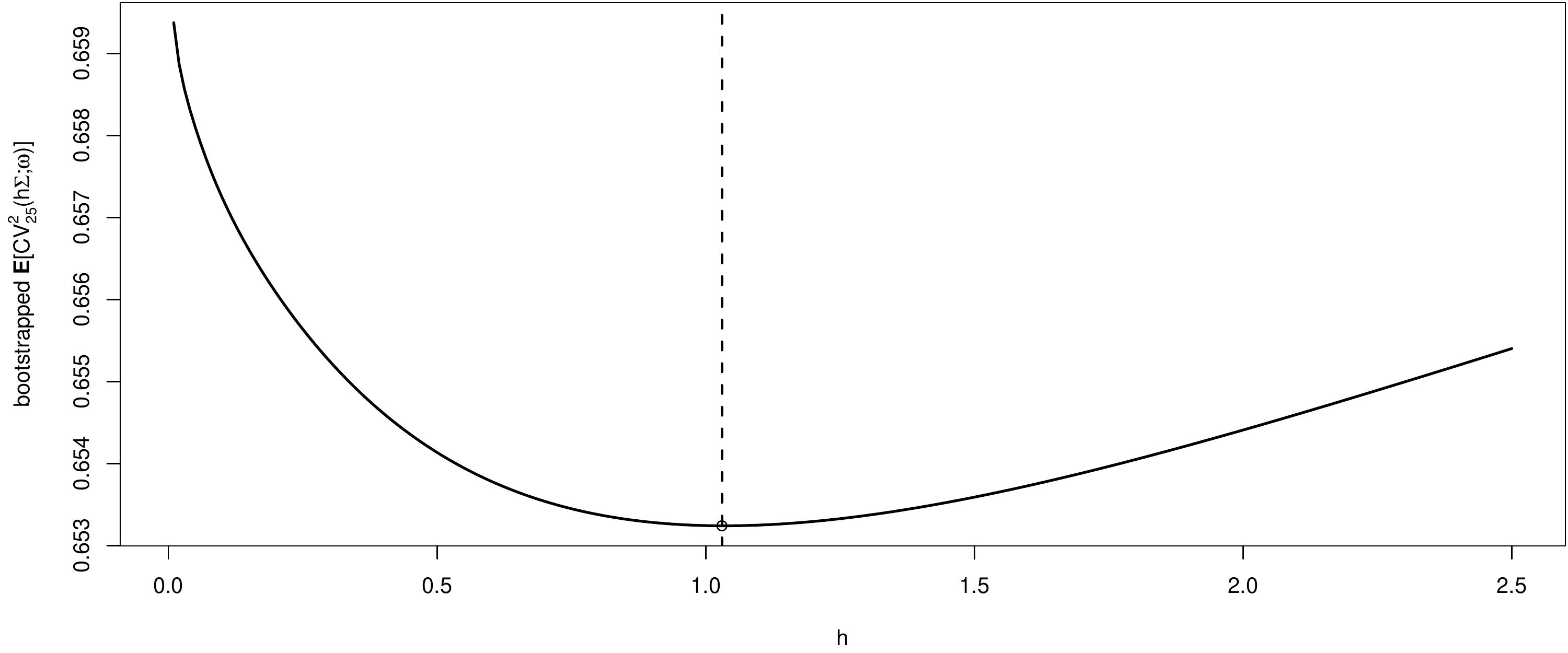}
\caption{Bootstrapped version of $\Eval{\CV_{25}^2(\BW \dispMat;\weightFunction)}$ based on the sample $\sample_1$ over a grid $\BW \in \{0.01,0.02,\ldots,2.5\}$. The approximation is based on $25$ independent bootstrap samples drawn from $\sample_1$. The minimum is indicated by a circle and the dashed vertical line.}
\label{fig_bandwidth_bootstrap}
\end{figure}

In our numerical experiments we find that this issue is more pronounced for small values of $\norig$.
To address this issue in a practical situation it is possible to use a bootstrap approach to generate artificial samples that can then be averaged.
Taking for example the sample $\sample_1$ that generated the black curve in Figure~\ref{fig_problemCrossValidation}, we generate $25$ new samples of size $\norig=25$ by resampling from $\sample_1$ with replacement.
Based on these $25$ new bootstrap samples we then compute an approximation to $\Eval{\CV_{25}^{2}(\BWMatrix;\weightFunction)}$ by averaging.
The resulting curve is shown in Figure~\ref{fig_bandwidth_bootstrap} and shows a preferable shape compared to the initial black curve in Figure~\ref{fig_problemCrossValidation}.
Although the optimal bandwidth $\BW$ taken from Figure~\ref{fig_bandwidth_bootstrap} does not match the population version shown in Figure~\ref{fig_crossValidation}, it is important to recall that the starting sample $\sample_1$ was problematic from the point of view of bandwidth selection by cross-validation to begin with due to its decreasing shape.
In this sense bootstrapping helped to obtain a reasonable bandwidth under difficult conditions.

While our bandwidth selection approach is specifically tailored towards estimating multivariate distribution functions, we compare it to the popular rule of thumb bandwidth selection for multivariate kernel density estimation.
In the multivariate case, Silverman's rule of thumb, see, e.g., \citealt[Chapter~4]{WandJones1995}, is given by setting
\begin{align}\label{equation_Silverman_bandwidth}
\BW(\dim,\norig) = \left(\frac{4}{\norig(\dim+2)}\right)^{2/(\dim+4)},
\end{align}
and then using sphering as before to arrive at $\BWMatrix_{\norig} = \BW(\dim,\norig) \empcovMat_{\norig}$.
While this choice can be justified when estimating multivariate normal densities, it is not theoretically justified for estimating multivariate distribution functions, even in cases like our example setup where all involved distributions are multivariate normal.
It is, however, computationally fast and easy to implement.

The dashed vertical gray line in Figure~\ref{fig_crossValidation} indicates the choice of $\BW$ when following Silverman's rule of thumb.
As indicated in Figure~\ref{fig_crossValidation}, this bandwidth choice leads to undersmoothing for $\norig=25$ and $\norig=50$.
On the contrary, for $\norig=100$ the rule of thumb bandwidth is slightly larger than the $\MISE_{\norig}^2$ optimal bandwidth.
It is not surprising that the rule of thumb bandwidth differs from the $\MISE_{\norig}^2$ optimal bandwidth.
The difference can on the one hand be attributed to different objectives, distribution versus density estimation, that both methods are trying to accomplish.
On the other hand, the weight function $\weightFunction$ is not part of the rule of thumb bandwidth selection procedure while it is explicitly necessary for the $\MISE_{\norig}^2$ procedure.

\subsection{Consistency}\label{section_bs_consistency}
In this section, we outline the difference between $\func\left(\Copula_{\bX}\right)$, $\func\left(\ExpCop\right)$ and $\func\left(\KDECop_{\norig}\right)$ from an asymptotic perspective, where $\KDECop_{\norig}$ is the random version of $\CopulaCondKDE$ defined in \eqref{eq_random_copula}. %
While a full development of the asymptotic theory is beyond the scope of this paper, we highlight the differences between the population and sample versions of the functionals and their relationship, creating a link to data augmentation as discussed in Section~\ref{section_smooth_bootstrap_copula_functionals}.
We then choose Spearman's rho as an example to highlight the challenges at hand when developing asymptotic theory for the smooth bootstrap.

In order to make the dependence of estimates on the respective bandwidth matrices clear we use the bandwidth matrix as an argument in the following.
For example, $\KDECop_{\norig}(\BWMatrix_{\norig})$ denotes the estimate $\KDECop_{\norig}$ based on the bandwidth matrix $\BWMatrix_{\norig}$.
In principle, the quantity of interest is the (random) approximation error
\begin{align*}
\Error_{\sample} = \metric{\func\left(\Copula_{\bX}\right)}{\func\left(\KDECop_{\norig}\(h_{\norig} \empcovMat_{\norig} \)\right)},
\end{align*}
where $\metric{\cdot}{\cdot}$ is an appropriate distance function (metric) chosen with regard to the functional $\func$.
Using the triangle inequality for $\metric{\cdot}{\cdot}$ we can incorporate our results concerning $\ExpCop$ into this discussion via
\begin{align}\label{equation_split_error}
\Error_{\sample}
&\leq
\metric{\func\left(\Copula_{\bX}\right)}{\func\left(\ExpCop\(\BW^*_{\norig} \dispMat \)\right) } +
\metric{\func\left(\ExpCop\(\BW^*_{\norig} \dispMat \)\right) }{
\func\left(\KDECop_{\norig}\(  \BW_{\norig} \empcovMat_{\norig}  \)\right) }
\\
&= \Error_1 + \Error_2, \nonumber
\end{align}
where $\BW^*_{\norig}$ is the non-random, $\MISE_{\norig}$ optimal bandwidth factor and $\BW_{\norig}$ is a data-driven and hence random choice for the bandwidth factor.
It is important to observe that the first error $\Error_1$ is non-random.
In the case of elliptical distributions and smoothing kernels, $\Error_1$ is a deterministic function of only the functional $\func$ and the bandwidth matrix $\BW^*_{\norig} \dispMat$, which introduces the differences of the characteristic generators $\charGen_{\bX}$ and $\charGen_{\bZ}$ as addressed in Section~\ref{section_kernel_smoothing}.
From our previous investigation we know that $\Error_1 = 0$ if either (i) the elliptical distributions for the data generating process and the kernel share the same dispersion matrix up to a scale factor and the functional $\func$ does not depend on the characteristic generator, cf. Section~\ref{section_dependence_distortion_elliptical}, or if (ii) the circumstances discussed in Section~\ref{section_exceptions} are met.
If $\Error_1 > 0$ in the elliptical setting of Section~\ref{section_kernel_smoothing}, the rate of convergence of $\charGen_{\bX} \to \charGen_{\bZ}$ discussed in Section~\ref{section_convergence_rates} is linked to the convergence rate $\metric{\func\left(\Copula_{\bX}\right)}{\func\left(\ExpCop\(\BW^*_{\norig} \dispMat\)\right) } \to 0$.

As an example, we consider Spearman's rho $\SpearmansRho$ of a bivariate random vector $\bX$ with an associated copula $\Copula_{\bX}$ which is given by
\begin{align}\label{eq_SpearmansRho}
\SpearmansRho\(\Copula_{\bX}\) = 12 \int_0^1 \int_0^1 \Copula_{\bX}(u,v) \d u \d v - 3;
\end{align}
see \citealt[Chapter~5]{Nelsen2006}.
The corresponding sample version $\func_{\norig}\(\bx_1,\ldots,\bx_{\norig}\)$ is given by the Pearson correlation coefficient of the ranks of the first and second components of $\bx_1,\ldots,\bx_{\norig}$.
As discussed in Section~\ref{section_dependence_distortion_elliptical}, similar alternatives given by Kendall's tau or Blomqvist's beta lead to $\Error_1 = 0$ in the considered setup.
Given that Spearman's rho takes values in $[-1,1]$, a suitable metric $\metric{\cdot}{\cdot}$ is in this case given by the absolute value $\metric{x}{y} = \abs{x-y}$.
For the error $\Error_{\sample}$ this leads to
\begin{align*}
\Error_{\sample}&= \metric{\func\left(\Copula_{\bX}\right)}{\func\left(\KDECop_{\norig}\(  \BW_{\norig} \empcovMat_{\norig} \)\right) }= \abs{\SpearmansRho\(\Copula_{\bX}\) - \SpearmansRho\(\KDECop_{\norig}\(  \BW_{\norig} \empcovMat_{\norig} \)\)}\\
&= 12 \abs{\int_0^1 \int_0^1 \Copula_{\bX}(u,v) - \KDECop_{\norig}\(  \BW_{\norig} \empcovMat_{\norig} \)(u,v) \d u \d v}\leq 12 \int_0^1 \int_0^1 \abs{ \Copula_{\bX}(u,v) - \KDECop_{\norig}\(  \BW_{\norig} \empcovMat_{\norig} \)(u,v) } \d u \d v\\
&\leq 12 \dKS\(\Copula_{\bX},\KDECop_{\norig}\(  \BW_{\norig} \empcovMat_{\norig} \)\);
\end{align*}
see \eqref{equation_def_KS} for the definition of the Kolmogorov-Smirnov distance $\dKS$.
We can follow \eqref{equation_split_error} to include $\ExpCop\(\BW^*_{\norig} \dispMat \)$ to obtain
\begin{align*}
\dKS\(\Copula_{\bX},\KDECop_{\norig}\(  \BW_{\norig} \empcovMat_{\norig} \)\) \leq \dKS\(\Copula_{\bX},\ExpCop\(\BW^*_{\norig} \dispMat \) \) + \dKS\(\ExpCop\(\BW^*_{\norig} \dispMat \),\KDECop_{\norig}\(  \BW_{\norig} \empcovMat_{\norig} \) \).
\end{align*}
In principle the bivariate Kolmogorov-Smirnov distance between distributions can be bounded in terms of the characteristic functions, see \cite{Sadikova1966} and \cite{HeubergerKropf2018} for which we give the details in Theorem~\ref{theorem_Berry_Essen_inequalty_2D}, and we followed this approach at the end of Section~\ref{section_convergence_rates}.
However, in this case, a direct application is too restrictive due to the need for bounded derivatives, a condition that is not met for a number of popular copula families.
As an alternative we utilize the invariance of the Kolmogorov-Smirnov distance under strictly increasing transforms.
For continuous distribution functions $\mdistI_1,\ldots,\mdistI_{\dim}$ supported on $\R$ we have
\begin{align*}
\dKS\(\Copula_{\bX},\KDECop_{\norig}\(  \BW_{\norig} \empcovMat_{\norig} \)\)
&= \sup_{\bu \in (0,1)^{\dim}} \abs{\Copula_{\bX}(\bu)- \KDECop_{\norig}\(  \BW_{\norig} \empcovMat_{\norig} \)(\bu)}\\
&= \sup_{\bx \in \R^{\dim}} \abs{\Copula_{\bX}(\mdistI_1(x_1),\ldots,\mdistI_{\dim}(x_{\dim}))-\KDECop_{\norig}\(  \BW_{\norig} \empcovMat_{\norig} \)(\mdistI_1(x_1),\ldots,\mdistI_{\dim}(x_{\dim}))}\\
&= \dKS\(\JointDist_{\bX},\KDET_{\norig}\),
\end{align*}
where $\JointDist_{\bX}$ and $\KDET_{\norig}$ are the joint distribution functions with the respective copulas and identical margins $\mdistI_1,\ldots,\mdistI_{\dim}$.
While showing that $\dKS\(\JointDist_{\bX},\KDET_{\norig}\) \almostsurly 0$, or $\dKS\(\JointDist_{\bX},\KDET_{\norig}\) \inprob 0$, is beyond the scope of this paper we outline one possible approach:
\cite{DevroyeWagner1979} give conditions under which the measure associated to the kernel density estimate converges to the unknown measure of the true underlying density in total variation distance.
For two measures $\mu$ and $\nu$ on $\R^{\dim}$ the total variation distance is defined as $\dTV{\mu}{\nu}=\sup_{B\in\SigAlg}\abs{\mu(B)-\nu(B)}$, where $\SigAlg$ is the Borel sigma algebra on $\R^{\dim}$.
In place of the measures $\mu$ and $\nu$ we will also use the distribution or density functions associated to them.
From the definitions we immediately have $\dKS(\distI,\distII) \leq \dTV{\distI}{\distII}$ for any two distributions functions $\distI$ and $\distII$.
To show that $\dTV{\kde_{\norig}}{\JointDens_{\bX}} \almostsurly 0$, and $\dTV{\kde_{\norig}}{\JointDens_{\bX}} \inprob 0$, \cite{DevroyeWagner1979} however rely, amongst other assumptions, on a diagonal bandwidth matrix in the definition of $\kde_{\norig}$.
We leave the adaptation to elliptical kernels with non-diagonal bandwidth matrices in our setting for further research.

As exemplified by our treatment of Spearman's, a detailed analysis of the asymptotic behaviour depends on the functional under consideration.
If, for example, the functional $\func$ is a level set of the underlying copula, the absolute value is not a suitable metric since we need to quantify the distance between two sets or their respective boundaries.
A suitable metric in this context is given by the Hausdorff distance between the sets enclosed by the contour lines.
We discuss this approach in our simulation studies in the next section.
\section{Simulation Study}\label{section_simulation_studies}
In this section, we illustrate Algorithm~\ref{algo_smoothed_copula_bootstrap} with several examples for two different functionals $\func$, namely copula level curves and copula based dependence measures.
In our examples, we compare $\func\left(\CopulaCondKDE\right)$ to $\func\left(\Copula\right)$, where we use the smooth bootstrap to approximate $\func\left(\CopulaCondKDE\right)$ via data augmentation, as discussed in Section~\ref{section_smooth_bootstrap_copula_functionals}.
In our simulations we use elliptical smoothing kernels but we do not restrict ourselves to elliptical data generating processes to highlight that the approach is not limited to the specific situation discussed in Section~\ref{section_dependence_distortion_elliptical}.

\subsection{Copula level curves and copula diagonals}
Inspired by the application in \cite{CoblenzDyckerhoffGrothe2017} we first focus on bootstrapping level curves for copulas.
For a copula $\Copula$ we define the sublevel set at level $t\in(0,1)$ as
\begin{align*}
\LevelSet_t(\Copula) = \{ \bu \in [0,1]^{\dim} : \Copula(\bu) \leq t \}.
\end{align*}
Sublevel sets of copulas have an interpretation as multivariate quantiles, see \cite{SalvadoriDuranteDeMicheleBernardiPetrella2016}, and are important for applications, e.g., in finance and hydrology.
To assess whether the sublevel sets of two copulas are close we measure their distance in terms of the Hausdorff distance.
For two subsets $A$ and $B$ of a metric space $(M,\metricd)$ the Hausdorff distance $\dH{\cdot}{\cdot}$ is defined as
\begin{align*}
\dH{A}{B} = \max\(\sup_{x\in A} \inf_{y\in B} \metric{x}{y}, \sup_{y\in B} \inf_{x\in A} \metric{x}{y}\),
\end{align*}
where in our simulations we use the standard Euclidean distance $\metric{\bx}{\by} = \xNorm{\bx-\by}{2}$ over the unit cube.

In our simulation we draw pseudorandom numbers $\{\bu_1,\ldots,\bu_{\norig}\}$ from a fixed bivariate copula $\Copula$ and record the distance $\dH{\LevelSet_t\(\Copula\)}{\widehat{\LevelSet}_t}$ for different values of $t$ over a number of independent simulations.
Here $\widehat{\LevelSet}_t$ is the estimated sublevel set based either on the original observations $\{\bu_1,\ldots,\bu_{\norig}\}$, $\bu_i \in [0,1]^2$ for $i\in\{1,\ldots,\norig\}$, only, or on the augmented sample $\{\bu_1^*,\ldots,\bu^*_{\naugm}\}$ produced by the smooth bootstrap.
Sufficient conditions for the convergence $\dH{\LevelSet_t\(\Copula\)}{\widehat{\LevelSet}_t}\almostsurly 0$ are discussed in \cite{CoblenzDyckerhoffGrothe2017}.

Concerning the bandwidth matrix in the smooth bootstrap we use the empirical variance-covariance matrix based on a transformation of $\{\bu_1,\ldots,\bu_{\norig}\}$ with the standard normal quantile function together with Silverman's rule of thumb smoothing parameter given in \eqref{equation_Silverman_bandwidth}.
We utilize this choice throughout the section due to the computational efficiency of the method in the simulation re-runs.
To transform the smooth bootstrap sample back into $[0,1]^2$ we use the marginal distribution functions of the associated kernel distribution mixture as outlined in Algorithm~\ref{algo_smoothed_copula_bootstrap}.
Finally, the multivariate normal density is used as smoothing kernel.

For a given sample, either $\{\bu_1,\ldots,\bu_{\norig}\}$ or $\{\bu_1^*,\ldots,\bu^*_{\naugm}\}$, the boundary of the sublevel set is estimated by computing the contour lines of the associated empirical copula
at level $t$, resulting in a piecewise linear approximation to
$\LevelSet_t(\Copula)$; to this end we apply \textsf{R}'s \lstinline{contourLines()} to the empirical copula constructed via \lstinline{empCopula()} of the \textsf{R} package \texttt{copula} of \cite{copula}.
For a given level $t$ it is, from a theoretical perspective, also clear that the copula contour lines ultimately need to pass through $(t,1)$ and $(1,t)$.
Due to the inherent randomness in the samples the estimated contour lines do not necessarily fulfill this constraint.
For points $(u_i,v_i)$ on the estimated contour line it is however possible to modify the results accordingly by replacing all values $u_i <t$ with $t$ and all values $v_i <t$ with $t$.
This modification ensures the validity of the boundary conditions and is utilised in our computations.
As a result of the algorithm we obtain (by adding the points $(0,0)$, $(1,0)$ and $(0,1)$) the vertices of a polygon that approximates $\LevelSet_t(\Copula)$, where the number of vertices depends on the chosen number of grid points used to discretize the x-axes between $t$ and $1$.
Concerning the sublevel sets of the true underlying copula $\LevelSet_t\(\Copula\)$ it is also necessary to discretize the boundary of $\LevelSet_t\(\Copula\)$ into a list of vertices over a sufficiently fine approximation grid.
Concerning the numerical computation of $\dH{\cdot}{\cdot}$ we implement the algorithm outlined in \cite{TahaHanbury2015}.
The algorithm efficiently computes the Hausdorff distance between two polygons which then applies directly to the estimated boundary of $\widehat{\LevelSet}_t$ and the discretized boundary of $\LevelSet_t\(\Copula\)$.

For the simulation setup we consider the following data.
We choose the underlying true copula in the Archimedean class of copulas, see \cite{Nelsen2006} for an overview.
This allows to obtain a closed-form expression for the boundary of $\LevelSet_t\(\Copula\)$ and hence allows to accurately discretize the boundary of the true underlying copula $\LevelSet_t\(\Copula\)$.
Specifically, we simulate from a Clayton copula $\Copula_{\theta}$ where we fix the model parameter $\theta$ in such a way that Kendall's tau takes specific values, $\tau(\Copula_{\theta}) \in \{-0.9,-0.8,-0.7,\ldots,0.9\}$.
We also consider $t \in \{0.1,0.2,\ldots,0.9\}$ and $\norig\in\{25,50,100\}$.
For the smooth bootstrap we generally set the sample size of the resulting (augmented) sample to $\naugm=5\,000$.
The difference between $\norig$ and $\naugm$ will (generally) lead to different discretization step sizes for the respective boundary approximations.
We repeat each simulation independently $M = 10\,000$ times.

\begin{figure}
\centering
\includegraphics[width=\textwidth]{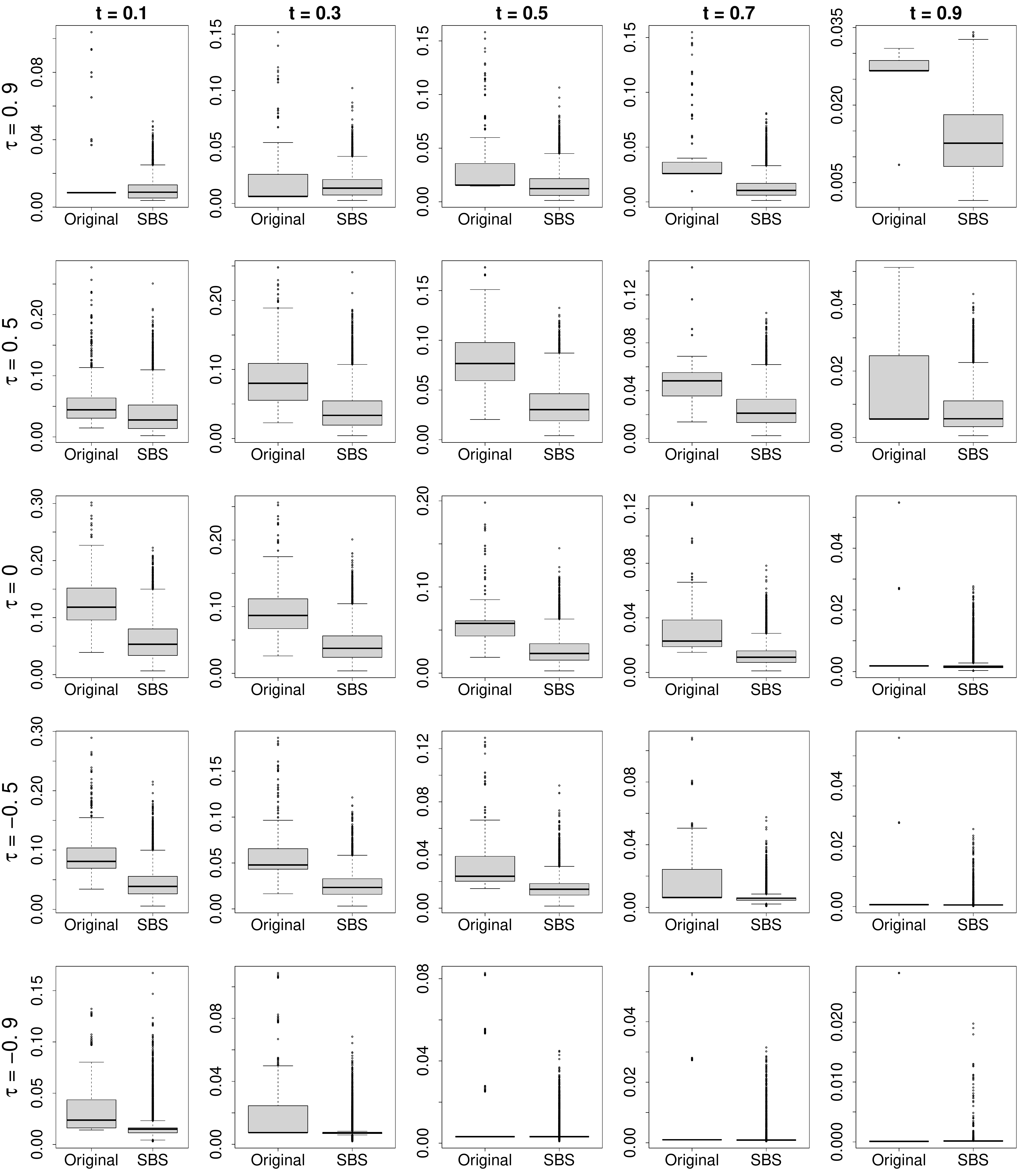}
\caption{Hausdorff distance simulation results between the true and estimated level sets for a Clayton copula $\Copula_{\theta}$ with $\theta$ such that $\tau(\Copula_{\theta}) \in \{-0.9,-0.5,0,0.5,0.9\}$, for levels $t\in\{0.1,0.3,0.5,0.7,0.9\}$. Original sample size $\norig=25$, augmented smooth bootstrap sample size $\naugm=5\,000$. Each boxplot is based on $M=10\,000$ independent reruns.}
\label{fig_HausdorffDistance_25}
\end{figure}

\begin{figure}
\centering
\includegraphics[width=\textwidth]{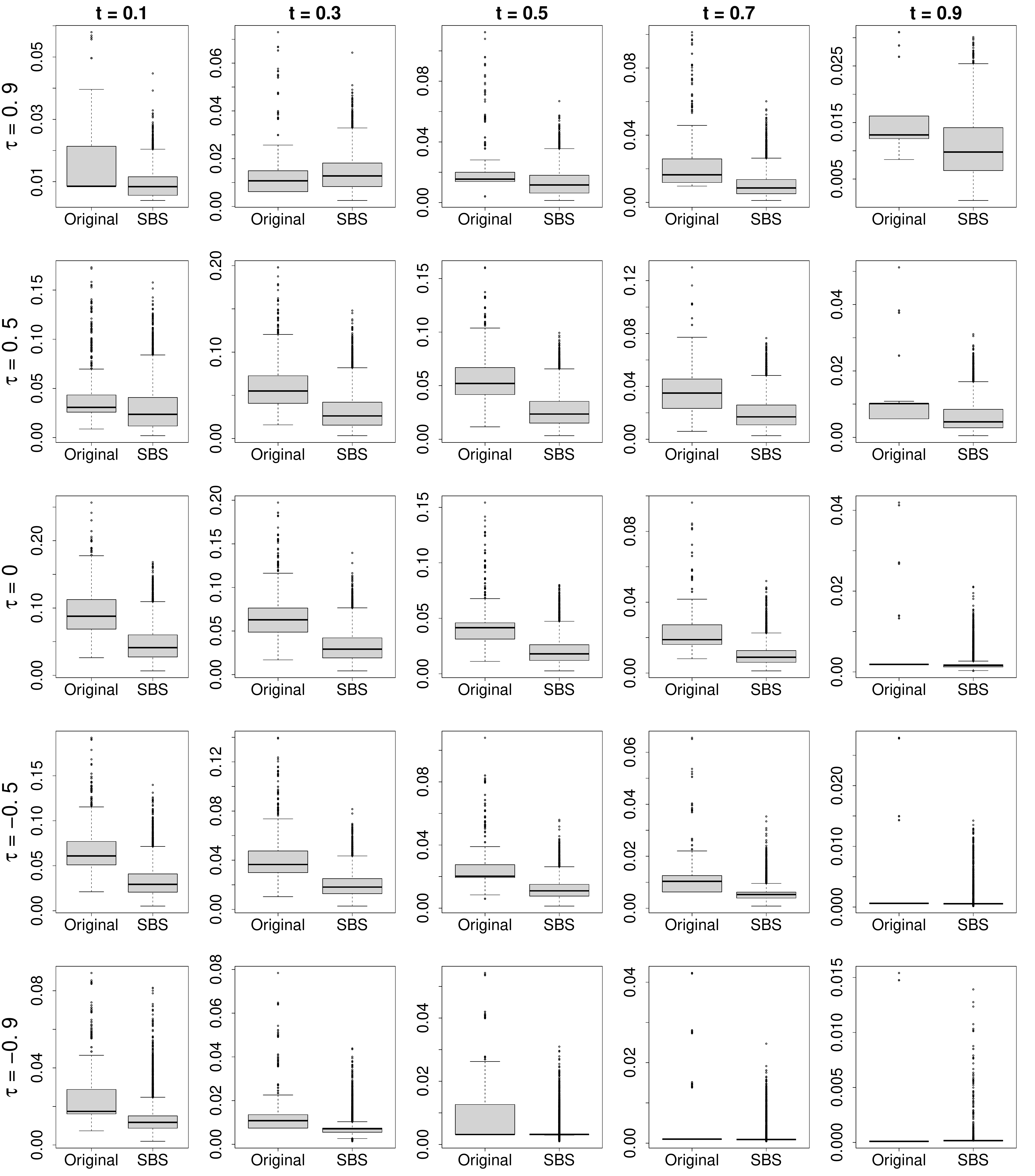}
\caption{Hausdorff distance simulation results between the true and estimated level sets for a Clayton copula $\Copula_{\theta}$ with $\theta$ such that $\tau(\Copula_{\theta}) \in \{-0.9,-0.5,0,0.5,0.9\}$, for levels $t\in\{0.1,0.3,0.5,0.7,0.9\}$. Original sample size $\norig=50$, augmented smooth bootstrap sample size $\naugm=5\,000$. Each boxplot is based on $M=10\,000$ independent reruns.}
\label{fig_HausdorffDistance_50}
\end{figure}

\begin{figure}
\centering
\includegraphics[width=\textwidth]{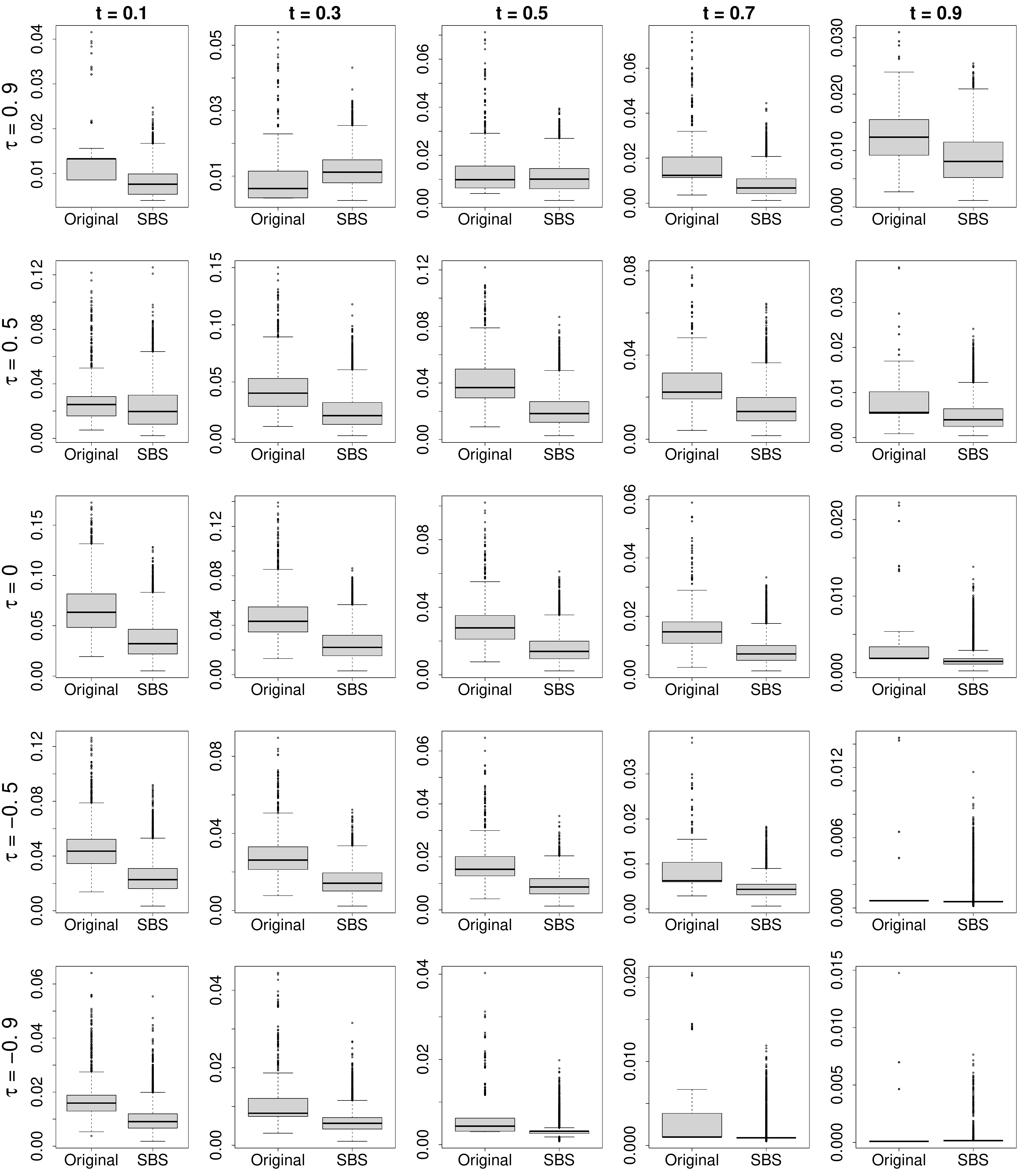}
\caption{Hausdorff distance simulation results between the true and estimated level sets for a Clayton copula $\Copula_{\theta}$ with $\theta$ such that $\tau(\Copula_{\theta}) \in \{-0.9,-0.5,0,0.5,0.9\}$, for levels $t\in\{0.1,0.3,0.5,0.7,0.9\}$. Original sample size $\norig=100$, augmented smooth bootstrap sample size $\naugm=5\,000$. Each boxplot is based on $M=10\,000$ independent reruns.}
\label{fig_HausdorffDistance_100}
\end{figure}

\begin{figure}
\centering
\includegraphics[width=0.6\textwidth]{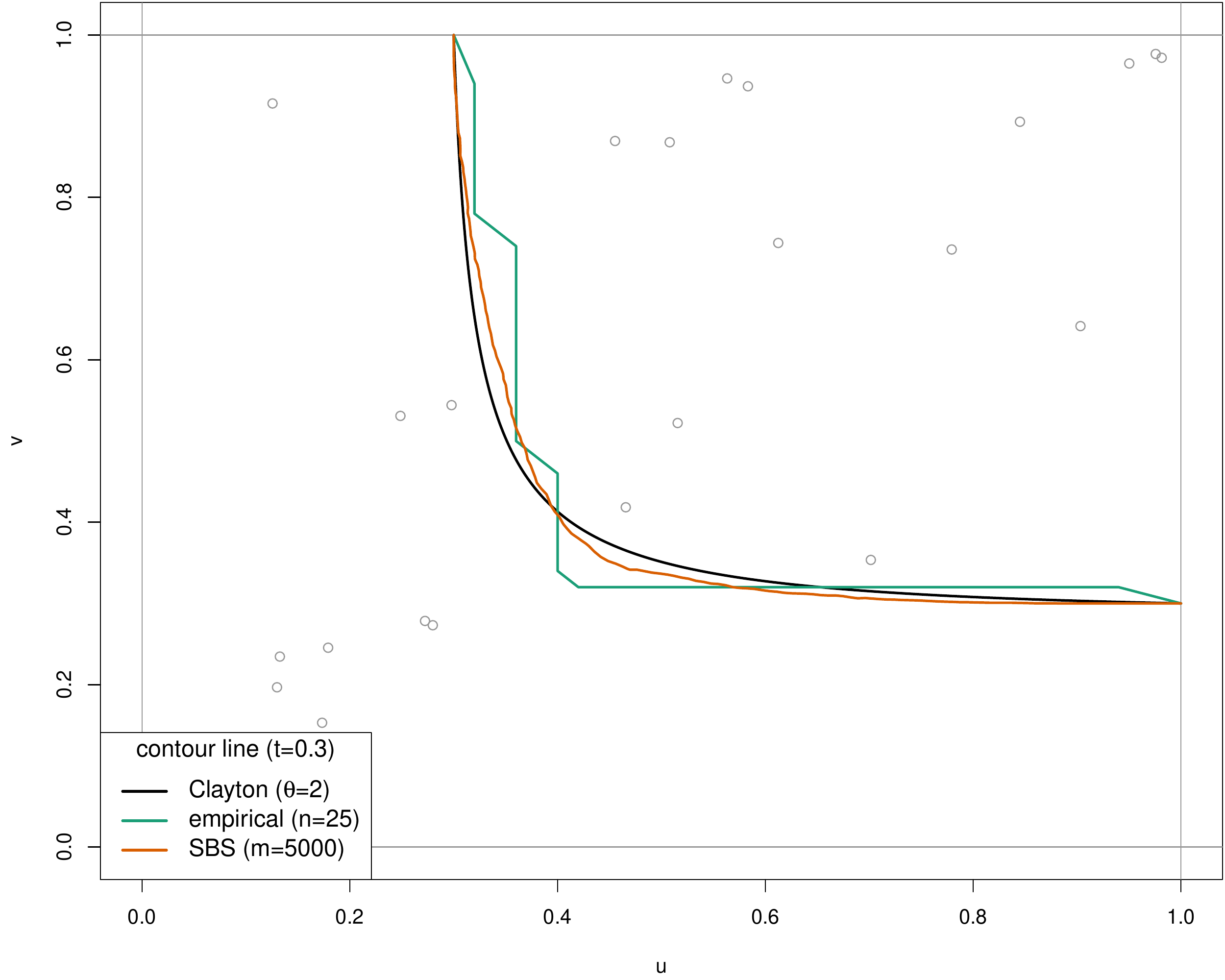}
\caption{Theoretical (black) and estimated contour lines for a Clayton copula with parameter $\theta=2$ at level $t=0.3$. Direct estimation (green) is based on $\norig=25$ sample points. For the smooth bootstrap (orange) $m=5\,000$ pseudo-observations are generated based on the initial sample.}
\label{fig_contour_lines}
\end{figure}

The resulting distances for $\norig=25$ are presented in the boxplots in Figure~\ref{fig_HausdorffDistance_25}.
The results for $\norig=50$ can be found in Figure~\ref{fig_HausdorffDistance_50}, while Figure~\ref{fig_HausdorffDistance_100} shows the results for $\norig=100$.
The figures clearly highlight the benefit of using the smooth bootstrap in this situation.
Without smoothing, the estimated contour lines based on the original data samples are too coarse.
On the one hand this makes them unusable in practice, see the discussion in \cite{CoblenzDyckerhoffGrothe2017}, on the other hand this leads to a significant distance from the theoretical target.
Using an augmented data sample constructed by a smooth bootstrap procedure on the other hand leads to an estimated curve that bears more resemblance with a contour line compared to the step-function like estimation result of the standard estimation.
Not only is the resulting curve more suitable for practical applications, but also it is closer to the theoretical target.
A visual representation of this situation can be found in Figure~\ref{fig_contour_lines}, where we depict in black the theoretical contour line of a Clayton copula at level $t = 0.3$ with parameter $\theta=2$ and an associated Kendall's tau of $\tau = 1/2$.
For a sample of size $\norig=25$ we also give the estimated contour lines based on the original sample (green) and the estimated contour lines based on the smooth bootstrap (orange).
As before, we use Silverman's rule of thumb, see \eqref{equation_Silverman_bandwidth}, in combination with the empirical variance-covariance matrix to construct the bandwidth matrix for the bivariate Gaussian kernel.
The empirical variance-covariance matrix is computed based on the data once they have been transformed into $\R^2$ via the standard normal quantile function.
As discussed, the smooth bootstrap contour lines provide a better approximation to the theoretical target by smoothing out the sharp kinks of the direct estimate which is clearly visible in Figure~\ref{fig_contour_lines}.

\begin{figure}
\centering
\includegraphics[width=0.8\textwidth]{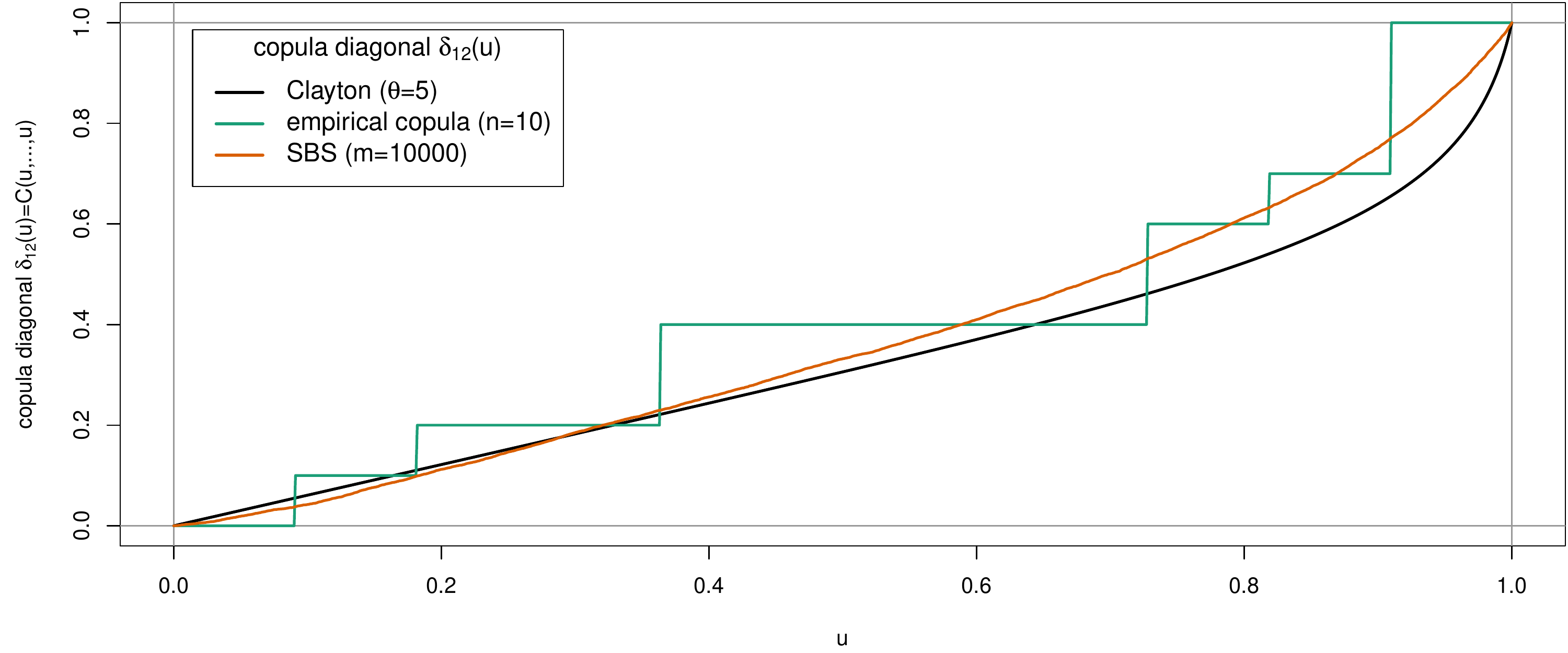}
\caption{Theoretical (black) and estimated copula diagonal for a twelve dimensional Clayton copula with parameter $\theta=5$. Direct estimation (green) via the empirical copula is based on $\norig=10$ sample points. For the smooth bootstrap (orange) $m=10\,000$ pseudo-observations are generated based on the initial sample which are then used to estimate the diagonal via the empirical copula.}
\label{fig_copula_diagonal}
\end{figure}

Similar improvements are visible in the estimation of the copula diagonal $\delta(u) = \Copula(u,\ldots,u)$ which is presented in Figure~\ref{fig_copula_diagonal} for a twelve dimensional Clayton copula with parameter $\theta=5$.
In small samples, $n=10$ in the example, the estimation based on the empirical copula is too coarse to be useful in a practical situation.
When using the smooth bootstrap to generate additional observations the empirical copula diagonal based on the enlarged sample, $m=10\,000$ in the example, is less similar to a step function and closer to the unknown copula diagonal.
As for the contour lines, the smooth bootstrap is advantageous when the target functional is a curve where the evaluation $\delta(u)$ needs to exhibit a smooth behavior for nearby values of $u$.
\subsection{Copula based dependence measures}\label{section_copula_based_dependence_measures}
\begin{figure}
\centering
\includegraphics[width=\textwidth]{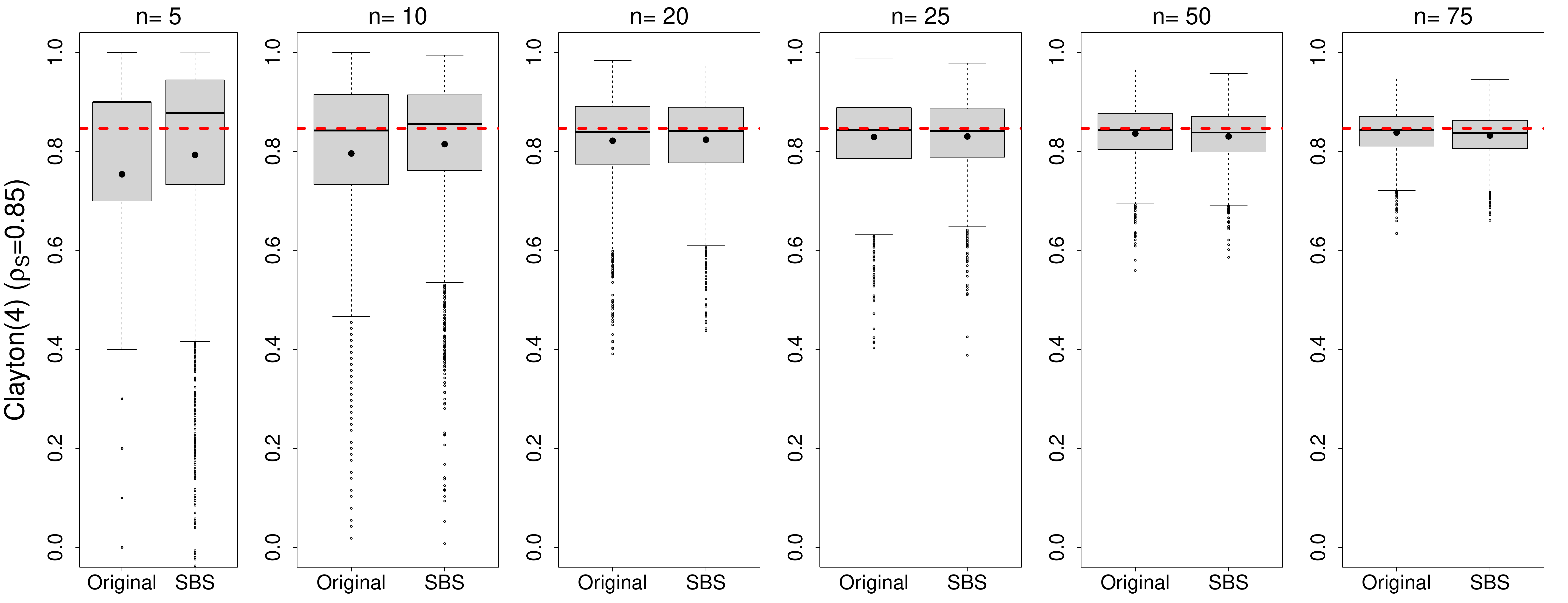}\\[4mm]
\includegraphics[width=\textwidth]{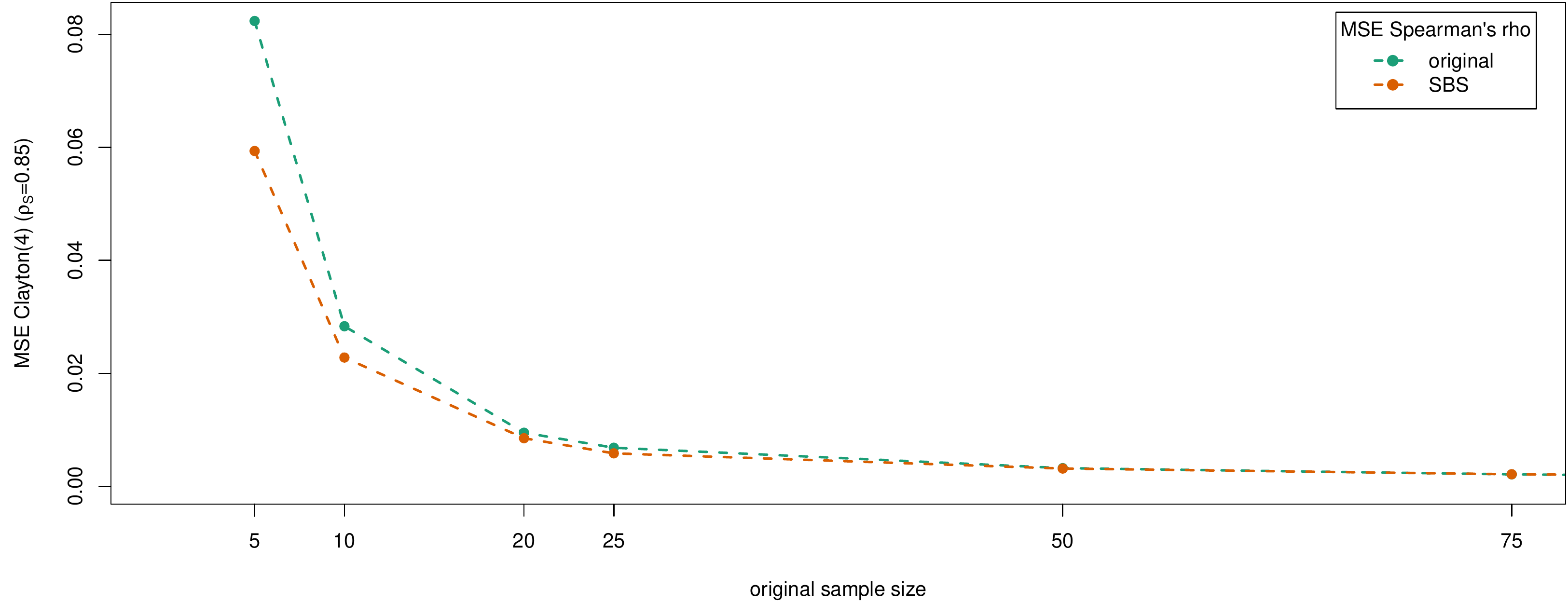}
\caption{Top: Estimated Spearman's rho for the bivariate Clayton($4$) copula.
The results are based on an original sample size of $\norig\in\{5,10,20,25,50,75\}$, while the augmented smooth bootstrap sample size is $\naugm=10\,000$.
Each boxplot is based on $M=2\,000$ independent reruns. The red dashed line indicates the theoretical value of $\rho_S$ while black dots indicate the means.\newline
Bottom: Mean squared error for estimation of Spearman's rho for the bivariate Clayton($4$) copula.
The results are based on an original sample size of $\norig\in\{5,10,20,25,50,75\}$, while the augmented smooth bootstrap sample size is $\naugm=10\,000$.}
\label{fig_Clayton_Spearman}
\end{figure}

\begin{figure}
\centering
\includegraphics[width=\textwidth]{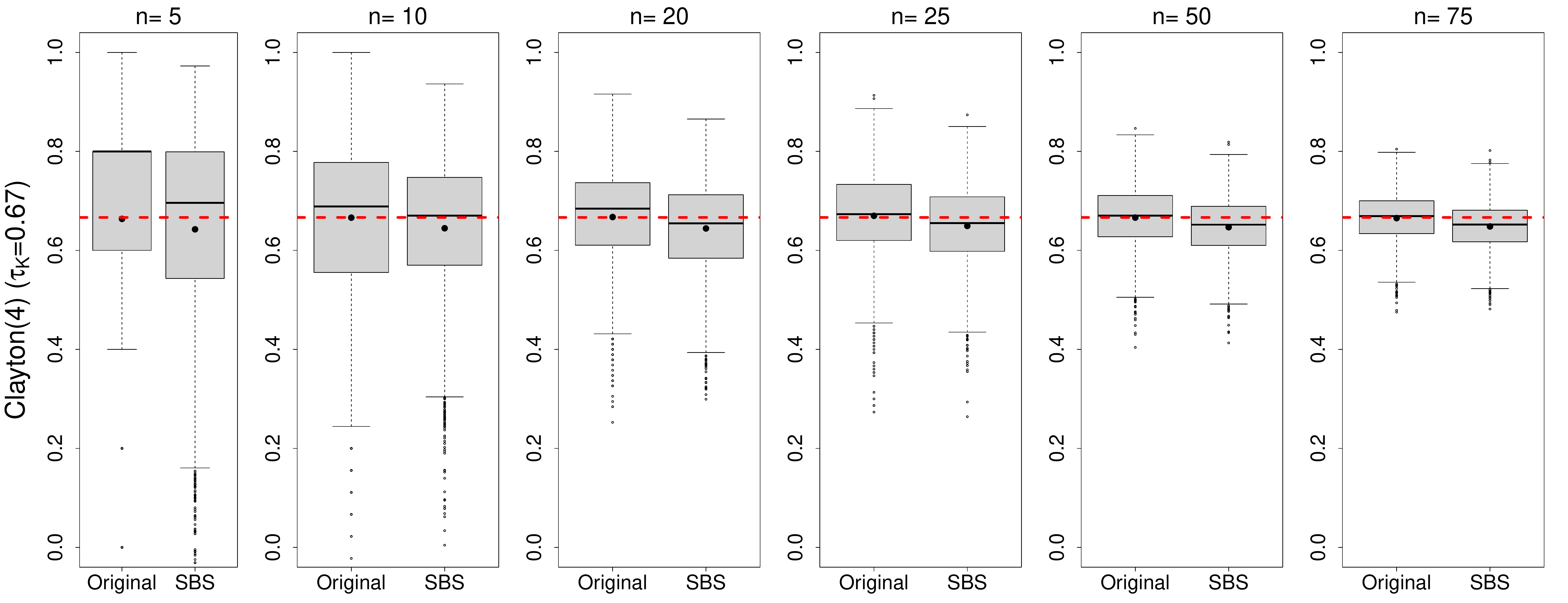}\\[4mm]
\includegraphics[width=\textwidth]{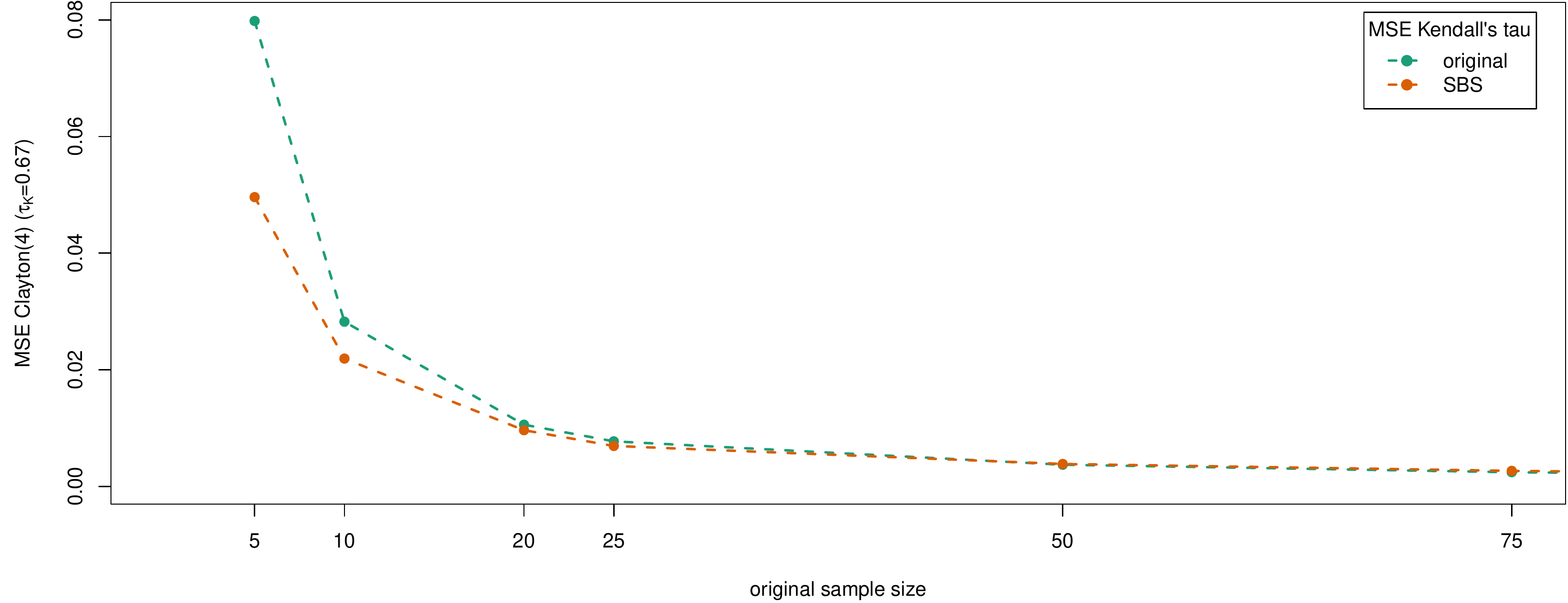}
\caption{Top: Estimated Kendall's tau for the bivariate Clayton($4$) copula.
The results are based on an original sample size of $\norig\in\{5,10,20,25,50,75\}$, while the augmented smooth bootstrap sample size is $\naugm=10\,000$.
Each boxplot is based on $M=2\,000$ independent reruns. The red dashed line indicates the theoretical value of $\tau_K$ while black dots indicate the means.\newline
Bottom: Mean squared error for estimation of Kendall's tau for the bivariate Clayton($4$) copula.
The results are based on an original sample size of $\norig\in\{5,10,20,25,50,75\}$, while the augmented smooth bootstrap sample size is $\naugm=10\,000$.}
\label{fig_Clayton_Kendall}
\end{figure}

As an alternative to copula contour lines we now consider the estimation of copula based dependence measures where we focus on Spearman's rho and Kendall's tau.
In our simulations we estimate Spearman's rho and Kendall's tau based on samples of size $\norig\in\{5,10,20,25,50,75,100\}$ where we consider Clayton, Student-$t$, Gumbel, Joe and Gaussian copulas.
In these copula families both dependence measures can be computed in closed form which allows us to compare our estimates to the true underlying values.
Based on the original observations we use the smooth bootstrap to generate $m=10\,000$ observations based on a Gaussian kernel where we use Silverman's rule of thumb to establish our bandwidth matrix.
Finally, we repeat the simulations $2\,000$ times.
For the Clayton copula with parameter $\theta=4$ this leads to the boxplots in Figure~\ref{fig_Clayton_Spearman} (top) and the mean squared error curves in Figure~\ref{fig_Clayton_Spearman} (bottom) for Spearman's rho, and the boxplots in Figure~\ref{fig_Clayton_Kendall} (top) and the mean squared error curves in Figure~\ref{fig_Clayton_Kendall} (bottom) for Kendall's tau.
The figures show that for small sample sizes the smooth bootstrap leads to an improved estimation in terms of the mean squared error for both, Spearman's rho and Kendall's tau.
For sample sizes larger than $75$ the advantage of the smooth bootstrap disappears.
The figures in Appendix~\ref{section_appendix_numerical_results} show that the same conclusions hold for our numerical experiments using the Student-$t$ copula with $\rho=0.9$ and $\nu=4$ degrees of freedom (see Figure~\ref{fig_student_Spearman} and \ref{fig_student_Kendall}), the Gumbel copula with parameter $\theta=4$ (see Figure~\ref{fig_Gumbel_Spearman} and \ref{fig_Gumbel_Kendall}), the Joe copula with parameter $\theta=4$ (see Figure~\ref{fig_Joe_Spearman} and \ref{fig_Joe_Kendall}) and the Gaussian copula with $\rho=0.9$ (see Figure~\ref{fig_Gauss_Spearman} and \ref{fig_Gauss_Kendall}).
Overall these findings seem to indicate that the smooth bootstrap improves the estimation of Spearman's rho and Kendall's tau for small sample sizes.
In the next section we conclude our results.
\section{Conclusion}\label{section_conclusion}
We investigate the distortion of the underlying dependence structure that arises as a side effect of the smooth bootstrap.
In the framework of elliptical distributions and elliptical smoothing kernels with a sphering type bandwidth matrix we provide the exact mechanism that leads to the distortion of the resulting elliptical copula.
Even though sphering is at first glance a strong restriction on the choice of a possible bandwidth matrix, it allows us to bypass otherwise necessary restrictions such as product kernels or diagonal bandwidth matrices.
While in our results the parameter matrix of the elliptical copula remains unchanged, the associated characteristic generator is distorted by a multiplicative factor related to the smoothing kernel.
We connect our investigation to the previous result of \cite{Bingham1972} to show that in general the pointwise convergence rate between the original and smoothed characteristic function is linked to the regular variation of the characteristic generator of the smoohting kernel.
Surprisingly we however also uncover situations where the underlying elliptical copula remains completely unaffected by the smooth bootstrap on the population level.
To complement this finding we discuss examples where the dependence distortion can be worked out in detail.
Furthermore, given that the parameter matrix remains unchanged, the dependence distortion introduced by kernel smoothing does not have any impact on a functional of the copula if the functional does not depend on the characteristic generator of the underlying elliptical copula.
Examples of such functionals are Kendall's tau and Blomqvist's beta.
However, in practical applications, even if it is known that the data generating process is elliptical, the estimation of $\dispMat$ may still impact our results and the estimation uncertainty connected to $\widehat{\dispMat}$ has to be taken into account.

From a practical perspective, we outline how the smooth bootstrap can be utilized to generate observations from the smoothed copula. Thus, it serves as a data augmentation scheme.
As well as stating an algorithm which can be used in this situation, we discuss details and options concerning the marginal transforms and application scenarios.
As a necessary part of the algorithm we generalise the univariate bandwidth selection procedure of \cite{BowmanHallPrvan1998} to the multivariate case.
This bandwidth selection procedure is not limited to diagonal bandwidth matrices and allows us to select an optimal full bandwidth matrix in terms of a weighted mean integrated squared error criterion.

In our simulation studies we utilize the smooth bootstrap to improve the estimation of copula contour lines in the bivariate and the copula diagonal in the multivariate case.
While our theoretical investigation is limited to elliptical distributions, our simulations show a vast improvement of the approximation measured in terms of the Hausdorff distance even though the utilised Clayton copula is not in class of elliptical copulas.
In a second set of simulations we consider copula based measures of dependence where we focus on Spearman's rho and Kendall's tau.
Our results show that the smooth bootstrap improves the estimation in small samples for a number of copulas in the elliptical and Archimedean class.

Based on our simulation results the smooth bootstrap can lead to an improved estimation of copula functionals for small sample sizes.
The procedure is especially advantageous if the target functional is smooth, such as level curves or the copula diagonal in the considered examples, but the estimator based on the initial (small) sample is too coarse.
Here an application of the smooth bootstrap leads to a virtually unlimited number of observations which, in the considered applications, leads to a dramatic improvement of the estimation.

\subsection*{Acknowledgements}
We thank an anonymous reviewer for helpful comments which improved the quality of the paper.
The third author gratefully acknowledges financial support of the Karlsruhe Institute of Technology (KIT) where part of this research was carried out.
The third author would like to thank NSERC for financial support for this work through Discovery
Grant RGPIN-2020-05784.
The fourth author acknowledges support from NSERC (RGPIN-2020-04897, RGPAS-2020-00093)

\bibliographystyle{plainnat}
\bibliography{paper}

\appendix
\section{Properties of characteristic functions}\label{section_properties_of_CF}
Here, we review properties of multivariate characteristic functions.
\begin{Definition}[$\dim$-dimensional Characteristic function]\label{definition_multivariate_cf}
The characteristic function of $\dim$-dimensional random vector $\bX$ is defined as
\begin{align*}
\CF_{\bX} \colon \R^{\dim} \to \C, \quad \bt \mapsto \CF_{\bX}(\bt) = \Eval{e^{\imu \bt^{\tr} \bX}},
\end{align*}
where $\imu \in \C$ is the imaginary unit with $\imu^2 = -1$.
\end{Definition}
In an abuse of notation we may also write $\CF_{\distI}$ and $\CF_{\densI}$ when $\bX$ has distribution $\distI$ and density $\densI$.
Among the properties of characteristic functions the following two theorems will be useful in the context of kernel smoothing.
A comprehensive treatment of multivariate characteristic functions can be found, e.g., in \cite{Sasvari2013}.
\begin{Theorem}[Characteristic function of convolutions, {\citealt[Theorem~1.1.3]{Sasvari2013}}]\label{theorem_cf_convolution}
If $\bX$ and $\bY$ are independent $\dim$-dimensional random vectors, then the characteristic function of their sum is $\CF_{\bX+\bY} = \CF_{\bX} \cdot \CF_{\bY}$.
\end{Theorem}
\begin{Theorem}[Characteristic function of affine transformations, {\citealt[Theorem~1.1.7]{Sasvari2013}}]\label{theorem_cf_affine}
Let $\bX$ be a $\dim$-dimensional random vector. Then the equation
\begin{align*}
\CF_{A\bX+\bb}(\bt) = e^{\imu \bt^{\tr}\bb} \cdot \CF_{\bX}\left(A^{\tr} \bt \right) , \bt \in \R^{n}
\end{align*}
holds for every linear mapping $A \colon \R^{\dim} \to \R^n$ and $\bb \in \R^n$.
\end{Theorem}
Due to the assumed independence of the underlying random vector and the smoothing kernel we can directly compute the characteristic function for our main objects under consideration.
Concerning the expected density function estimator, we immediately obtain that
$
\CF_{\expkde}(\bt) = \CF_{\bX}(\bt)\CF_{\bY_{\BWMatrix_{\norig}}}(\bt)
$
due to the independence of $\bX$ and $\bY_{\BWMatrix_{\norig}}$; see Theorem~\ref{theorem_cf_convolution}.
The characteristic function of $\bY_{\BWMatrix_{\norig}}$ is obtained in terms of the characteristic function of $\bY$, or equivalently $\kernel$, as
$
\CF_{\bY_{\BWMatrix_{\norig}}}(\bt) = \CF_{\kernel}\left(\BWMatrix_{\norig}^{1/2}\bt\right),
$
see Theorem~\ref{theorem_cf_affine}, which overall leads to
\begin{align}\label{eq_cf_kernel_population}
\CF_{\expkde}(\bt) = \CF_{\bX}(\bt)\CF_{\kernel}\left(\BWMatrix_{\norig}^{1/2}\bt\right).
\end{align}

Via Theorem~\ref{theorem_cf_affine} also the characteristic function of the conditional density estimate $\condkde$ can be computed as
\begin{align}\label{eq_cf_kernel_sample}
\CF_{\condkde}(\bt) = \CF_{\kernel}\left(\left(\BWMatrix_{\norig}^{1/2}\right)^{\tr}\bt\right)\frac{1}{\norig}\sum^{\norig}_{i=1} e^{\imu \bt^{\tr}\bx_i},
\end{align}
from which we can recover the characteristic function of the expected kernel density estimate given in \eqref{eq_cf_kernel_population} when taking the expectation with respect to the underlying random vector $\bX$.

Characteristic functions will also play an important role in determining the distance between two distribution functions.
The Kolmogorov-Smirnov distance between two $\dim$-dimensional distribution functions is denoted by
\begin{align}\label{equation_def_KS}
\dKS\left(\distI,\distII\right) = \sup_{\bx \in \R^{\dim}} \abs{\distI(\bx)-\distII(\bx)}.
\end{align}
The distance $\dKS$ can be bounded by the average scaled difference of the associated characteristic functions.
In the univariate case this smoothing inequality is linked to the Berry--Esseen theorem and for $T > 0$ takes the form
\begin{align*}
\dKS\left(\mdistI,\mdistII\right) \leq \frac{1}{\pi} \int_{-T}^T \abs{\frac{\CF_X(t)-\CF_Y(t)}{t}} \d t + \frac{24}{\pi T} \sup_{x\in\R}\abs{G'(x)},
\end{align*}
see, for example, \citealt[Lemma~2, page 538]{Feller1971}.
In the multivariate case extensions are available.
To keep the notation to a minimum we present the bivariate case where we use the presentation of \cite{HeubergerKropf2018}.
Here the cut-off interval $[-T,T]$ is generalized to a ball.
Alternatively, a representation in terms of a cube $[-T,T]^{2}$ is also possible; see \cite{Sadikova1966}.

\begin{Theorem}[Smoothing inequality; {\citealt[Theorem~2]{HeubergerKropf2018}}]\label{theorem_Berry_Essen_inequalty_2D}
Denote by $\bX \sim \distI_{\bX}$ and $\bY \sim \distI_{\bY}$ two $2$-dimensional random vectors.
Assume that $\distI_{\bY}$ is differentiable.
Let $T > 0$ be fixed, then
\begin{align*}
\dKS\left(\distI_{\bX},\distI_{\bY}\right) &\leq \frac{2}{(2\pi)^2}\int_{\xNorm{\bt}{2} \leq T} \abs{\frac{ \CF_{\bX}(\bt) - \CF_{X_1}(t_1)\CF_{X_2}(t_2) - \CF_{\bY}(\bt) + \CF_{Y_1}(t_1)\CF_{Y_2}(t_2)}{\prod_{i=1}^{2} t_i }} \d\bt\\
&+ \frac{2}{\pi} \int_{\abs{t_1}\leq T} \abs{\frac{\CF_{X_1}(t_1)-\CF_{Y_1}(t_1)}{t_1}} \d t_1 + \frac{2}{\pi} \int_{\abs{t_2}\leq T} \abs{\frac{\CF_{X_2}(t_2)-\CF_{Y_2}(t_2)}{t_2}} \d t_2
\\
&+ 2\frac{\xNorm{\frac{\partial \distI_{\bY}}{\partial y_1}}{\infty}+\xNorm{\frac{\partial \distI_{\bY}}{\partial y_2}}{\infty}}{T} \left(\frac{12}{\pi} + \sqrt[3]{\frac{32}{\pi\left(1-\left(\frac{3}{4}\right)^{1/2}\right)}}\right)\\
&+ \frac{4 \xNorm{\mdensI_{Y_1}}{\infty}}{T}\left(\frac{12}{\pi} + \sqrt[3]{\frac{128}{\pi}}\right)
+ \frac{4 \xNorm{\mdensI_{Y_2}}{\infty}}{T}\left(\frac{12}{\pi} + \sqrt[3]{\frac{128}{\pi}}\right).
\end{align*}
\begin{proof}
This is the bivariate case of \citealt[Theorem~2]{HeubergerKropf2018}.
\end{proof}
\end{Theorem}
\section{Elliptically distributed random vectors and copulas}\label{section_elliptical_random_vectors}
For our considerations the class of elliptical distributions will play a central role.
Textbook introductions can be found in \cite{FangKotzNg1990} and \cite{McNeilFreyEmbrechts2015}.
Before going to elliptical distributions we first introduce spherical distributions as a necessary stepping stone.
\begin{Definition}[Spherical distribution; {\citealt[Definition~3.18]{McNeilFreyEmbrechts2015}}]\label{definition_spherical_distribution}
A $\dim$-dimensional random vector $\bY = (Y_1,\ldots,Y_{\dim})^{\tr}$ has a spherical distribution if, for every orthogonal matrix $U \in \R^{\dim\times\dim}$, $UU^{\tr} = U^{\tr}U = \IdMat_{\dim}$, we have $U \bX \equalindist \bX$.
\end{Definition}
Spherical distributions can equivalently be characterized via their characteristic functions or by randomly scaling a uniform distribution on the unit sphere.
\begin{Theorem}[Equivalent characterization of spherical distributions; {\citealt[Theorem~3.19 and Theorem~3.22]{McNeilFreyEmbrechts2015}}]\label{theorem_characterization_spherical_distribution}
Denote by $\bY$ a $\dim$-dimensional random vector.
Then the following are equivalent:
\begin{enumerate}
\item
$\bY$ has a spherical distribution,
\item
there exists a real valued function $\charGen \colon [0,\infty) \to [-1,1]$ with $\charGen(0) = 1$ such that the characteristic function of $\bY$ is given by
$
\CF_{\bY}(\bt) = \charGen\left(\bt^{\tr}\bt \right),
$
\item
$\bY$ has the stochastic representation
$
\bY \equalindist \radialPart \sphericalPart,
$
where $\sphericalPart$ is uniformly distributed on the unit sphere $\{\bx \in \R^{\dim} : \xNorm{\bx}{2} = 1\}$ and $\radialPart$ is an almost surly non-negative random variable independent of $\sphericalPart$.
\end{enumerate}
The random variable $\radialPart$ specific to $\bY$ is called the radial distribution of $\bY$.
The function $\charGen$ is called the characteristic generator of $\bY$.
We denote the distribution of $\bY$ by $\Spherical_{\dim}(\charGen)$.
The subclass of spherical distributions for which $\Prob{\radialPart = 0} = 0$ is denoted by $\Spherical_{\dim}^+(\charGen)$.
\end{Theorem}
In Table~\ref{table_characteristic_generator} we collect the characteristic generators and the value of their derivative at $0$ for popular spherical and elliptical models.
\begin{table}
  \centering
  \begin{tabular}{l c c}
    \hline\hline
    Distribution & $\charGen(u)$ & $\lim_{u \to 0^+} \charGen'(u)$ \\
    \hline
    Gauss & $\exp(-u/2)$ & $-1/2$ \\
    Laplace & $(1+u/2)^{-1}$ & $-1/2$ \\
    Student $t$ ($\df=1$) & $\exp\left(-\sqrt{u}\right)$ & $-\infty$ \\
    Student $t$ ($\df=2$) & $\Bessel_1\left(\sqrt{2x}\right)\sqrt{2x}$  & $-\infty$ \\
    Student $t$ ($\df=4$) & $\Bessel_2\left(\sqrt{4x}\right)2x$  & $-1$\\
    Student $t$ ($\df>2$) & $\frac{\Bessel_{\df/2}\left(\sqrt{\df x}\right)\left(\sqrt{\df x}\right)^{\df/2}}{\Gamma\left(\df/2\right)2^{\df/2-1}}$ & $-\df/(2\df-4)$\\
    \hline\hline
  \end{tabular}
  \caption{Characteristic generator $\charGen$ and value of its derivative at
    $0$ for popular spherical and elliptical models.}
  \label{table_characteristic_generator}
\end{table}
Based on Definition~\ref{definition_spherical_distribution} we can now go on to define elliptical distributions.
\begin{Definition}[Elliptical distribution; {\citealt[Definition~3.26]{McNeilFreyEmbrechts2015}}]\label{definition_elliptical_distribution}
A $\dim$-dimensional random vector $\bX$ has an elliptical distribution if
\begin{align}\label{eq_elliptical_distribution_definition}
\bX \equalindist \bMu + A \bY,
\end{align}
where $\bY \sim \Spherical_{\dim}(\charGen)$ and $A \in \R^{\dim\times k}$ and $\bMu \in \R^{\dim}$ are a matrix and vector of constants, respectively.
The distribution of $\bX$ is denoted by $\Ellip_{\dim}(\bMu,\dispMat,\charGen)$ where $\dispMat = A A^{\tr}$ is called the dispersion matrix.
The radial distribution of $\bX$ is the radial distribution $\radialPart$ associated to $\bY$ in \eqref{eq_elliptical_distribution_definition}.
\end{Definition}
Given the stochastic representation of spherical random vectors an elliptical random vector naturally has the representation
\begin{align*}
\bX \equalindist \bMu + \radialPart A \sphericalPart
\end{align*}
where $\radialPart$ is the associated radial distribution, $A$ is a non-random matrix and $\sphericalPart$ is uniformly distributed on the unit sphere.
Based on the characteristic function of spherical random vectors, the characteristic function of an elliptical random vector $\bX \sim \Ellip_{\dim}(\bMu,\dispMat,\charGen)$ is given by
\begin{align*}
\CF_{\bX}(\bt) = e^{\imu \bt^{\tr}\bMu}\charGen\left( \bt^{\tr} \Sigma \bt \right).
\end{align*}
If the radial distribution of an elliptical random vector has a finite second moment, the mean vector and covariance matrix admit the following convenient expressions.
\begin{Theorem}[Moments of elliptical random vectors; {\citealt[Theorem~2.17, p.~43]{FangKotzNg1990}}]\label{theorem_elliptical_moments}
If $\bX \sim \Ellip_{\dim}(\bMu,\dispMat,\charGen)$ and $\Eval{\radialPart^2} < \infty$, then $\Eval{\bX} = \bMu$ and $\cov{\bX} = \frac{\E[\radialPart^2]}{\rank(\dispMat)}\dispMat = -2 \charGen'(0) \dispMat$.
\end{Theorem}
\begin{Remark}[Re-parameterization of elliptical random vectors]\label{remark_reparameterization}
The parameterization of an elliptical random vector is non-unique since the dispersion matrix and the characteristic generator can be rescaled.
From the characteristic function representation it is clear that the pairs $(\dispMat,\charGen)$ and $(c \dispMat, \charGen(\cdot/c))$ lead to the same distribution for every $c > 0$.
If $\bX \sim \Ellip_{\dim}(\bMu,\dispMat,\charGen)$ we can consequently set $\widetilde{\dispMat} = -2 \charGen'(0) \dispMat$ and $\widetilde{\charGen}(u) = \charGen(u/ (-2 \charGen'(0)) )$ to obtain a parameterization $\bX \sim \Ellip_{\dim}\left(\bMu,\widetilde{\dispMat},\widetilde{\charGen}\right)$ such that $\cov{\bX} = \widetilde{\dispMat}$.
\end{Remark}
Elliptically distributed random vectors are especially well behaved when considering affine transformations and sums as shown in the following theorems.
\begin{Theorem}[Linear combinations of elliptical random vectors; {\citealt[Theorem~2.16, p.~43]{FangKotzNg1990}}]\label{theorem_elliptical_linear_combinations}
Denote by $\bB$ a $k\times\dim$ matrix and by $\bb \in \R^k$ a vector.
If $\bX \sim \Ellip_{\dim}(\bMu,\dispMat,\charGen)$ with $\rank(\dispMat) = k$ then $\bB \bX + \bb \sim \Ellip_{k}(\bB \bMu + \bb,\bB \dispMat \bB^{\tr},\charGen)$.
\end{Theorem}
As an application of Theorem~\ref{theorem_elliptical_linear_combinations} we can obtain the marginal distributions of an elliptical random vector by considering $X_j = e_j^{\tr}\bX$, where $e_j = (0,\ldots,1,\ldots,0)^{\tr}$ denotes the $j$th unit vector.
We then obtain
\begin{align}\label{eq_univariate_margins_elliptical}
X_j \sim \Ellip_1(\bMu_j,\dispMat_{jj},\charGen).
\end{align}
Summation of elliptical random vectors produces again an elliptical random vector under certain conditions.
Our first theorem in this direction is due to \cite{HultLindskog2002} and relaxes the commonly requested independence assumption by allowing for non-independent radial distributions.
\begin{Theorem}[{\citealt[Theorem~4.1]{HultLindskog2002}}]\label{theorem_summation_elliptical_random_vectors}
Let $\radialPart_1$ and $\radialPart_2$ be non-negative random variables and let $\bX_1 = \bMu_1 + \radialPart_1 A_1 \sphericalPart_1 \sim \Ellip_{\dim}(\bMu_1,\dispMat,\charGen_1)$ and $\bX_2 = \bMu_2 + \radialPart_2 A_2 \sphericalPart_2 \sim \Ellip_{\dim}(\bMu_2,\dispMat,\charGen_2)$, where the random vectors $(\radialPart_1,\radialPart_2)$, $\bZ_1$ and $\bZ_2$ are mutually independent.
Then $\bX_1+\bX_2 \sim \Ellip_{\dim}(\bMu_1 + \bMu_2,\dispMat,\charGen_3)$.
Moreover, if $\radialPart_1$ and $\radialPart_2$ are independent, then $\charGen_3(u) = \charGen_1(u) \charGen_2(u)$.
\end{Theorem}
In our context a slight generalization of Theorem~\ref{theorem_summation_elliptical_random_vectors} for non-equal dispersion matrices is needed.
This is essentially a combination of Theorem~\ref{theorem_summation_elliptical_random_vectors} and \citealt[Lemma~1]{LindskogMcNeilSchmock2003}.
\begin{Corollary}\label{corollary_summation_elliptical_random_vectors}
Let $\radialPart_1$ and $\radialPart_2$ be non-negative random variables and let $\bX_1 = \bMu_1 + \radialPart_1 A_1 \sphericalPart_1 \sim \Ellip_{\dim}(\bMu_1,\dispMat,\charGen_1)$ and $\bX_2 = \bMu_2 + \radialPart_2 A_2 \sphericalPart_2 \sim \Ellip_{\dim}(\bMu_2,c \dispMat,\charGen_2)$, where the random vectors $(\radialPart_1,\radialPart_2)$, $\bZ_1$ and $\bZ_2$ are mutually independent and $c > 0$.
Then $\bX_1+\bX_2 \sim \Ellip_{\dim}(\bMu_1 + \bMu_2,\dispMat,\charGen_c)$ for some characteristic generator $\charGen_c$.
Moreover, if $\radialPart_1$ and $\radialPart_2$ are independent, then $\charGen_c(u) = \charGen_1(u) \charGen_2(c u)$.
\begin{proof}
Denote by $\CF_1$ and $\CF_2$ the conditional characteristic functions
\begin{align*}
\CF_1(\bt) &= \Eval{e^{\imu (\bMu_1 + \radialPart_1 A_1 \sphericalPart_1)^{\tr}\bt}\Big|\radialPart_1 = r_1} = e^{\imu \bt^{\tr}\bMu_1}\charGen_1\left(r_1^2 \bt^{\tr}\dispMat\bt\right),\\
\CF_2(\bt) &= \Eval{e^{\imu (\bMu_2 + R_2 A_2 \sphericalPart_2)^{\tr}\bt}\Big|\radialPart_1 = r_1} =  e^{\imu \bt^{\tr}\bMu_2}\charGen_2^{r_1}\left(\bt^{\tr}c \dispMat \bt\right),
\end{align*}
where $\charGen_1$ is the characteristic generator of $A_1 \sphericalPart_1$ and $\charGen_2^{r_1}$ is the characteristic generator of $\radialPart_2 A_2 \sphericalPart_2$ given $\radialPart_1 = r_1$.
Following the same steps as in the proof of \citealt[Theorem~4.1]{HultLindskog2002}, we obtain
\begin{align*}
\CF_{\bX_1+\bX_2}(\bt) = e^{\imu \bt^{\tr}(\bMu_1 + \bMu_2)}\Eval{\charGen_1\left(\radialPart_1^2 \bt^{\tr}\dispMat\bt\right)\charGen_2^{\radialPart_1}\left(c \bt^{\tr} \dispMat \bt\right)},
\end{align*}
showing that $\bX_1+\bX_2 \sim \Ellip_{\dim}(\bMu_1 + \bMu_2,\dispMat,\charGen_c)$ with characteristic generator
\begin{align*}
  \charGen_c(u) = \int_0^{\infty} \charGen_1\left(r_1^2 u\right)\charGen_2^{r_1}\left(c u\right)\d \mdistI_{\radialPart_1}(r_1),
\end{align*}
where $\mdistI_{\radialPart_1}$ is the marginal distribution of $\radialPart_1$.
If $\radialPart_1$ and $\radialPart_2$ are independent we directly get $\charGen_c(u) = \charGen_1(u) \charGen_2(c u)$ without conditioning on $\radialPart_1 = r_1$.
\end{proof}
\end{Corollary}
If a spherical or elliptical random vector is absolutely continuous with respect to the Lebesgue measure, the density takes a particular form.
\begin{Theorem}[Density generator; {\citealt[Equation~(3.46)]{McNeilFreyEmbrechts2015}}]\label{theorem_density_generator}
If a spherical random vector $\bY \sim \Spherical_{\dim}(\charGen)$ is absolutely continuous with respect to the Lebesgue measure, the density $\mdensI_{\bY}$ for $\by \in \R^{\dim}$ takes the form
$
\mdensI_{\bY}(\by) = \dGen_{\bY}(\by^{\tr}\by),
$
where $\dGen_{\bY}$ is a positive function $\dGen_{\bY} \colon [0,\infty) \to [0,\infty), t \mapsto \dGen_{\bY}(t)$.
The function $\dGen_{\bY}$ is called the \emph{density generator} of $\bY$.\newline
If an elliptical random vector $\bX \sim \Ellip_{\dim}(\bMu,\dispMat,\charGen)$ is absolutely continuous with respect to the Lebesgue measure, the density $\mdensI_{\bX}$ for $\bx \in \R^{\dim}$ takes the form
$
\mdensI_{\bX}(\bx) = \frac{1}{\sqrt{\det(\dispMat)}}\dGen_{\bY}\left((\bx-\bMu)^{\tr}\dispMat^{-1}(\bx-\bMu)\right),
$
where $\dGen_{\bY}$ is the density generator of the spherical distribution $\bY$ associated to $\bX$.
\end{Theorem}

Based on Sklar's theorem every elliptically distributed random vector gives rise to an associated copula.
Due to the fact that the mean and variances are only marginal attributes they do not play a role when focusing on the dependence structure inherent to a given elliptical random vector.
Hence it is sufficient to neglect location parameters and to only consider correlation matrices in the following definition.
For a one-dimensional distribution function $\mdistI$ we denote its quantile function by $\mdistI^{\ginv}$.
\begin{Definition}[Elliptical copula]\label{definition_elliptical_copula}
Denote by $\corrMat$ a correlation matrix and by $\charGen$ a characteristic generator.
The \emph{elliptical copula} $\Copula_{\corrMat,\charGen}$ is given by the copula associated to the $\dim$-dimensional random vector $\bX \sim \Ellip_{\dim}(\bZero,\corrMat,\charGen)$ by virtue of Sklar's theorem
\begin{align*}
\Copula_{\corrMat,\charGen}(u_1,\ldots,u_{\dim}) = \Prob{X_1 \leq \mdistI^{\ginv}(u_1),\ldots,X_{\dim} \leq \mdistI^{\ginv}(u_{\dim})},
\end{align*}
where $\mdistI = \Ellip_1(0,1,\charGen)$ is the marginal distribution common to $X_1,\ldots,X_{\dim}$.
\end{Definition}
\begin{Remark}\label{remark_same_elliptical_copula}
Given that copulas are invariant under strictly increasing transformations, different elliptical random vectors (and their parameterizations) can give rise to the same elliptical copula.
Specifically, the copula $\Copula_{\bX}$ of an elliptical random vector $\bX\sim \Ellip_{\dim}(\bMu,\dispMat,\charGen)$ is the same as the elliptical copula $\Copula_{\corr{\bX},\charGen}$.
This can readily be seen by considering a random vector $\bY$ with elliptical copula
$\Copula_{\corr{\bX},\charGen}$, i.e., $\bY \sim \Ellip_{\dim}(\bZero,\corr{\bX},\charGen)$ and defining $\bm{D} = \diag(\sd[X_1],\ldots,\sd[X_{\dim}]) / \sqrt{-2\charGen'(0)}$.
This yields $\bm{D}\bY + \bMu \sim \Ellip_{\dim}(\bMu,\dispMat,\charGen)$ and hence $\bm{D}\bY + \bMu \equalindist \bX$.
We therefore have $\Copula_{\bX} = \Copula_{\corr{\bX},\charGen}$, since the transformation applied to $\bY$ is strictly increasing in every component.\newline\noindent
Considering for example a multivariate normal random vector $\bX \sim \Ellip_{\dim}(\bZero,c\corrMat,\charGen)$, where $\corrMat$ is a correlation matrix and $c>0$, we have a characteristic generator $\charGen(u) = \exp(-u/2)$ with $-2\charGen'(0)=1$.
Adjusting the margins by $\widetilde{D} \bX$, where $\widetilde{D} = \diag(1/\sqrt{c},\ldots,1/\sqrt{c})$, corrects the correlation structure to $\corr{\widetilde{D} \bX} = \corrMat$.
This shows that the random vectors $\bX$ and $\widetilde{D} \bX$ share the same Gaussian copula.
Alternatively, the copula can also be obtained by the multivariate probability integral transform which in this case takes the form $\left(\Phi\left(\frac{X_1}{\sqrt{c}}\right),\ldots,\Phi\left(\frac{X_{\dim}}{\sqrt{c}}\right)\right)$.
\end{Remark}

\section{Numerical results}\label{section_appendix_numerical_results}
This section contains the numerical results concerning the estimation of Spearman's rho and Kendall's tau using the smooth bootstrap discussed in Section~\ref{section_copula_based_dependence_measures}.
\begin{figure}
\centering
\includegraphics[width=\textwidth]{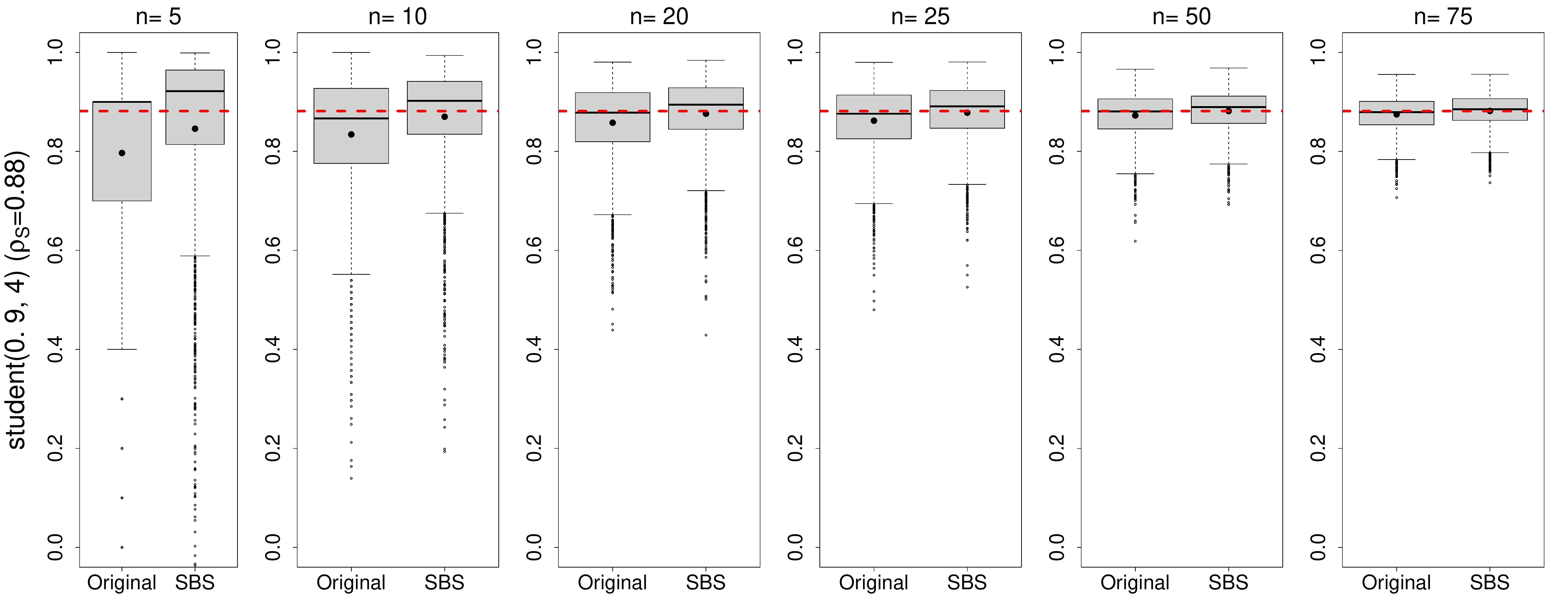}\\[4mm]
\includegraphics[width=\textwidth]{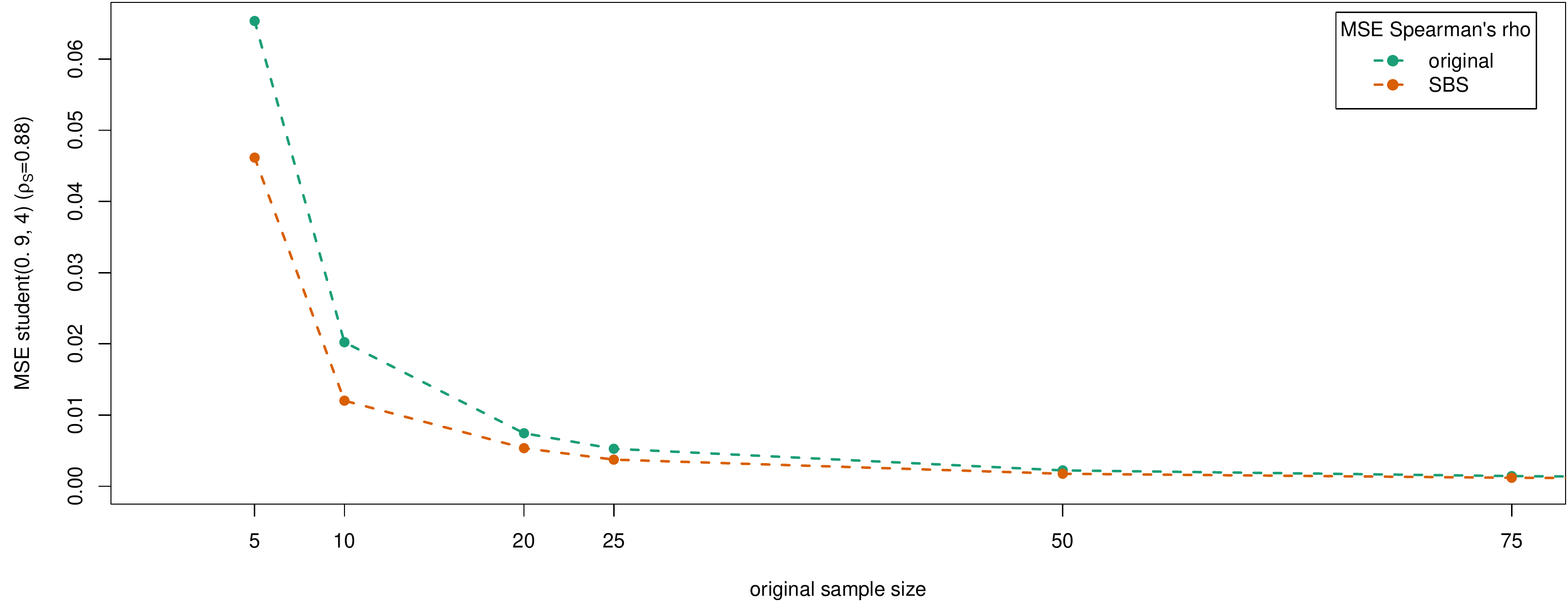}
\caption{Top: Estimated Spearman's rho for the bivariate Student-$t$ copula with $\rho=0.9$ and $\nu=4$ degrees of freedom.
The results are based on an original sample size of $\norig\in\{5,10,20,25,50,75\}$, while the augmented smooth bootstrap sample size is $\naugm=10\,000$.
Each boxplot is based on $M=2\,000$ independent reruns. The red dashed line indicates the theoretical value of $\rho_S$ while black dots indicate the means.\newline
Bottom: Mean squared error for estimation of Spearman's rho.
The results are based on an original sample size of $\norig\in\{5,10,20,25,50,75\}$, while the augmented smooth bootstrap sample size is $\naugm=10\,000$.}
\label{fig_student_Spearman}
\end{figure}
\begin{figure}
\centering
\includegraphics[width=\textwidth]{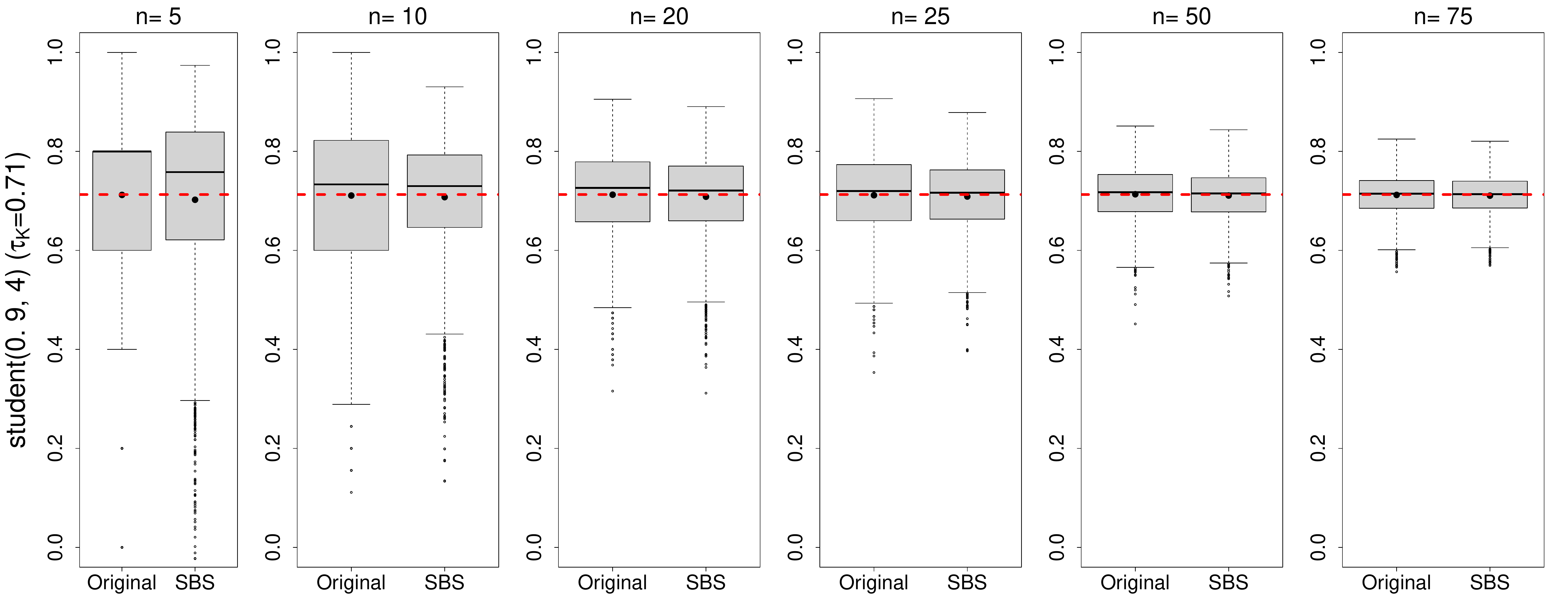}\\[4mm]
\includegraphics[width=\textwidth]{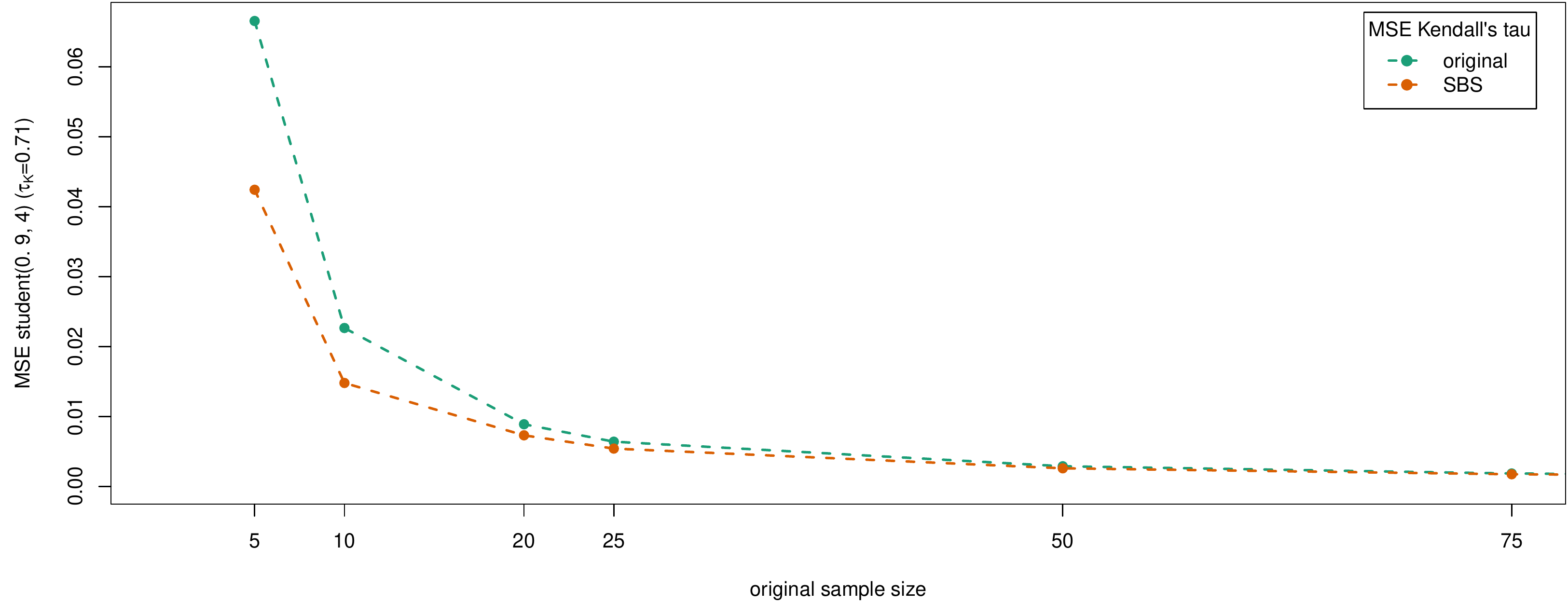}
\caption{Top: Estimated Kendall's tau for the bivariate Student-$t$ copula with $\rho=0.9$ and $\nu=4$ degrees of freedom.
The results are based on an original sample size of $\norig\in\{5,10,20,25,50,75\}$, while the augmented smooth bootstrap sample size is $\naugm=10\,000$.
Each boxplot is based on $M=2\,000$ independent reruns. The red dashed line indicates the theoretical value of $\tau_K$ while black dots indicate the means.\newline
Bottom: Mean squared error for estimation of Kendall's tau.
The results are based on an original sample size of $\norig\in\{5,10,20,25,50,75\}$, while the augmented smooth bootstrap sample size is $\naugm=10\,000$.}
\label{fig_student_Kendall}
\end{figure}
\begin{figure}
\centering
\includegraphics[width=\textwidth]{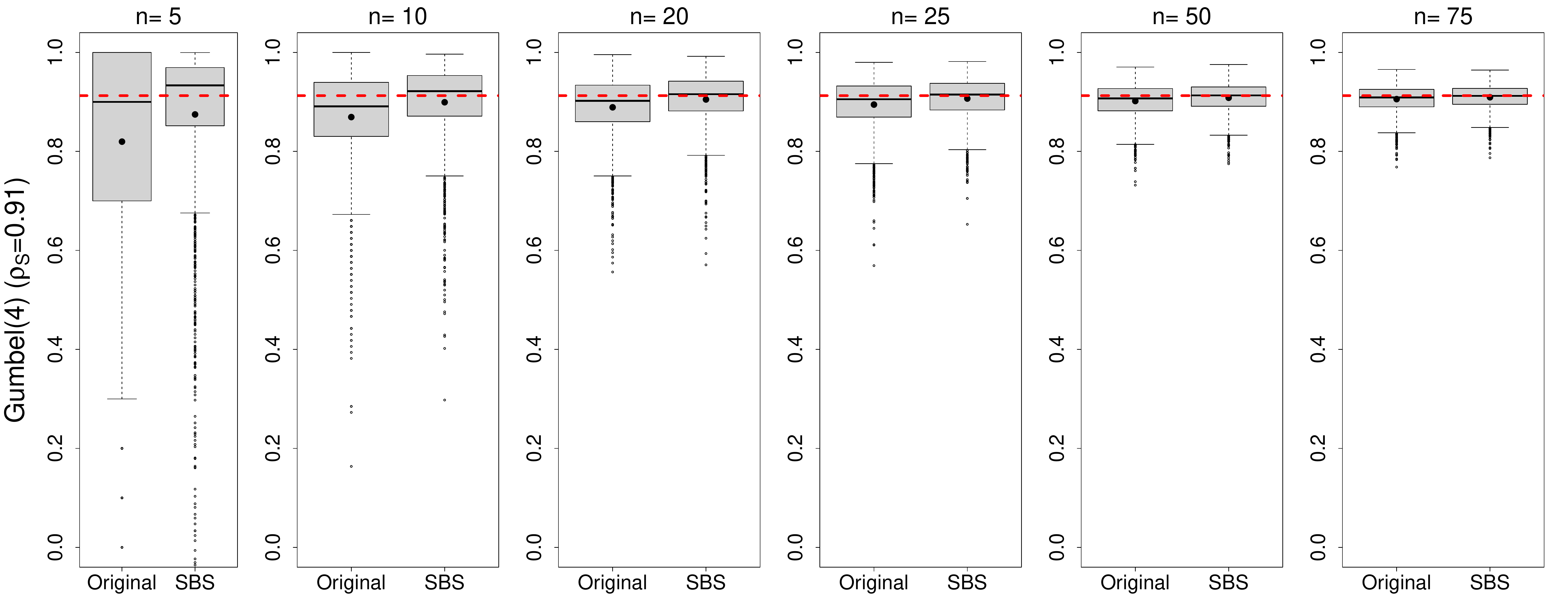}\\[4mm]
\includegraphics[width=\textwidth]{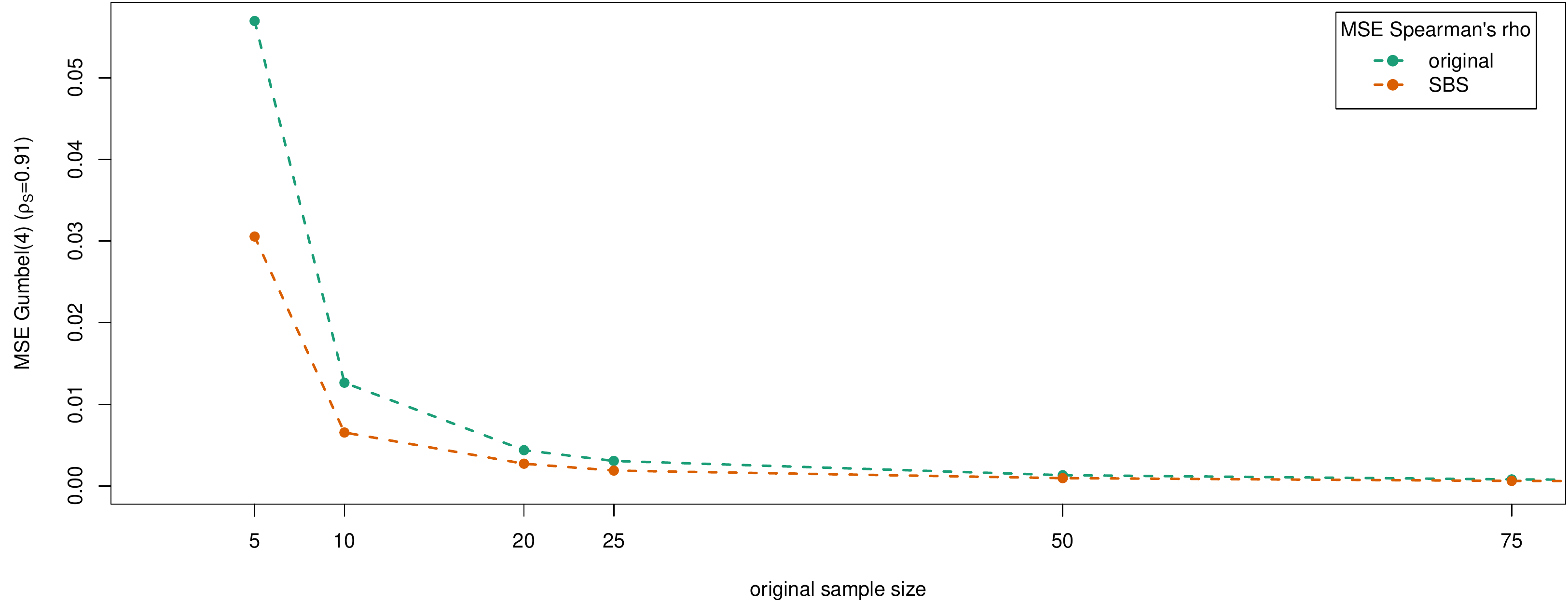}
\caption{Top: Estimated Spearman's rho for the bivariate Gumbel copula with parameter $\theta=4$.
The results are based on an original sample size of $\norig\in\{5,10,20,25,50,75\}$, while the augmented smooth bootstrap sample size is $\naugm=10\,000$.
Each boxplot is based on $M=2\,000$ independent reruns. The red dashed line indicates the theoretical value of $\rho_S$ while black dots indicate the means.\newline
Bottom: Mean squared error for estimation of Spearman's rho.
The results are based on an original sample size of $\norig\in\{5,10,20,25,50,75\}$, while the augmented smooth bootstrap sample size is $\naugm=10\,000$.}
\label{fig_Gumbel_Spearman}
\end{figure}
\begin{figure}
\centering
\includegraphics[width=\textwidth]{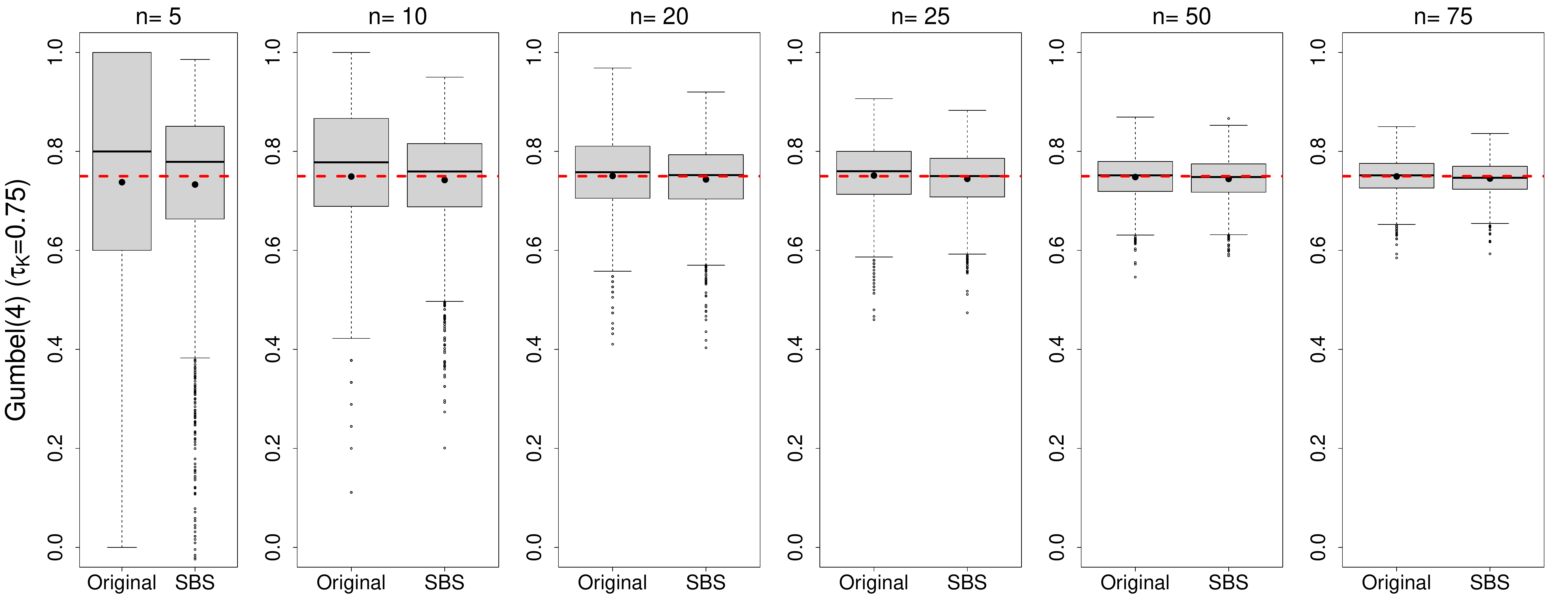}\\[4mm]
\includegraphics[width=\textwidth]{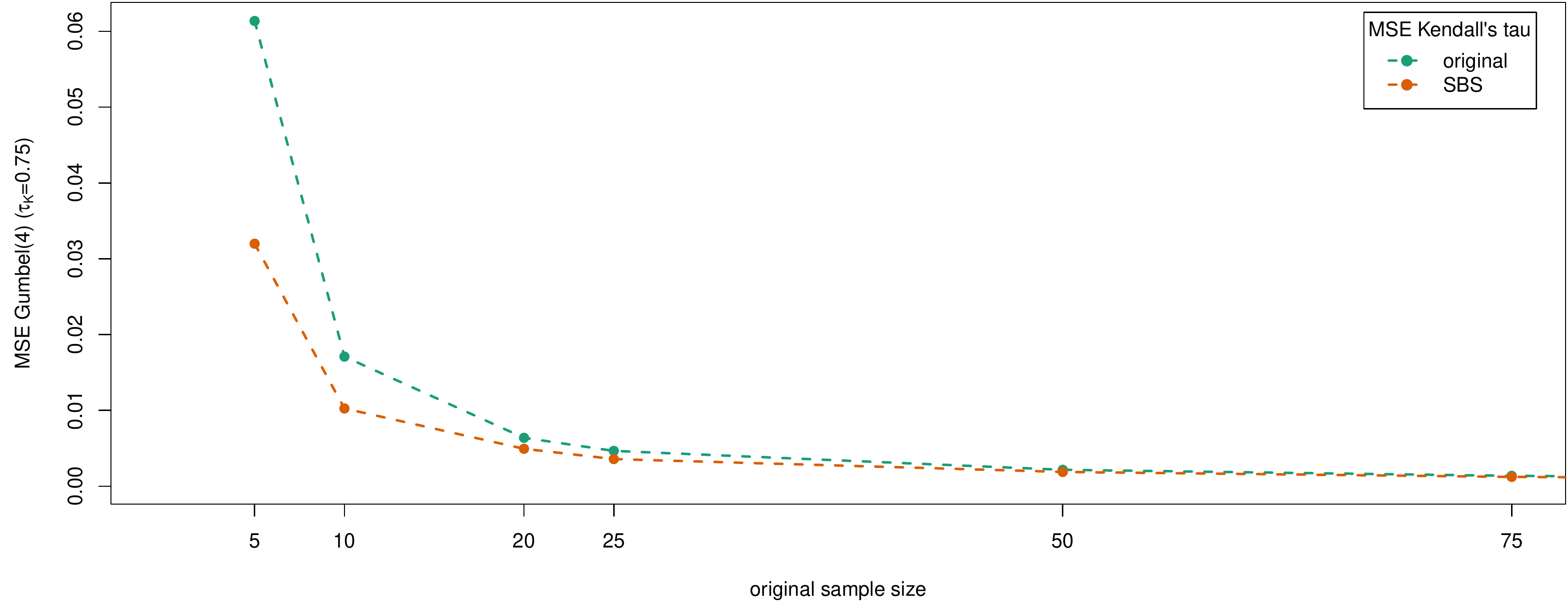}
\caption{Top: Estimated Kendall's tau for the bivariate Gumbel copula with parameter $\theta=4$.
The results are based on an original sample size of $\norig\in\{5,10,20,25,50,75\}$, while the augmented smooth bootstrap sample size is $\naugm=10\,000$.
Each boxplot is based on $M=2\,000$ independent reruns. The red dashed line indicates the theoretical value of $\tau_K$ while black dots indicate the means.\newline
Bottom: Mean squared error for estimation of Kendall's tau.
The results are based on an original sample size of $\norig\in\{5,10,20,25,50,75\}$, while the augmented smooth bootstrap sample size is $\naugm=10\,000$.}
\label{fig_Gumbel_Kendall}
\end{figure}
\begin{figure}
\centering
\includegraphics[width=\textwidth]{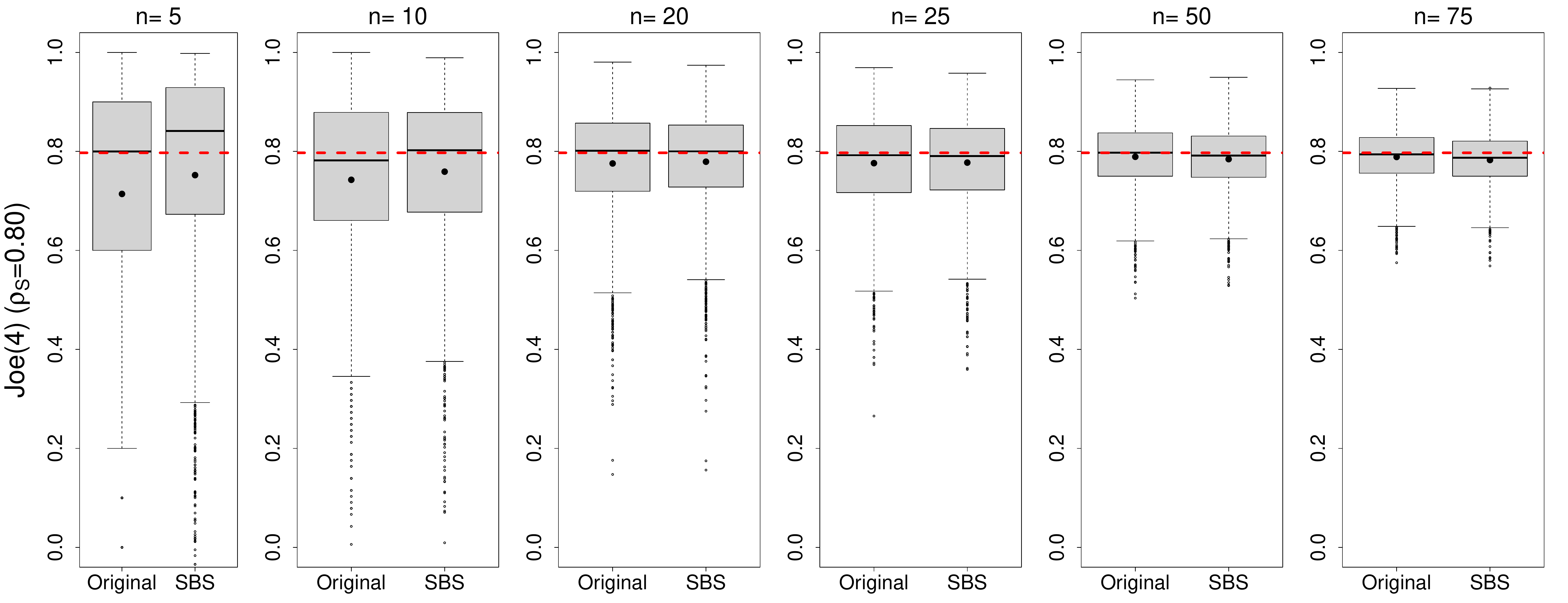}\\[4mm]
\includegraphics[width=\textwidth]{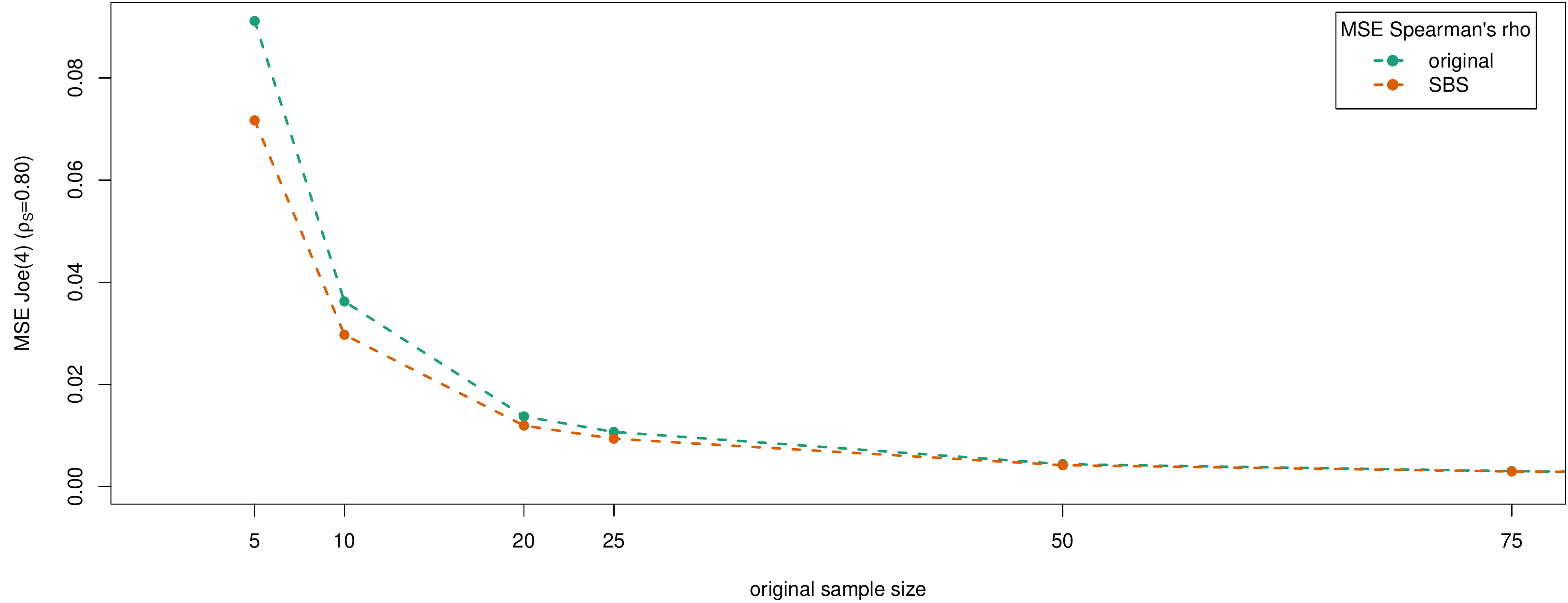}
\caption{Top: Estimated Spearman's rho for the bivariate Joe copula with parameter $\theta=4$.
The results are based on an original sample size of $\norig\in\{5,10,20,25,50,75\}$, while the augmented smooth bootstrap sample size is $\naugm=10\,000$.
Each boxplot is based on $M=2\,000$ independent reruns. The red dashed line indicates the theoretical value of $\rho_S$ while black dots indicate the means.\newline
Bottom: Mean squared error for estimation of Spearman's rho.
The results are based on an original sample size of $\norig\in\{5,10,20,25,50,75\}$, while the augmented smooth bootstrap sample size is $\naugm=10\,000$.}
\label{fig_Joe_Spearman}
\end{figure}
\begin{figure}
\centering
\includegraphics[width=\textwidth]{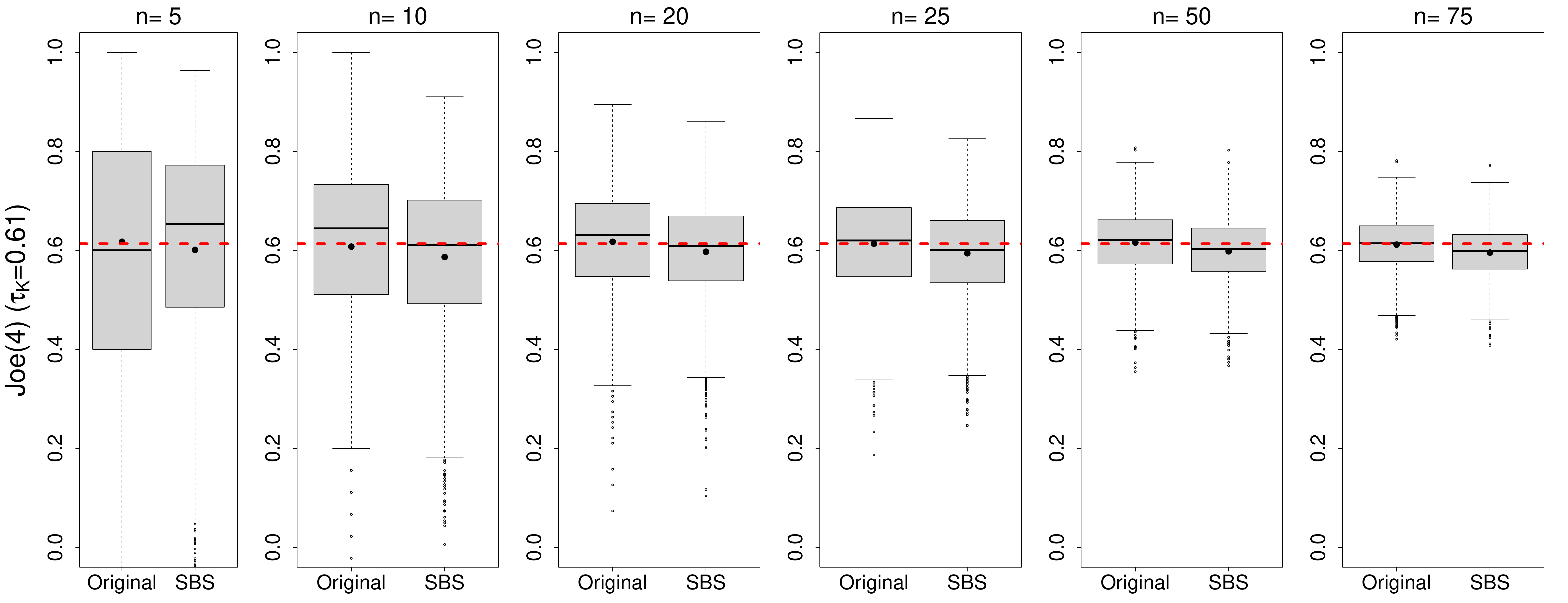}\\[4mm]
\includegraphics[width=\textwidth]{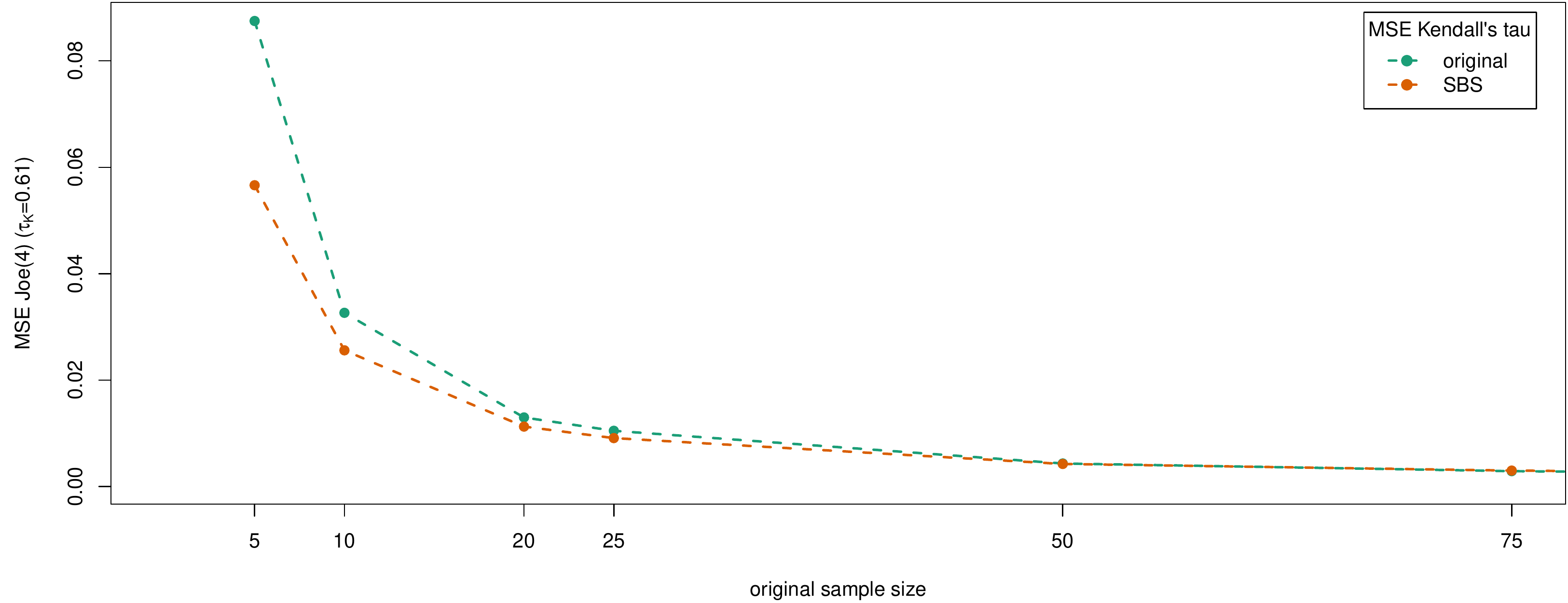}
\caption{Top: Estimated Kendall's tau for the bivariate Joe copula with parameter $\theta=4$.
The results are based on an original sample size of $\norig\in\{5,10,20,25,50,75\}$, while the augmented smooth bootstrap sample size is $\naugm=10\,000$.
Each boxplot is based on $M=2\,000$ independent reruns. The red dashed line indicates the theoretical value of $\tau_K$ while black dots indicate the means.\newline
Bottom: Mean squared error for estimation of Kendall's tau.
The results are based on an original sample size of $\norig\in\{5,10,20,25,50,75\}$, while the augmented smooth bootstrap sample size is $\naugm=10\,000$.}
\label{fig_Joe_Kendall}
\end{figure}
\begin{figure}
\centering
\includegraphics[width=\textwidth]{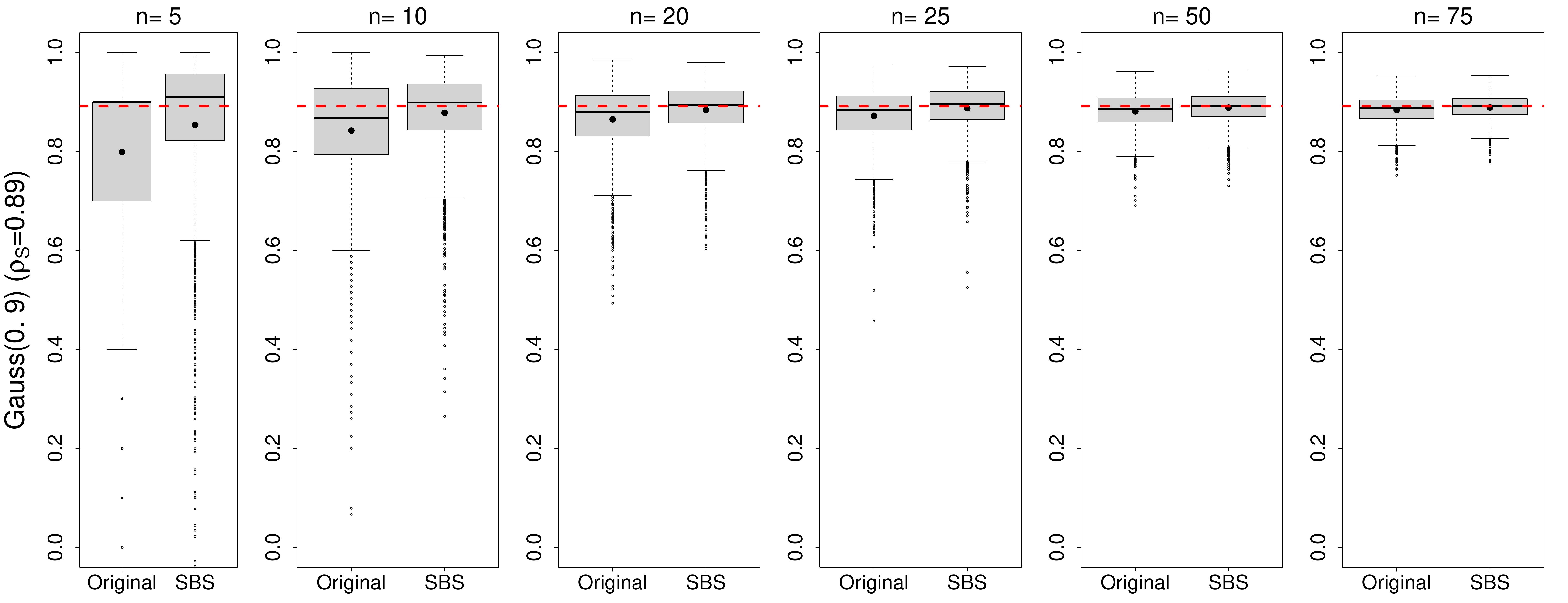}\\[4mm]
\includegraphics[width=\textwidth]{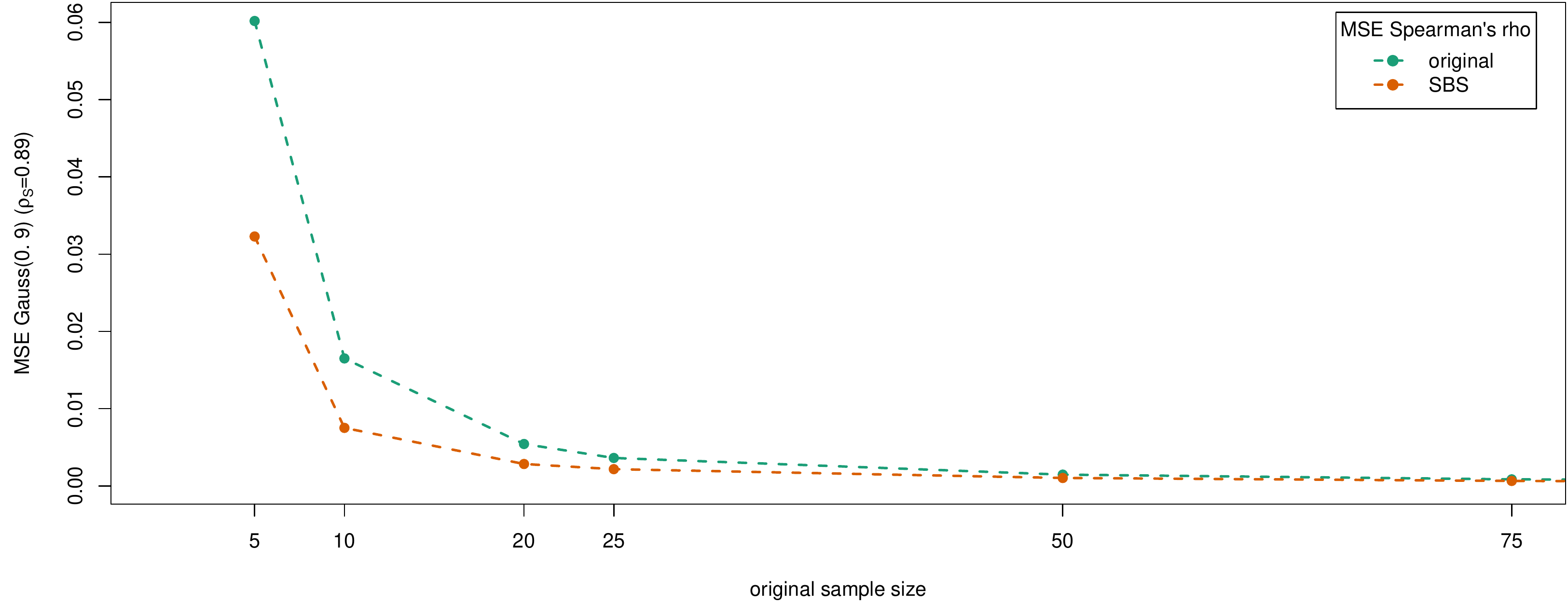}
\caption{Top: Estimated Spearman's rho for the bivariate Gaussian copula with $\rho=0.9$
The results are based on an original sample size of $\norig\in\{5,10,20,25,50,75\}$, while the augmented smooth bootstrap sample size is $\naugm=10\,000$.
Each boxplot is based on $M=2\,000$ independent reruns. The red dashed line indicates the theoretical value of $\rho_S$ while black dots indicate the means.\newline
Bottom: Mean squared error for estimation of Spearman's rho.
The results are based on an original sample size of $\norig\in\{5,10,20,25,50,75\}$, while the augmented smooth bootstrap sample size is $\naugm=10\,000$.}
\label{fig_Gauss_Spearman}
\end{figure}
\begin{figure}
\centering
\includegraphics[width=\textwidth]{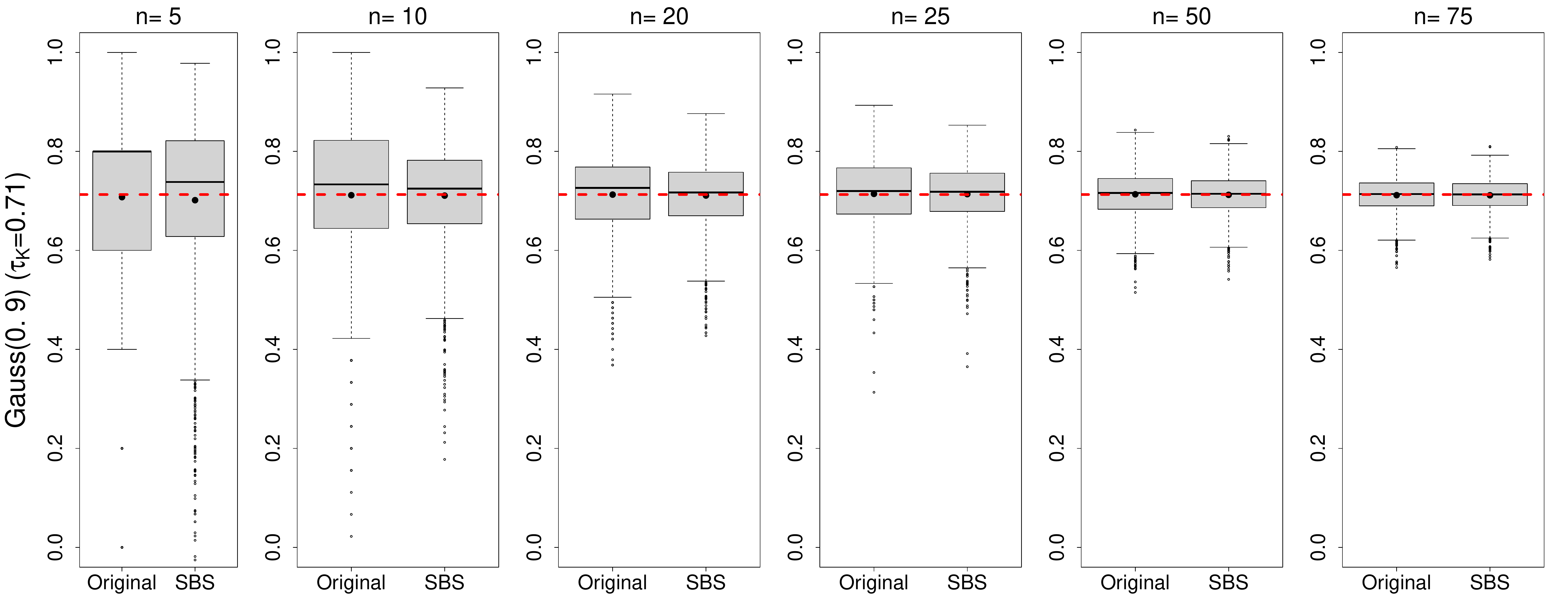}\\[4mm]
\includegraphics[width=\textwidth]{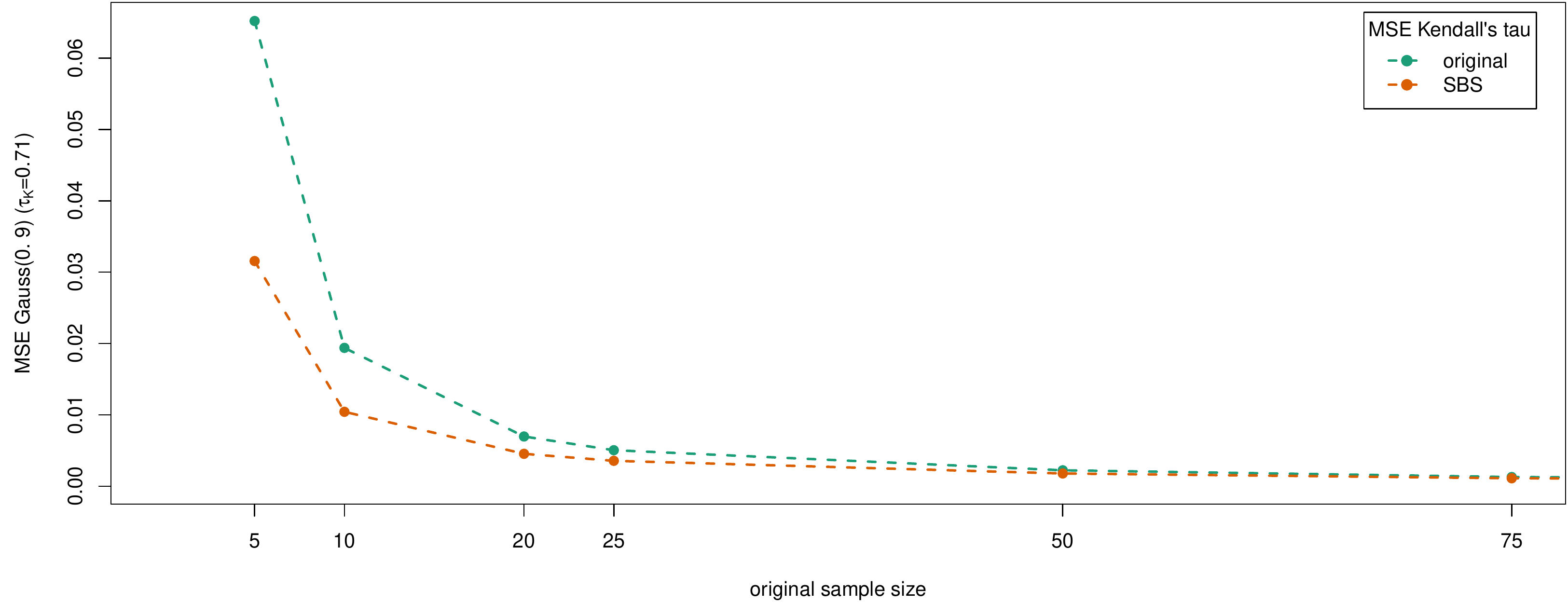}
\caption{Top: Estimated Kendall's tau for the bivariate Gaussian copula with $\rho=0.9$
The results are based on an original sample size of $\norig\in\{5,10,20,25,50,75\}$, while the augmented smooth bootstrap sample size is $\naugm=10\,000$.
Each boxplot is based on $M=2\,000$ independent reruns. The red dashed line indicates the theoretical value of $\tau_K$ while black dots indicate the means.\newline
Bottom: Mean squared error for estimation of Kendall's tau.
The results are based on an original sample size of $\norig\in\{5,10,20,25,50,75\}$, while the augmented smooth bootstrap sample size is $\naugm=10\,000$.}
\label{fig_Gauss_Kendall}
\end{figure}

\end{document}